\documentclass[useAMS,usenatbib]{mn2e}
\usepackage{graphicx}

\title[Scale-dependent non-Gaussianities in the WMAP data]
       {Scale-dependent non-Gaussianities in the WMAP data as
         identified by using surrogates and scaling indices}
         
\author[C. R\"ath et al.]
       {C. ~R\"ath$^1$\thanks{E-mail: cwr@mpe.mpg.de}, A. J. ~Banday$^{2,3,4}$, G. ~Rossmanith$^1$, H. ~Modest$^1$, 
        R. ~S\"utterlin$^1$,  \newauthor  K. M. ~G\'{o}rski$^{5,6}$, J. Delabrouille$^7$ and  G. E. ~Morfill$^1$\\
        $^1$ Max-Planck Institut f\"ur extraterrestrische Physik, Giessenbachstr. 1, 85748 Garching, Germany\\
        $^2$ Universit\'e de Toulouse; UPS-OMP; IRAP;  Toulouse, France \\
        $^3$ CNRS; IRAP; 9 Av. colonel Roche, BP 44346, F-31028 Toulouse cedex 4, France\\
        $^4$ Max-Planck-Institut f\"ur Astrophysik, Karl-Schwarzschild-Str. 1, 85741 Garching, Germany\\
        $^5$ Jet Propulsion Laboratory, California Institute of Technology, Pasadena, CA 91109, USA \\
        $^6$ Warsaw University Observatory, Aleje Ujazdowskie 4, 00 - 478 Warszawa, Poland \\
        $^7$ CNRS, Laboratoire APC, 10, Rue Alice Domon et L\'{e}onie Duquet, 75205 Paris, France}
        
\begin{document}

\date{Accepted ...
      Received .....;
      in original form ....}
\pagerange{\pageref{firstpage}--\pageref{lastpage}} \pubyear{2009}

\maketitle

\label{firstpage}

\begin{abstract}

We present a model-independent  investigation of the  
{ \it Wilkinson Microwave Anisotropy Probe (WMAP)}  data with respect to 
scale-independent and scale-dependent non-Gaussianities (NGs). To this end, we
employ the method of constrained randomization.  For  generating so-called 
surrogate maps a well-specified shuffling scheme
is applied to the Fourier phases of the original data, which allows 
to test for the presence of higher order correlations (HOCs) also and especially 
on well-defined scales.  \\
Using scaling indices as test statistics for the HOCs in the maps 
we find highly significant signatures for non-Gaussianities when considering all scales.
We test for NGs in four different $l-$bands $\Delta l$, namely in the 
bands $\Delta l = [2, 20]$, $\Delta l = [20, 60]$,
$\Delta l = [60, 120]$ and $\Delta l = [120, 300]$.
We find highly significant signatures for both non-Gaussianities and 
ecliptic hemispherical asymmetries for the 
interval $\Delta l = [2, 20]$ covering the large scales.
We also obtain highly significant deviations from Gaussianity for the band  $\Delta l = [120, 300]$.
The result for the full $l$-range can then easily be interpreted as a superposition of the 
signatures found in the bands $\Delta l = [2, 20]$ and  $\Delta l = [120, 300]$.
We find remarkably similar results when analyzing different ILC-like maps 
based on the WMAP three, five and seven year data.
We perform a set of tests to investigate whether and to what extend the detected anomalies 
can be explained by systematics.
While none of these tests can convincingly rule out the intrinsic nature of the anomalies for the
low $l$ case, the ILC map making procedure and/or residual noise in the maps can also lead 
to NGs at small scales.\\
Our investigations prove that there {\it are} phase correlations in the 
WMAP data of the CMB. In the absence of an explanation in terms
of Galactic foregrounds or known systematic artefacts, the signatures
at low $l$ must so far be taken to be cosmological at high
significance. These findings would strongly disagree with 
predictions of isotropic cosmologies with single field slow roll inflation.\\  
The task is now to elucidate the origin of the phase correlations  and to understand the physical
processes leading to these scale-dependent non-Gaussianities -- 
if it turns out that systematics as cause for them must be ruled out.

\end{abstract}


\section{Introduction}

The Cosmic Microwave Background (CMB) radiation represents the oldest observable signal 
in the Universe. Since this relic radiation has its origin just $380000$
years after the Big Bang when the CMB photons were last scattered off electrons, this radiation
is one of the most important sources of information to gain more knowledge 
about the very early Universe. Estimating the linear correlations of the temperature fluctuations 
in the CMB as measured e.g. with the WMAP satellite by means of the power spectrum 
has yielded very precise determinations of the  parameters of the standard $\Lambda$CDM cosmological model 
like the age, the geometry and the matter and energy content of the Universe \citep{komatsu09a,komatsu10a}.\\ 
Analyzing CMB maps by means of the power spectrum represents an enormous compression
of information contained in the data from approx. $10^6$ temperature values to roughly $1000$ numbers for the
power spectrum.   
It has often been pointed out  (\cite{komatsu09b} and references therein) that
this data compression is lossless and thus fully justified, if and {\it only if} the statistical distribution of the 
observed fluctuations is a Gaussian distribution with random phases.
Any information that is contained in the phases and the correlations among them, is not encoded in the 
power spectrum, but has to be extracted from measurements of higher-order correlation (HOC).
Thus, the presence of phase correlations may be considered as an unambiguous evidence of 
non-Gaussianity (NG). Otherwise, non-Gaussianity can only be defined by the negation of Gaussianity.\\
Primordial NG represents one way to test theories of inflation with the ultimate 
goal to constrain the shape of the potential of the inflaton field(s) and their possible (self-)interactions.
While the simplest single field slow roll inflationary scenario predicts that fluctuations are nearly Gaussian \citep{guth81a,linde82a,albrecht82a},
a variety of more complex models predict deviations from Gaussianity  \citep{linde97a,peebles97a,bernardeau02a,acquaviva03a}.
Models in which the Lagrangian is a general function of the 
inflaton and powers of its first derivative \citep{armendariz99a,garriga99a} 
can lead to scale-dependent non-Gaussianities, if the sound speed varies during inflation. 
Similarly, string theory models that give rise to large non-Gaussianity 
have a natural scale dependence \citep{chen05a,loverde08a}. 
Also, NGs put strong constraints on alternatives to the inflationary paradigm \citep{buchbinder08a,lehners08a}.\\
Given the plethora of conceivable scenarios for the very early Universe, it is worth first checking what is in the data 
in a model-independent way.
Further, such a model-independent approach has a large discovery potential to detect yet 
unexpected fingerprints of nonlinear physics in the early universe. 
Thus, a detection of possibly scale-dependent non-Gaussianity being encrypted in the phase correlations 
in the WMAP data would be of great interest. 
While a detection of non-Gaussianity could be indicative of an experimental systematic 
effect or of residual foregrounds,  it could also point to new cosmological physics.\\ 
The investigations of deviations from Gaussianity  in the CMB   (see \cite{komatsu09a} 
and references therein) and claims for the detection of 
non-Gaussianitiy and a variety of other anomalies like hemispherical asymmetries, lack of power at large angular scales, 
alignment of multipoles, detection of the Cold Spot etc.  (see e.g. 
\cite{park04a,eriksen04a, hansen04a, vielva04a, eriksen05a,eriksen07a,deoliveiracosta04a,raeth07a,mcewen08a,rossmanith09a,copi09a,copi10a, hansen09a,yoho10a})
have been made, where the statistical significance of some of the detected signatures is still subject to discussion  \citep{zhang10a, bennett10a}.
These studies have in common that the level of non-Gaussianity 
is assessed  by comparing the results for the measured data with
simulated CMB-maps which were generated on the 
basis of the standard cosmological model and/or specific assumptions 
about the nature of the non-Gaussianities as parametrized with e.g. 
the scalar, scale-independent parameter $f_{nl}$.
Other studies focused on the detection of signatures in the distribution of Fourier phases \citep{chiang03a,coles04a,naselsky05a,chiang07a}
representing deviations from the random phase hypothesis for Gaussian random fields.
These model-independent tests also revealed signatures of NGs.
Pursuing this approach one can go one step further and investigate 
possible phase correlations and  their relation to the morphology of the CMB maps 
by means of so-called surrogate maps.\\
This technique of surrogate data  sets \citep{theiler92} was originally developed for 
nonlinear time series analysis.
In this field of research complex systems like the climate, stock-market, 
heart-beat variability, etc. are analyzed (see e.g. \cite{bunde02a} and references therein).
For those systems a full modeling is barely or not possible. Therefore, statistical methods of constrained randomization 
involving surrogate data sets  were developed to infer some information about the nature of the underlying 
physical process in a completely data-driven, i.e. model-independent way. One of the first and most basic 
question here is whether a (quasiperiodic) process is completely linear or whether  also weak nonlinearities 
can be detected in the data.
The basic formalism to answer this question is to compute statistics sensitive to HOCs
for the original data set and for an ensemble of surrogate data sets,
which mimic the linear properties of the original time series while wiping out all phase correlations.
If the computed measure for the
original data is significantly different from the values obtained
for the set of surrogates, one can infer that the data contain HOCs.\\
Extensions of this formalism to three-dimensional  galaxy distributions \citep{raeth02a} and two-dimensional 
simulated flat CMB maps \citep{raeth03a} have been proposed and discussed. 
By introducing a more sophisticated two-step surrogatization scheme for full-sky CMB observations it has become possible 
to also test for scale-dependent NG in a model-independent way \citep{raeth09a}.  
Probing NG on the largest scales ($l<20$) yielded highly significant signatures for both NG and ecliptic hemispherical asymmetries.\\
In this paper, we apply the method of constrained randomization to the WMAP five year and seven year data in order 
to test for scale-independent and scale-dependent non-Gaussianity up to $l=300$ as encoded in the Fourier phase correlations.
Further, this work fully recognises the need to rule out foregrounds and systematic artefacts as the
origin of the detections (as advised by \cite{bennett10a}).
Therefore, a large part of our analyses is dedicated to various checks on systematics 
to single out possible causes of the detected anomalies.\\
The paper is organized as follows: 
In Section 2 we briefly describe the observational and simulated data we use in our study.
The method of constrained randomization is reviewed in some detail in Section 3. 
Scaling indices, which we use as test statistic, and the statistics derived out of them are 
discussed in Section 4. In Section 5 we present our results and 
we draw our conclusions in Section 6.
  
\section{Data Sets}

We used the seven years foreground-cleaned internal linear combination (ILC) map  \citep{gold10a}
generated and provided by the WMAP team\footnote{http://lambda.gsfc.nasa.gov} (in the following: ILC7).
For comparison we also included the map produced by \citet{delabrouille09a}, namely the 
five years needlet-based ILC map, which has been shown to be significantly less contaminated by
foreground and noise than other existing maps obtained from WMAP data (in the following: NILC5).\\
To check for systematics we also analyzed the following set of maps:\\
1) {\it Uncorrected ILC map}\\
The ILC map is a weighted linear combination of the 5 frequency channels that recovers the CMB signal. The weights are derived 
by requiring minimum variance in a given region of the sky under the constraint that the sum of the weights is unity. 
Such weights, however, cannot null an arbitrary foreground signal with a non-blackbody frequency spectrum, 
thus some residuals due to Galactic emission will remain. The WMAP team attempts to correct for 
this "bias" with an estimation of the residual signal based on simulations and a model of the 
foreground sky. Our uncorrected map (UILC7 in the following) is simply the ILC without applying this correction, 
computed from the weights provided in  \cite{gold10a} and the 1-degree smoothed WMAP data.\\  
2){\it Asymmetric beam map}\\ 
Beam asymmetries may result in statistically anisotropic CMB maps . To asses these effects
on the signatures  of scale-dependent NGs and their (an-)isotropies we make use 
of the publicly available CMB sky simulations including the effects of asymmetric beams \citep{wehus09a}. Specifically we 
analyse a simulated map of the V1-band, because this band is considered to have the
least foreground contamination.\\
3) {\it Simulated coadded VW-band map}\\
To make sure that neither systematic effects are induced by the method of constrained randomization nor the WMAP-like 
beam and noise properties lead to systematic deviations from Gaussianity we include in our analysis a co-added VW-map
as obtained using the standard  $\Lambda$CDM best fit power spectrum and WMAP-like beam and noise properties. 
Note that this map did not undergo the ILC-map making procedure.\\
4)  {\it Simulated ILC map}\\
Simulated sky maps result from processing a simulated differential time-ordered data (TOD) stream 
through the same calibration and analysis pipeline that is used for the flight data. The TOD is generated 
by sampling a reference sky that includes both CMB and Galactic foregrounds with the actual flight pointings, 
and adding various instrumental artefacts. We have then processed the individual resulting data into 7 separate 
simulated yearly ILC maps, plus a 7-year merge. It is worth noting that, if the yearly frequency-averaged maps are 
combined into ILCs using the  \cite{gold10a}  7-year weights per region, then the resulting ILCs show clear Galactic plane 
residuals. This reflects the fact that the simulated data has a different CMB realisation to the observed sky, and may 
additionally represent a mismatch between the simulated foreground properties and the true sky in the Galactic 
plane. Instead, we analyse the 7-year merged simulated data to compute the ILC weights for the simulations, 
then apply to all yearly data sets separately. However, the derived weights are quite different from the WMAP7 
ones, which would imply different noise properties in the simulated ILC data compared to the real data. 
Care should be exercised for any results that are sensitive to the specific noise pattern.\\
5) {\it Difference ILC map}\\
Finally, we consider the difference map (year 7 - year 6) from yearly ILC-maps computed using the same 
weights and regions as the 7-year data set from   \cite{gold10a}. 
No debiasing has been applied. With this map we estimate what effect possible 
ILC-residuals may have on the detection of NGs.     

\begin{figure*}
\centering

\includegraphics[width=3.4cm, keepaspectratio=true]{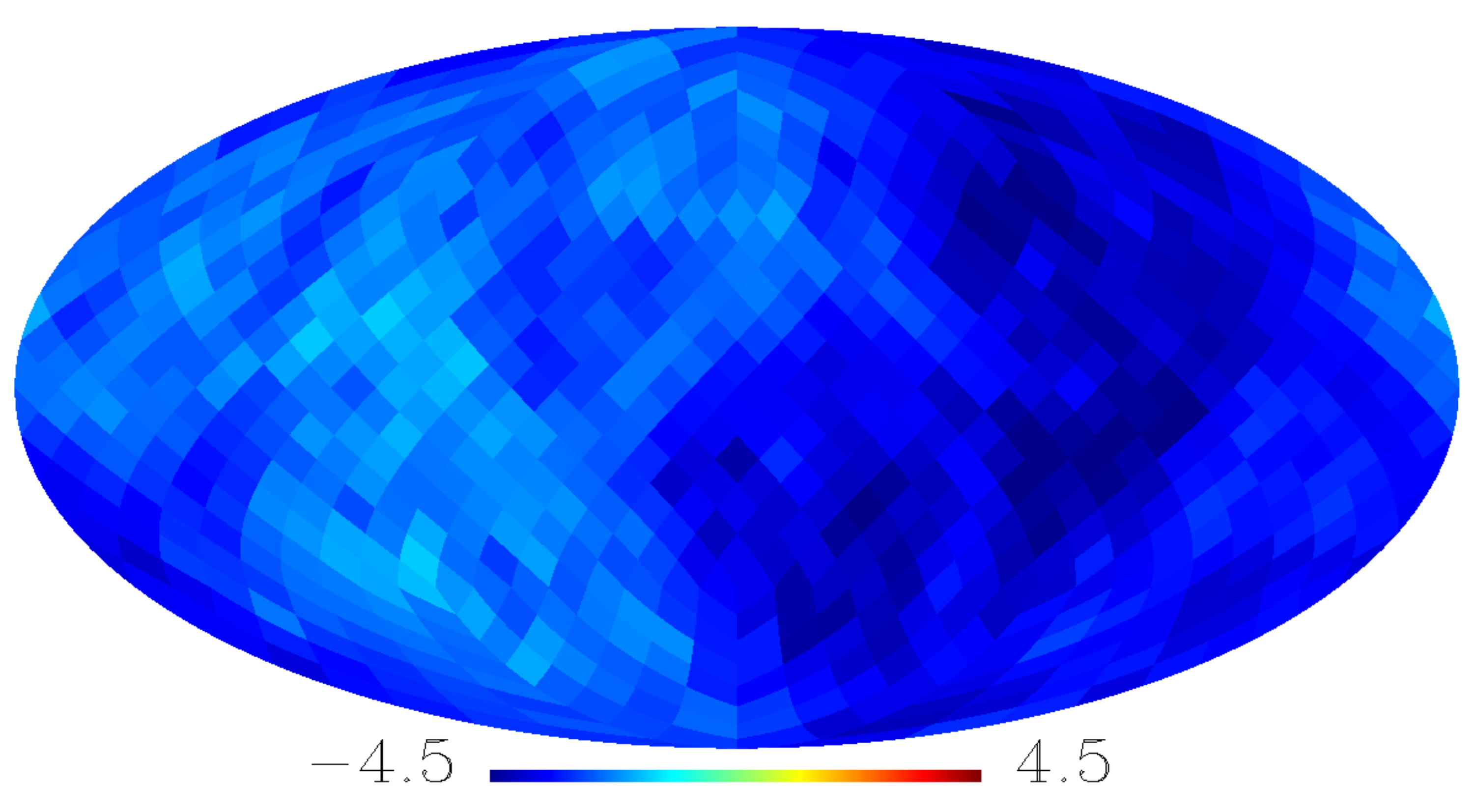} \hspace{0.25cm}
\includegraphics[width=3.4cm, keepaspectratio=true]{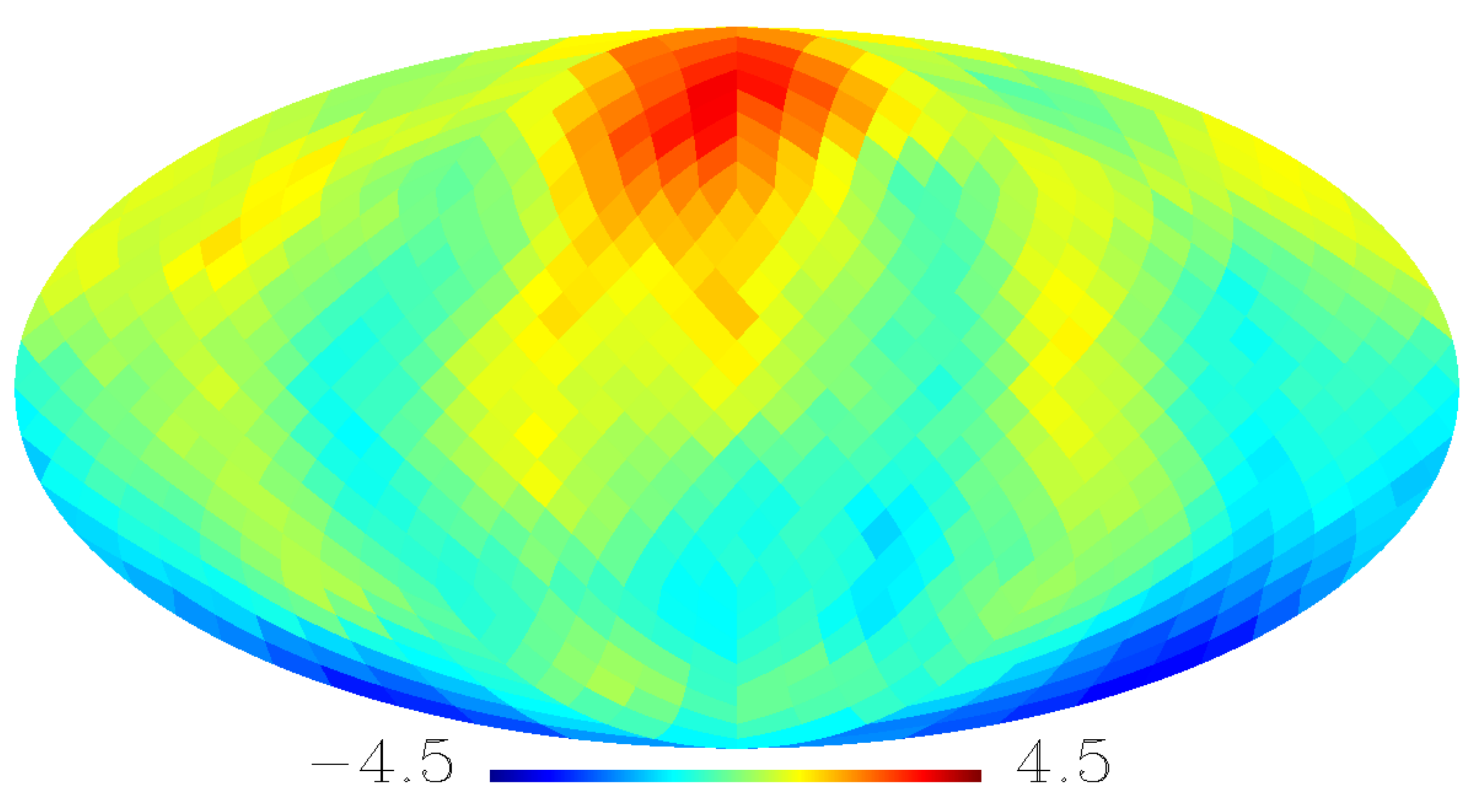}
\includegraphics[width=3.4cm, keepaspectratio=true]{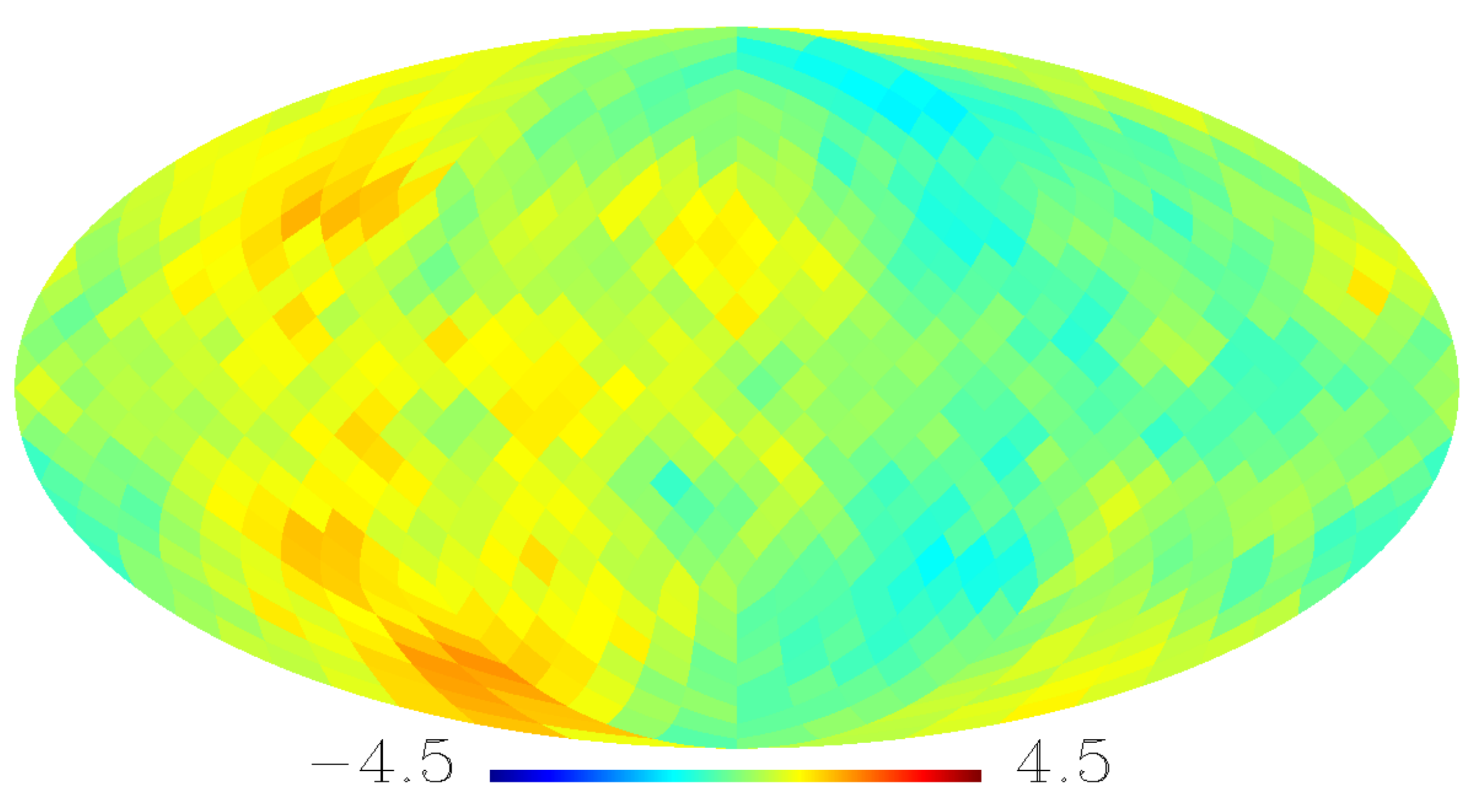}
\includegraphics[width=3.4cm, keepaspectratio=true]{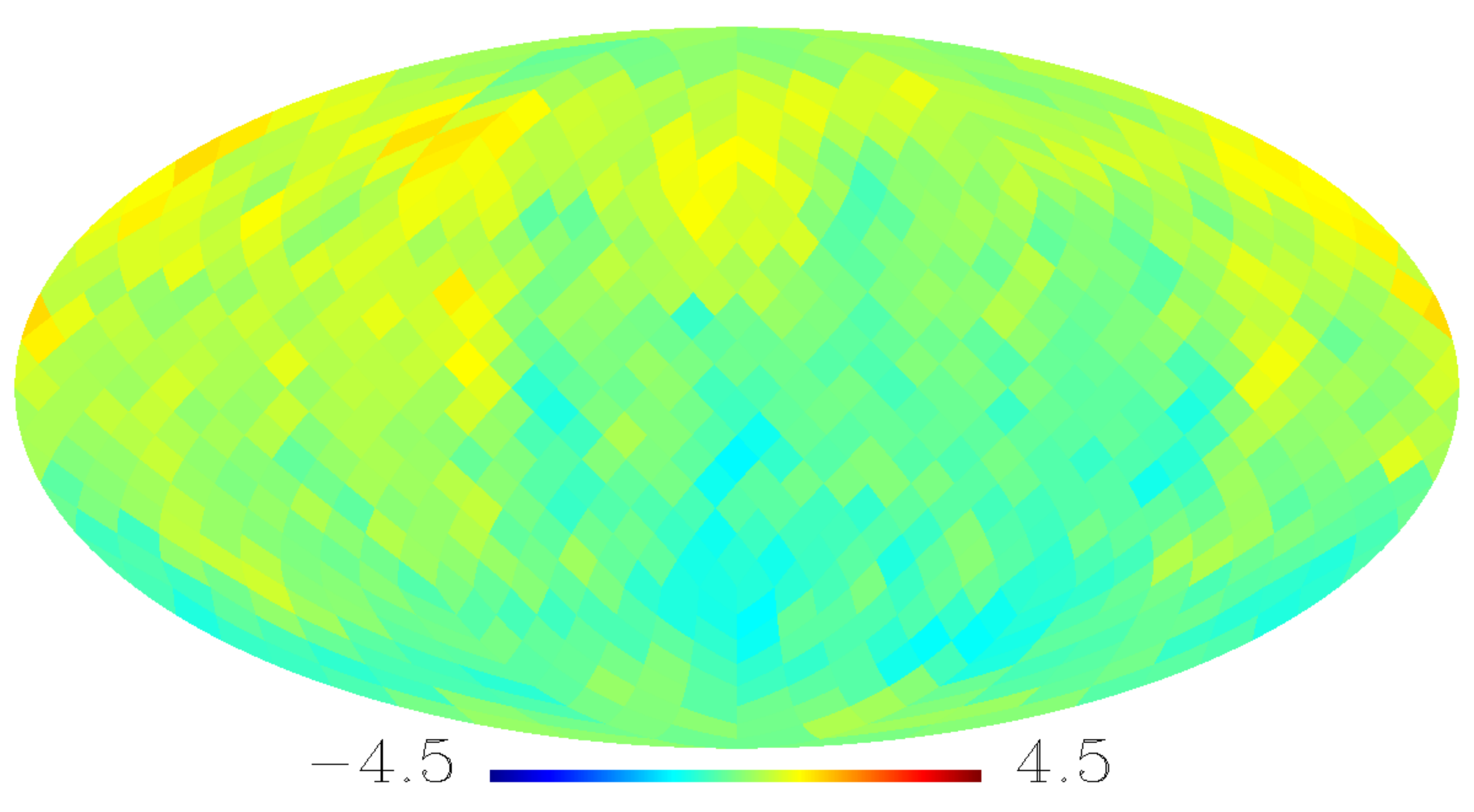}
\includegraphics[width=3.4cm, keepaspectratio=true]{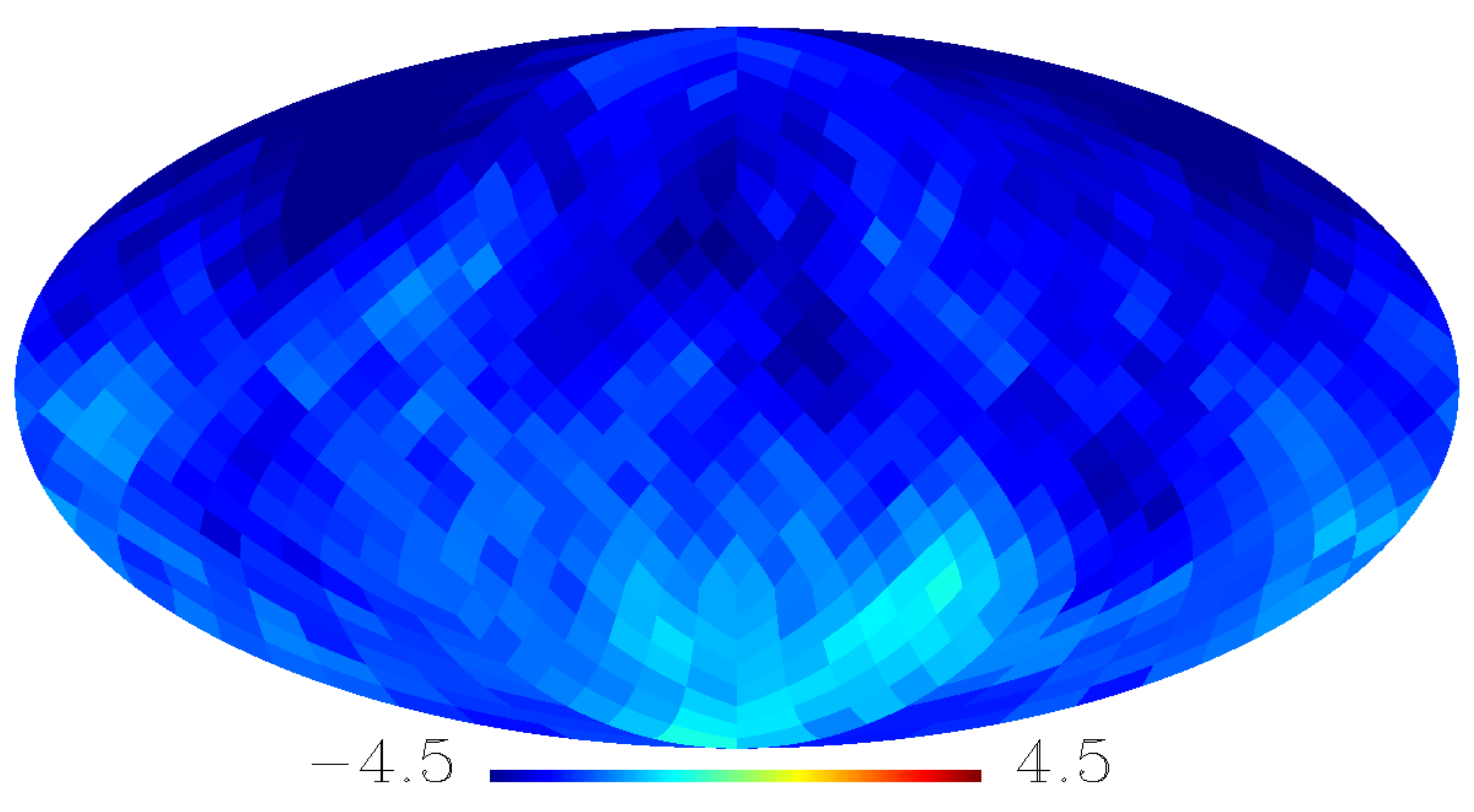}
\
\includegraphics[width=3.4cm, keepaspectratio=true]{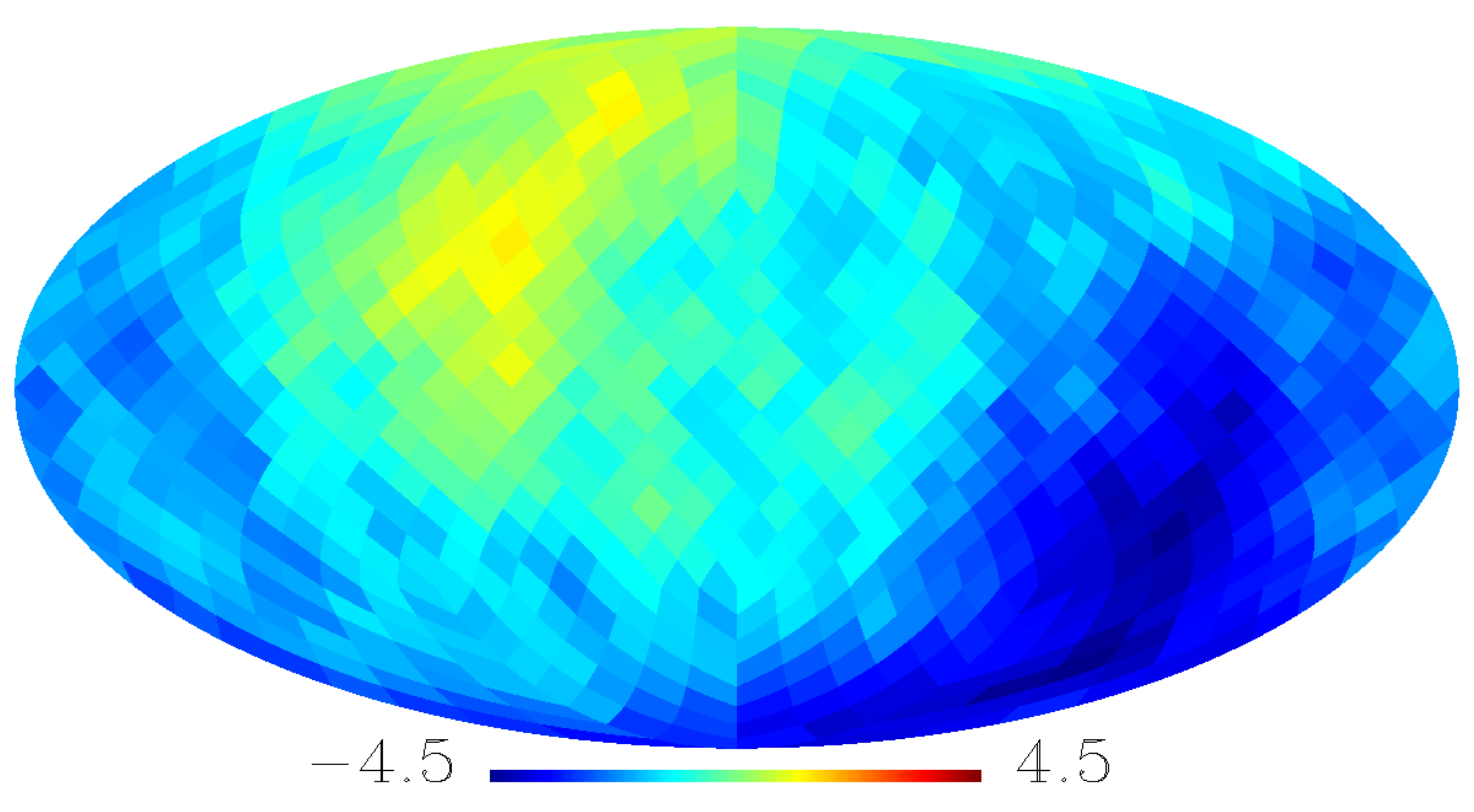}  \hspace{0.25cm}
\includegraphics[width=3.4cm, keepaspectratio=true]{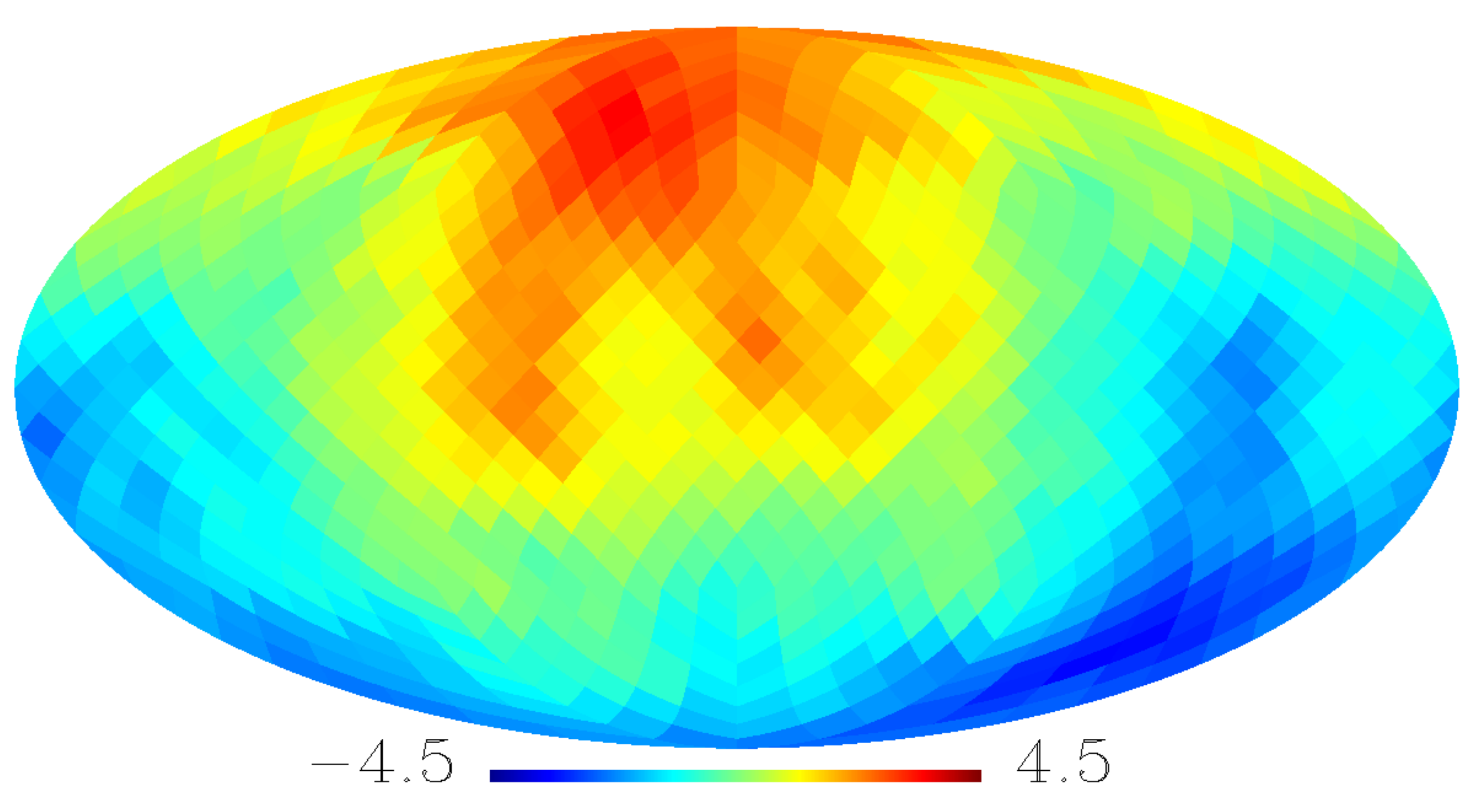}
\includegraphics[width=3.4cm, keepaspectratio=true]{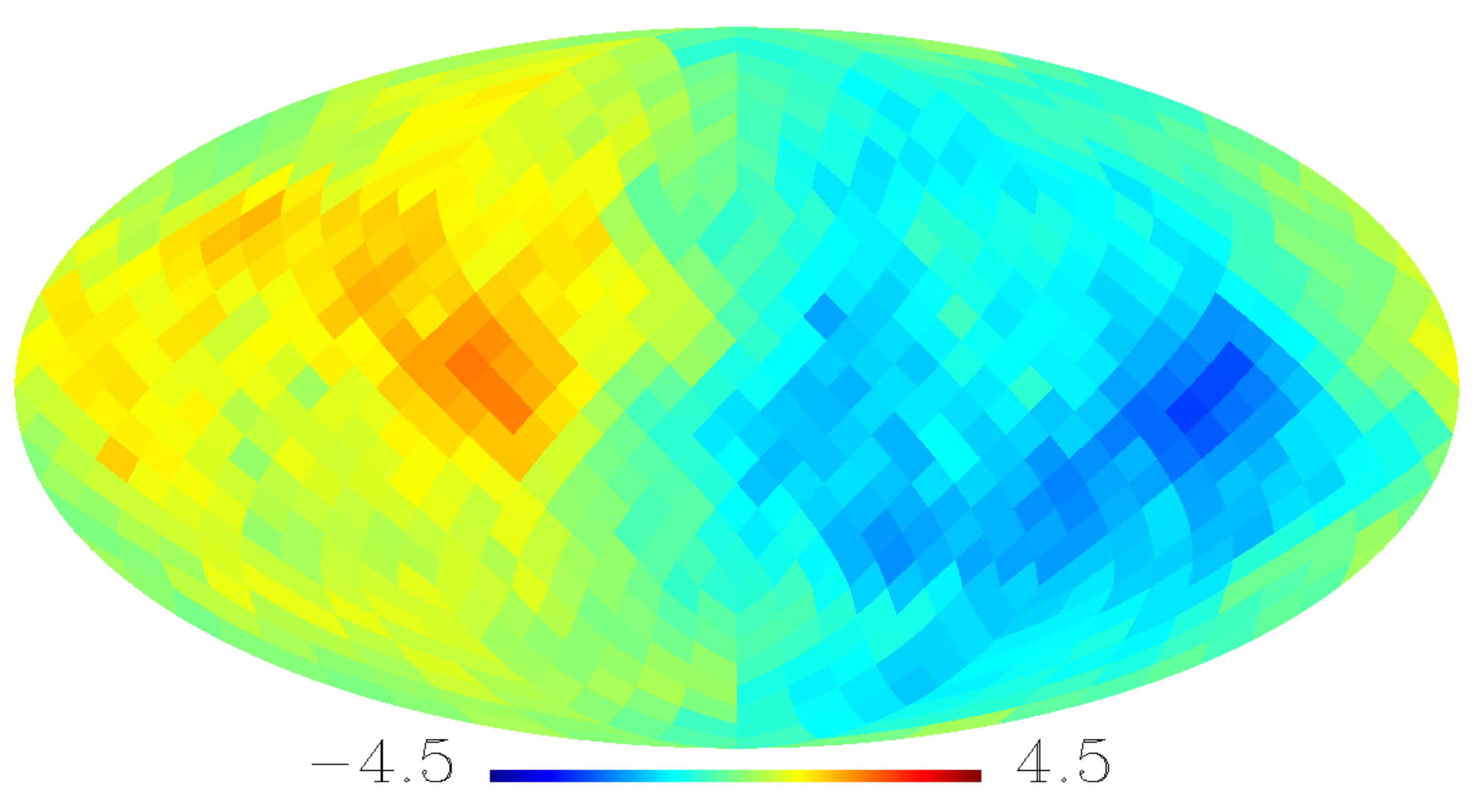}
\includegraphics[width=3.4cm, keepaspectratio=true]{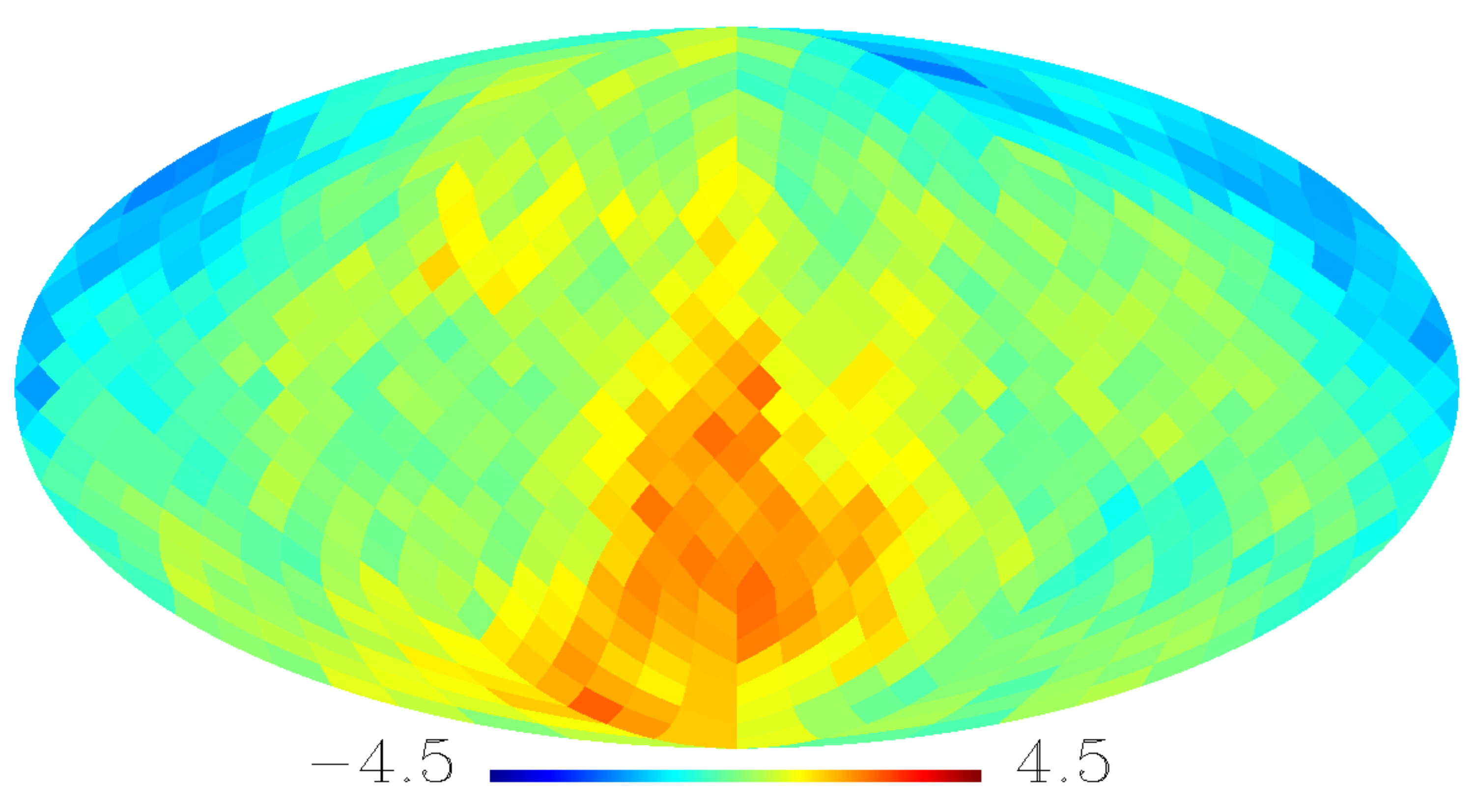}
\includegraphics[width=3.4cm, keepaspectratio=true]{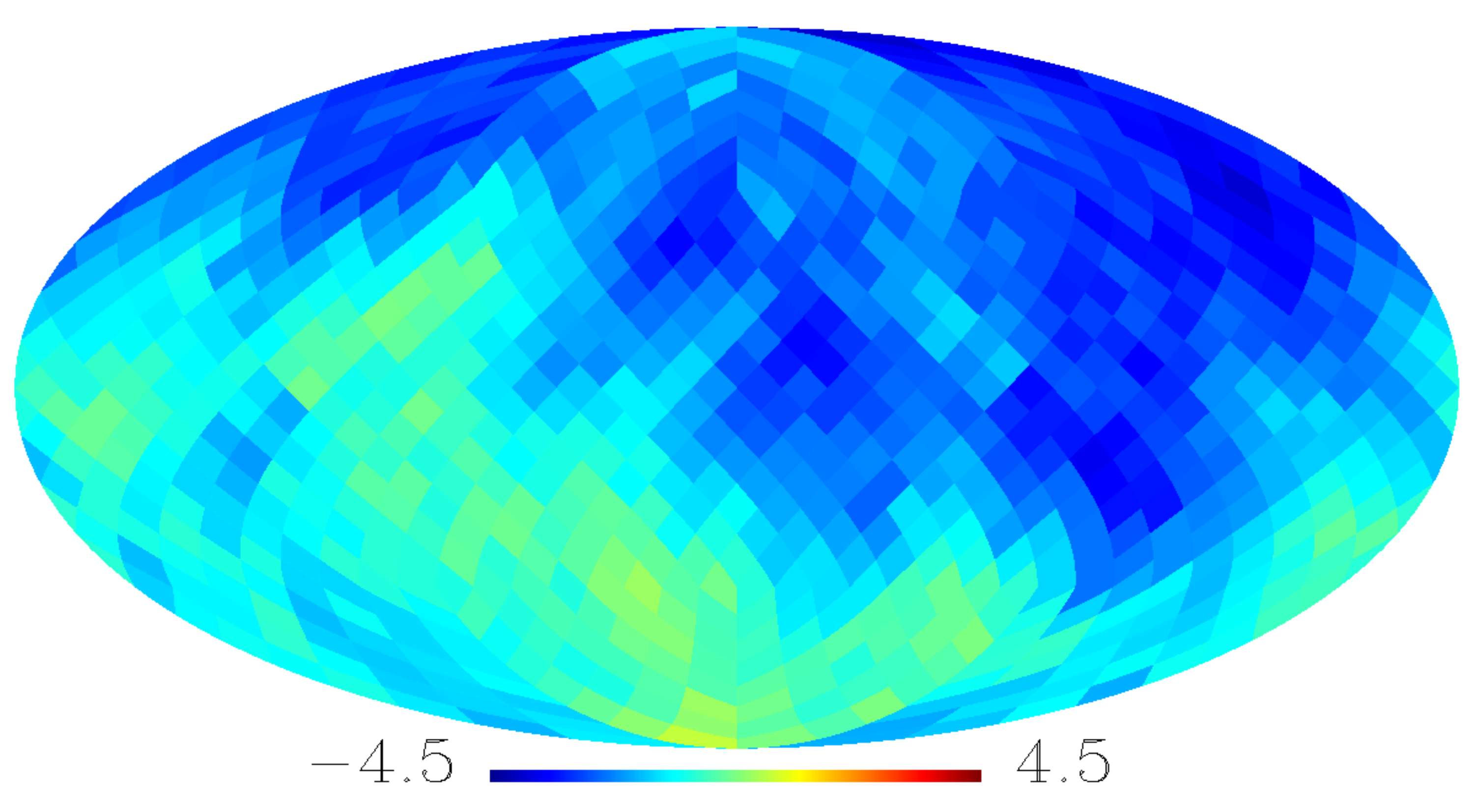}

\includegraphics[width=3.4cm, keepaspectratio=true]{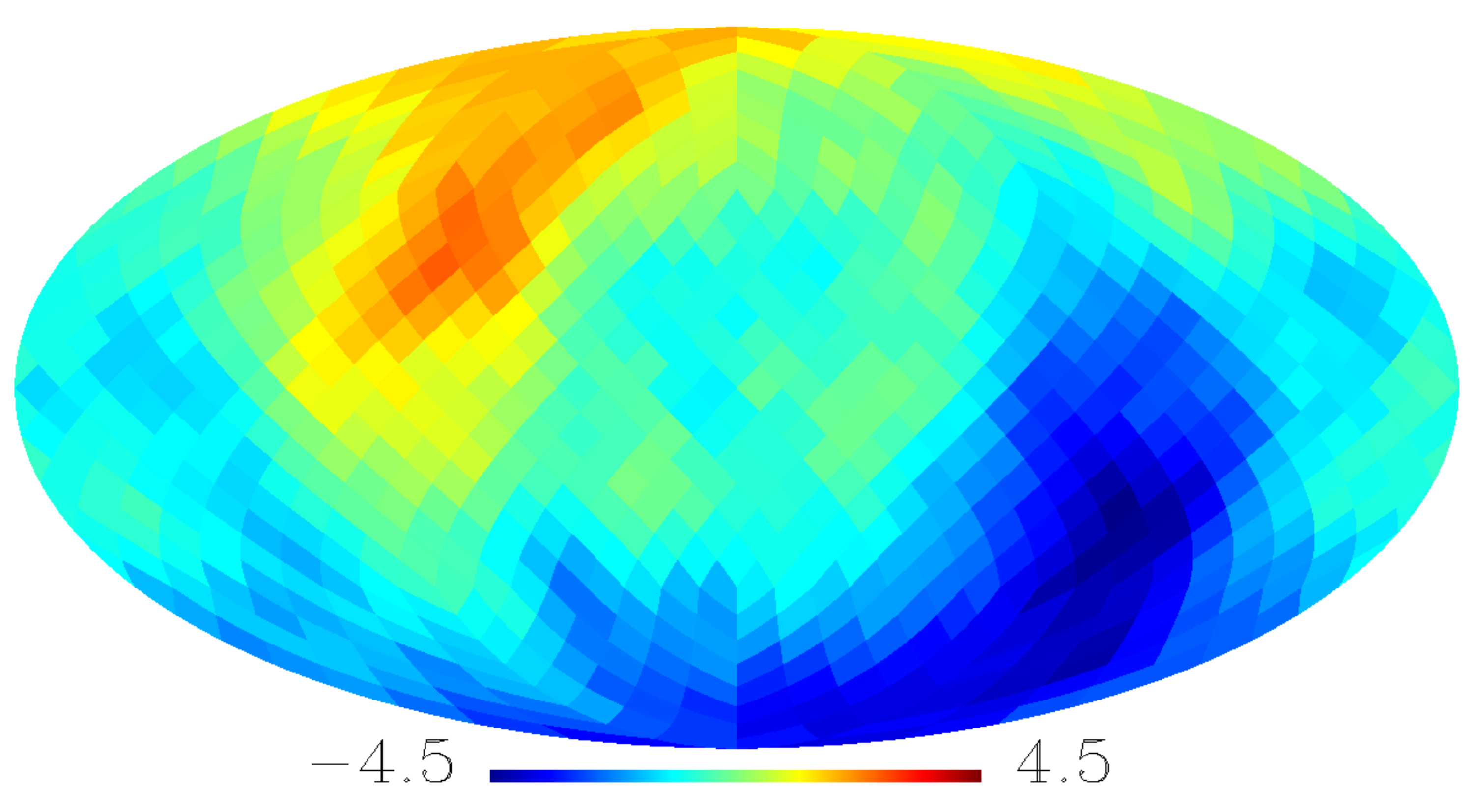}  \hspace{0.25cm}
\includegraphics[width=3.4cm, keepaspectratio=true]{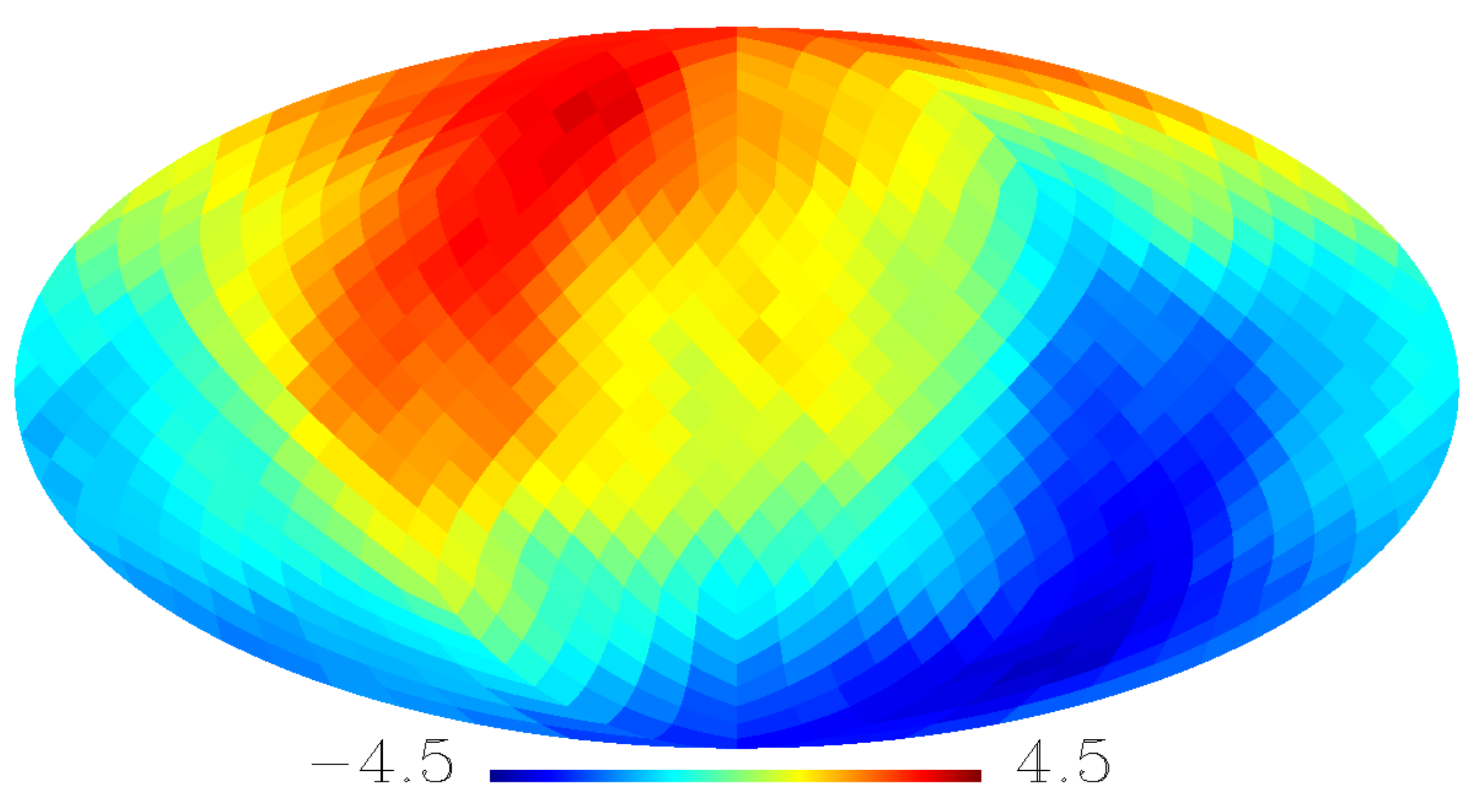}
\includegraphics[width=3.4cm, keepaspectratio=true]{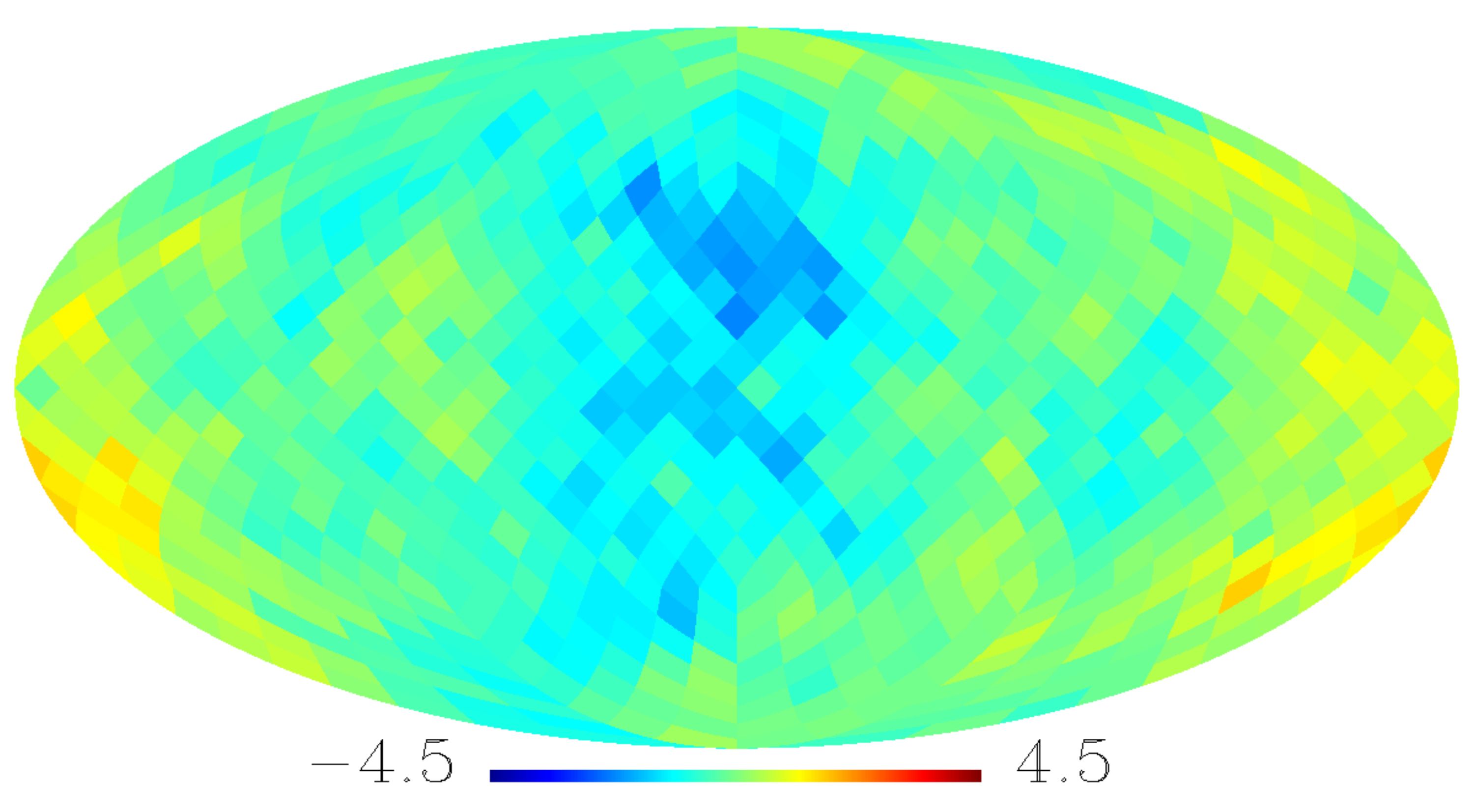}
\includegraphics[width=3.4cm, keepaspectratio=true]{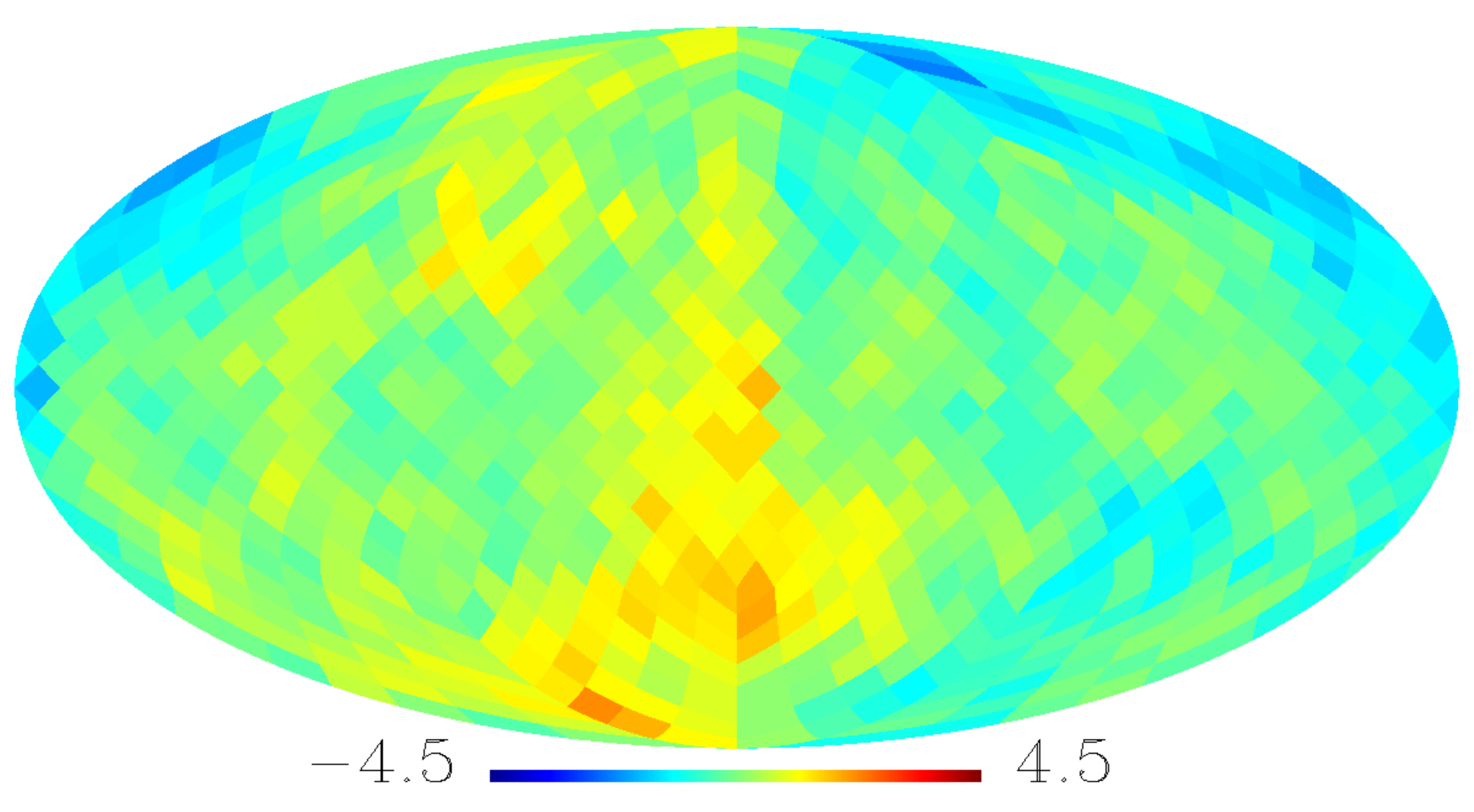}
\includegraphics[width=3.4cm, keepaspectratio=true]{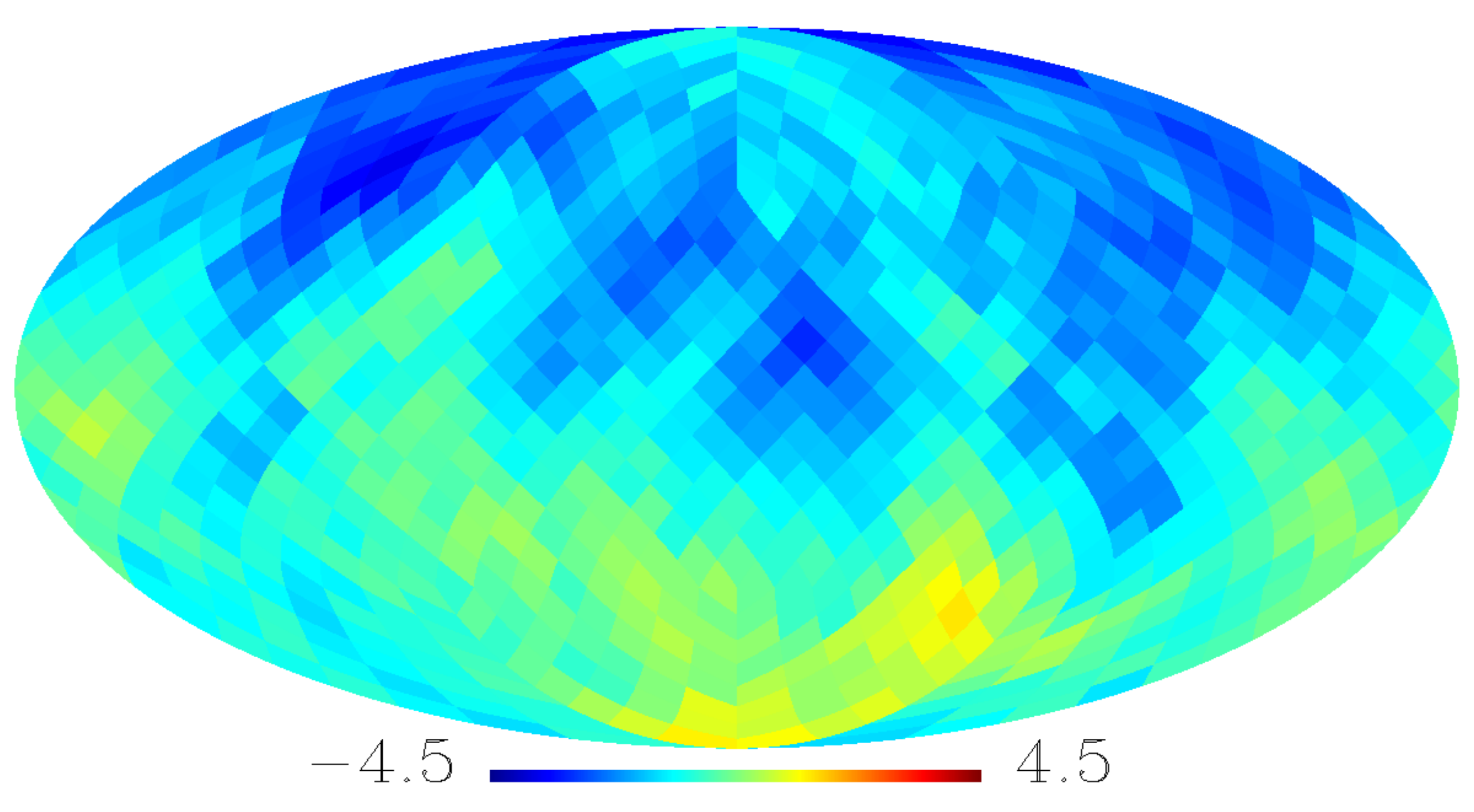}

\caption{Deviations $S(\langle \alpha (r_{k}) \rangle)$ of the rotated hemispheres for three scales $r_{k}, k=2,6,10$ 
(from top to bottom) for the ILC7 map and for (from left to right) the shuffling intervals $\Delta l = [2,1024]$, 
$\Delta l = [2,20]$, $\Delta l = [20,60]$, $\Delta l = [60,120]$ and
$\Delta l = [120,300]$. The expected correspondence between the shuffling range $\Delta l$ and the scales $r_k$ of the 
scale-dependent higher order statistics $\langle \alpha(r_k) \rangle $, for which 
the largest deviations are detected, becomes apparent.
While the ecliptic hemispherical asymmetries for $\Delta l = [2,20]$ are most pronounced for the 
largest scaling range $r_{10}$ (second column), the deviation $S$
becomes largest for $r_{2}$ when shuffling the phases of the smallest scales   $\Delta l = [120,300]$ (rightmost column).} \label{fig3}
\end{figure*}

\section{Generating Surrogate Maps}

To test for scale-dependent non-Gaussianities in a model-independent way
we apply a two-step procedure that has been proposed and 
discussed in \cite{raeth09a}.  Let us describe the various steps for generating
surrogate maps in more detail:\\
Consider a CMB map $T(\theta,\phi)$,
where $T(\theta,\phi)$ is Gaussian distributed and its Fourier 
transform. The Fourier coefficients $a_{lm}$ can be written as
$a_{lm} = | a_{lm} | e^{i \phi_{lm}} $ 
with  $\phi_{lm}=\arctan \left( {\rm Im}(a_{lm}) / {\rm Re}(a_{lm} )  \right)$.
The linear or Gaussian properties of the underlying random field 
are contained in the absolute values $ | a_{lm} | $, 
whereas all HOCs -- if present -- 
are encoded in the phases $\phi_{lm}$ and the correlations among them.
Having this in mind, a versatile approach for testing for scale dependent non-Gaussianities 
relies on a scale-dependent 
shuffling procedure of the phase correlations followed by a statistical comparison 
of the so-generated surrogate maps. \\ 
However, the Gaussianity of the temperature distribution and the 
randomness of the set of Fourier phases in the sense that they are uniformly 
distributed in the interval $[- \pi, \pi]$, are a necessary 
prerequisite for the application of the surrogate-generating 
algorithm, which we propose in the following.
To fulfill these two conditions, we perform the following preprocessing steps.
First, the maps are remapped onto a Gaussian distribution in 
a rank-ordered way. This means that the amplitude distribution of the original temperature map 
in real space is replaced by a Gaussian distribution in a way that the rank-ordering is preserved, 
i.e. the lowest value of the original distribution is replaced with the lowest value of the
Gaussian distribution etc.
By applying this remapping we automatically focus on HOCs induced by the 
spatial correlations in the data while excluding any effects coming from 
deviations of the temperature distribution from a Gaussian one. \\ 
To ensure the randomness of the set of 
Fourier phases we performed a rank-ordered  remapping of the phases 
onto a set of uniformly distributed ones followed
by an inverse Fourier transformation. 
These two preprocessing steps only have marginal influence to the maps.
The main effect is that the
outliers in the temperature distribution are removed.
Due to the large number of temperature values (and phases) we 
did not find any significant dependence of the specific Gaussian (uniform)  
realization used for remapping of the temperatures (phases).
The resulting maps may already be considered as a surrogate map 
and we named it zeroth order surrogate map.
The first and second order surrogate maps are obtained as follows:\\
We first generate a first order surrogate map, 
in which any phase correlations 
for the scales, which are not of interest, 
are randomized.  This is achieved by a random shuffle of the phases  $\phi_{lm}$ 
for $l \notin  \Delta l = [l_{min}, l_{max}], 0 < m \le l$ and
by performing an inverse Fourier transformation.\\
In a second step, $N$ ($N=500$ throughout this study) 
realizations of second order surrogate maps 
are generated for the first order surrogate map, in which the remaining 
phases $\phi_{lm}$  with $l \in  \Delta l$,$0 < m \le l$ are shuffled, while the 
already randomized phases for the scales, which are not under consideration, 
are preserved.
Note that the  Gaussian properties  of 
the maps, which are given by $| a_{lm} |$,
are {\it exactly} preserved in all surrogate maps.\\
So far, we have applied the method  of surrogates only to 
the $l$-range  $\Delta l = [2, 20]$. In this paper we will repeat
the investigations for this $l$-interval but using newer CMB maps.
Furthermore,  we extend the analysis to smaller scales. Namely, we
consider three more $l$-intervals $\Delta l = [20, 60]$,
$\Delta l = [60, 120]$ and $\Delta l = [120, 300]$.
The choice of $60$ as $l_{min}$ and  $l_{max}$ is somewhat 
arbitrary, whereas the $l_{min}=120$ and  $l_{max}=300$ for 
the last $l$-interval was selected in such a way that the first peak in the 
power spectrum is covered. Going to even higher $l$'s doesn't make much 
sense, because the ILC7 map is smoothed to $1$ degree FWHM.
Some other maps which we included in our study -- especially NILC5 -- are not 
smoothed and we could in principle go to higher $l$'s. But to allow for a consistent 
comparison of the results obtained with the different observed and simulated input maps
we restrict ourselves to only investigate $l$-intervals up to  $l_{max}=300$ in this study.    \\
Besides this two-step procedure aiming at a dedicated 
scale-dependent search of non-Gaussianity, we also 
test for non-Gaussianity using surrogate maps
without specifying certain scales. In this case there are no
scales, which are not of interest, and the first step in the surrogate map
making procedure becomes dispensable. 
The zeroth order surrogate map is to be considered here as first 
order surrogate and the second order surrogates are generated by
shuffling all phases with $0 < m \le l$ for all available $l$'s, i.e. in our 
case $\Delta l = [2, 1024]$.\\
Finally, for calculating scaling indices to test for higher order correlations 
the surrogate maps were degraded to $N_{side}=256$ and residual monopole 
and dipole contributions were subtracted. 
The  statistical comparison of the two classes of surrogates will reveal, 
whether possible HOCs on certain scales have left traces
in the first order surrogate maps, which were then deleted in the second
order surrogates.  
Before the results of such a comparison of the surrogate maps are shown in 
detail, we review the formalism of scaling indices. 

\section{Weighted Scaling Indices and Test Statistics}

As test statistics for detecting and assessing possible scale-dependent 
non-Gaussianities in the CMB data
weighted scaling indices are calculated \citep{raeth02a, raeth03a}. 
The basic ideas of the scaling index method (SIM) stem from the calculation of the dimensions of 
attractors in nonlinear time series analysis \citep{grassberger83a}. 
Scaling indices essentially represent  one way to estimate the local scaling properties of a point set in an 
arbitrary  $d$-dimensional embedding space. The technique offers the possibility of revealing local structural 
characteristics of a given point distribution. Thus, point-like, string-like and sheet-like structures  
can be discriminated from each other and from a random background. The alignment of e.g. string-like structures
can be detected by using a proper metric for calculating the distances between the  points \citep{raeth08a, suetterlin09a}.\\
Besides the countless applications in time series analysis the use of
scaling indices has been extended to the field of image processing for texture discrimination \citep{raeth97a} 
and feature extraction \citep{jamitzky01a, raeth08a} tasks. 
Following further this line we performed several non-Gaussianity  studies of the CMB  based on WMAP data 
using scaling indices in recent years \citep{raeth03a, raeth07a,raeth09a, rossmanith09a}. \\
Let us review the formalism for calculating this test statistic for  assessing HOCs:\\
In general, the SIM is a mapping that calculates for every point $\vec{p}_i, i=1,\ldots, N_{pix}$ of a point set $P$ a single value, 
which depends on the spatial position of $\vec{p}_i$ relative to the group of other nearby points, in which the 
point under consideration is embedded in. 
Before we go into the details of assessing the local scaling properties, let us first of all outline the 
steps of generating a point set $P$ out of observational CMB-data.
To be able to apply the SIM on the spherical CMB data, we have to transform the pixelised sky $S$ 
with its pixels at positions $(\theta_i,\phi_i)$, $i=1,...,N_{pix}$, on the unit sphere to a point-distribution in an artificial embedding space. 
One way to achieve this is by transforming each temperature value $T(\theta_i,\phi_i)$ to a radial jitter around a sphere of radius $R$
at the position of the pixel center $(\theta_i,\phi_i)$. 
Formally, the three-dimensional  position vector of the point $\vec{p}_i$ reads as
\begin{eqnarray}
x_i &= & (R+dR) \cos(\phi_i) \sin(\theta_i) \\
y_i &= & (R+dR) \sin(\phi_i) \sin(\theta_i) \\
z_i &= & (R+dR) \cos(\theta_i)
\end{eqnarray}
with
\begin{equation} 
dR = a \left( \frac{T(\theta_i,\phi_i) - \langle T \rangle}{\sigma_T} \right) \;.
\end{equation}
Hereby, $R$ denotes the radius of the sphere while $a$ describes an adjustment parameter. 
The mean temperature and its standard deviation are characterised by $\langle T \rangle$ 
and $\sigma_T$, respectively. By the use of the normalisation we obtain for $dR$ zero mean 
and a standard deviation of $a$. Both $R$ and $a$ should be chosen properly to ensure a high 
sensitivity of the SIM with respect to the temperature fluctuations at a certain spatial scale. 
For the analysis of WMAP-like CMB data, it turned out that this requirement is provided using $R=2$ for the radius of the sphere and coupling the 
adjustment parameter $a$ to the value of the below introduced scaling range parameter $r$ via $a=r$ \citep{raeth07a}.
Now that we obtained our point set $P$, we can apply the SIM. 
For every point $\vec{p}_i$ we calculate the local weighted cumulative point distribution which is defined as
\begin{equation}
\rho(\vec{p}_i,r) = \sum_{j=1}^{N_{pix}} s_r(d(\vec{p}_i,\vec{p}_j))
\end{equation}
with $r$ describing the scaling range, while $s_r(\bullet)$ and $d(\bullet)$ denote a shaping function 
and a distance measure, respectively. The scaling index $\alpha(\vec{p}_i,r)$ is then defined as the 
logarithmic derivative of $\rho(\vec{p}_i,r)$ with respect to $r$:
\begin{equation}
\alpha(\vec{p}_i,r) = \frac{\partial \log \rho(\vec{p}_i,r)}{\partial \log r} \;.
\end{equation}
As mentioned above,  $s_r(\bullet)$ and $d(\bullet)$ can in general  be chosen arbitrarily. 
For our analysis we use a quadratic gaussian shaping function $s_r(x) = e^{-(\frac{x}{r})^2}$ 
and an isotropic euclidian norm  $d(\vec{p}_i,\vec{p}_j) = \| \vec{p}_i - \vec{p}_j \|$ as distance measure.
With this specific choice of $s_r(\bullet)$ and $d(\bullet)$
we obtain the following analytic formula for the scaling indices
\begin{equation}\label{AlphaFormel}
\alpha(\vec{p}_i,r) = \frac{\sum_{j=1}^{N_{pix}} 2\big(\frac{d_{ij}}{r}\big) e^{-\big(\frac{d_{ij}}{r}\big)^2}}
                                    {\sum_{j=1}^{N_{pix}} e^{-\big(\frac{d_{ij}}{r}\big)^2}} \;,
\end{equation}
where we used the abbreviation $d_{ij} := d(\vec{p}_i,\vec{p}_j)$.
As becomes obvious from equation (\ref{AlphaFormel}), the calculation of scaling indices 
depends on the scale parameter $r$. Therefore, we can investigate the structural configuration in 
the underlying CMB-map in a scale-dependent manner.
For our analysis, we use the ten scaling range parameters $r_k = 0.025, 0.05, ..., 0.25$, $k=1,2,...10$,
which (roughly) correspond to sensitive $l$-ranges 
from $\Delta l =[83;387], \Delta l =[41;193], \ldots, \Delta l =[8;39]$
\citep{rossmanith09a}.\\
In order to quantify the degree of agreement between the surrogates 
of different orders with respect to their signatures left in distribution of 
scaling indices, we calculate the mean
\begin{equation}
\langle \alpha(r_k) \rangle = \frac{1}{N_p}\sum_{i=1}^{N_p} \alpha(\vec{p}_i,r_k)
\end{equation}
and the standard deviation
\begin{equation}
\sigma_{\alpha (r_k)} = \left( \frac{1}{N_p-1}\sum_{i=1}^{N_p} (\alpha(\vec{p}_i,r_k)-  \langle \alpha(r_k) \rangle)^2 \right)^{1/2}
\end{equation}
of the scaling indices $\alpha_i$ derived from $N_p$ considered pixels for the different scaling ranges $r_k$. 
$N_p$ becomes the number of all pixels $N_{pix}$ for a full sky analysis. 
To investigate possible spatial variations of signatures of NG and to be able to measure asymmetries 
we also consider the moments as derived from the pixels belonging to rotated hemispheres.
In these cases the number $N_p$ of the pixels halves 
and their positions defined by the corresponding $\phi$- and $\theta$-intervals vary according to the part of the sky being considered. 
Furthermore, we combine these two test statistics by using $\chi^2$ statistics.
There is an ongoing discussion, whether a diagonal  $\chi^2$  statistic or the ordinary  $\chi^2$ statistic, which takes into account 
correlations among the different random variables through the covariance matrix is the better suited measure. On the one hand it is of course
important to take into account correlations among the test statistics,  on the other hand it has been argued \citep{eriksen04a} that the calculation of 
the  inverse covariance matrix may become numerically unstable when the correlations among the variables are strong making
the ordinary  $\chi^2$  statistic sensitive to fluctuations  rather than to absolute deviations. 
Being aware of this we calculated both  $\chi^2$  statistics, namely the scale dependent diagonal  $\chi^2$ combining the mean and 
the standard deviation at a given scale $r_k$,
and the scale-independent $\chi^2$ combining the mean or/and the standard deviation calculated at all scales $r_k, k=1,\ldots, 10$
(see \citet{rossmanith09a}).\\
Further, we calculate the corresponding ordinary  $\chi^2$  statistics, which is obtained by summing 
over the full inverse correlations matrix $\bf C^{-1}$. In general, this is expressed by the bilinear form
\begin{equation}
\chi^2  = ( \vec{M} - \langle \vec{M} \rangle)^T   {\bf C^{-1}}  (\vec{M} - \langle \vec{M} \rangle),
\end{equation}
where the test statistics to be combined are comprised in the vector $\vec{M}$ and  $\bf C$ is obtained by cross correlating the 
elements of  $\vec{M}$. Specifically, for obtaining the scale dependent   
$\chi^2_{{\rm full}, \langle \alpha(r_k) \rangle,\sigma_{\alpha(r_k)}}$ combining the mean and 
the standard deviation at a given scale $r_k$  the vector $\vec{M}^T$ becomes 
$\vec{M}^T = ( M_1, M_2)$ with
$M_1 = \langle \alpha (r_k) \rangle$, $M_2=\sigma_{\alpha (r_k)}$.\\
Similarly, the full scale-independent $\chi^2$  statistics $\chi^2_{{\rm full}, \langle \alpha \rangle}$,
$\chi^2_{{\rm full}, \sigma_{\alpha}}$ and 
$\chi^2_{{\rm full}, \langle \alpha \rangle,\sigma_{\alpha}}$ are derived from the vectors $\vec{M}^T$ consisting of  
$\vec{M}^T=( \langle \alpha (r_1) \rangle, \ldots,  \langle \alpha (r_{10}) \rangle)$,
$\vec{M}^T=( \sigma_{\alpha (r_1)}, \ldots, \sigma_{\alpha (r_{10})})$ and
$\vec{M}^T=( \langle \alpha (r_1) \rangle, \ldots,  \langle \alpha (r_{10}) \rangle, \sigma_{\alpha (r_1)}, \ldots, \sigma_{\alpha (r_{10})})$, respectively.
For all our investigations we calculated both $\chi^2$ statistics and found out that 
the results are only marginally dependent from the chosen $\chi^2$ statistics. 
Thus, in the following we will only list explicit numbers for the 
full $\chi^2$ statistics, if not stated otherwise, because this measure yielded 
overall slightly more conservative results.

\section{Results}


\begin{figure}
\centering
\includegraphics[width=8.5cm, keepaspectratio=true]{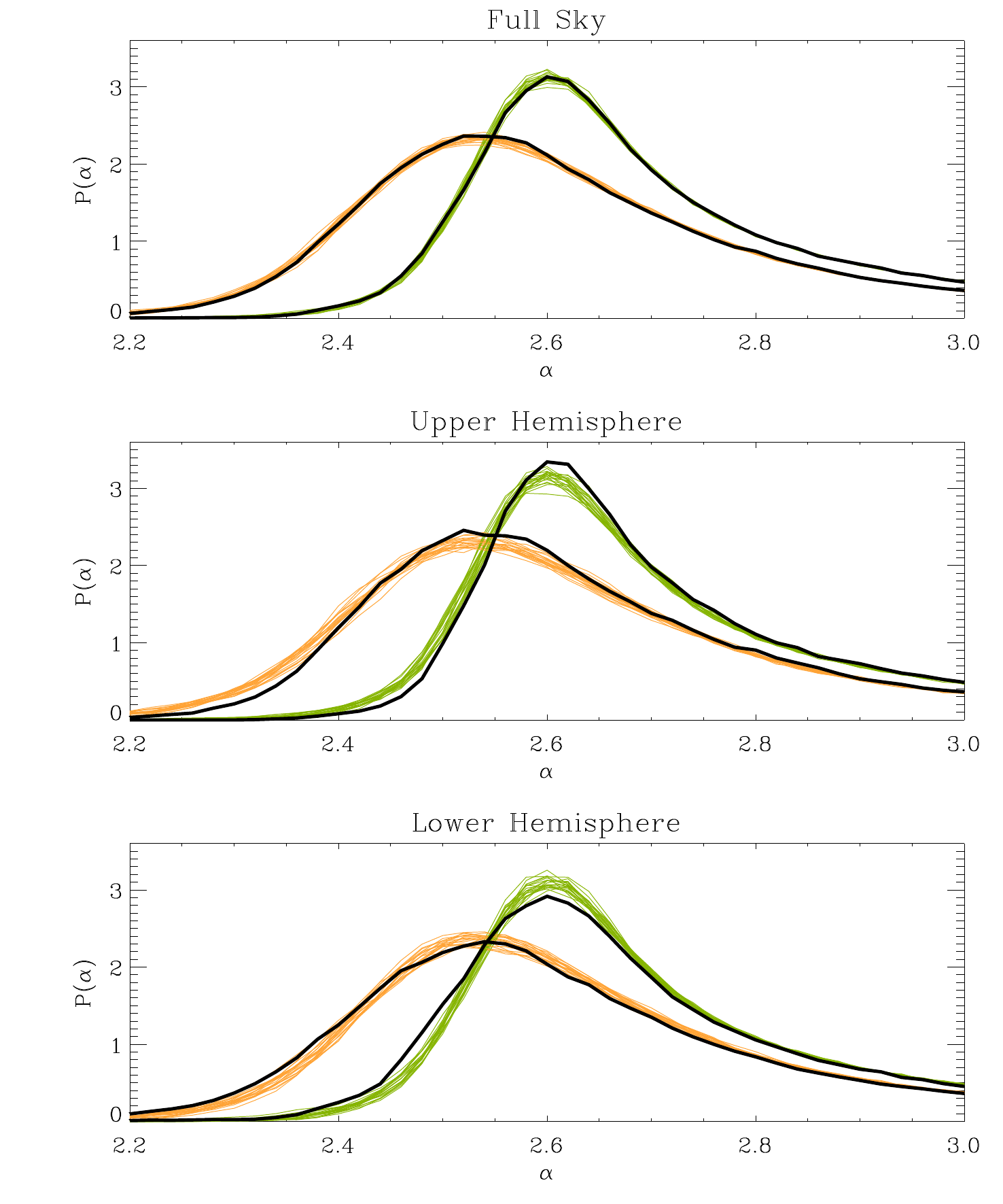}
\caption{Probability density $P(\alpha)$ of the first and second order surrogates 
for the scaling indices calculated for the largest scaling range $r_{10}$ and
for the $l-$interval $\Delta l = [2,20]$. Yellow (green) curves denote the densities for $20$ realizations of 
second order surrogates derived from the  ILC7  (NILC5) map. The black lines are the corresponding
first order surrogates. The reference frame for defining the upper and lower hemispheres 
is chosen such that the difference $\Delta S = S_{up} - S_{low}$ becomes maximal for $\langle \alpha \rangle$
of the respective map and respective scale $r$.} \label{fig4}
\end{figure}

\begin{figure}
\centering
\includegraphics[width=8.5cm, keepaspectratio=true]{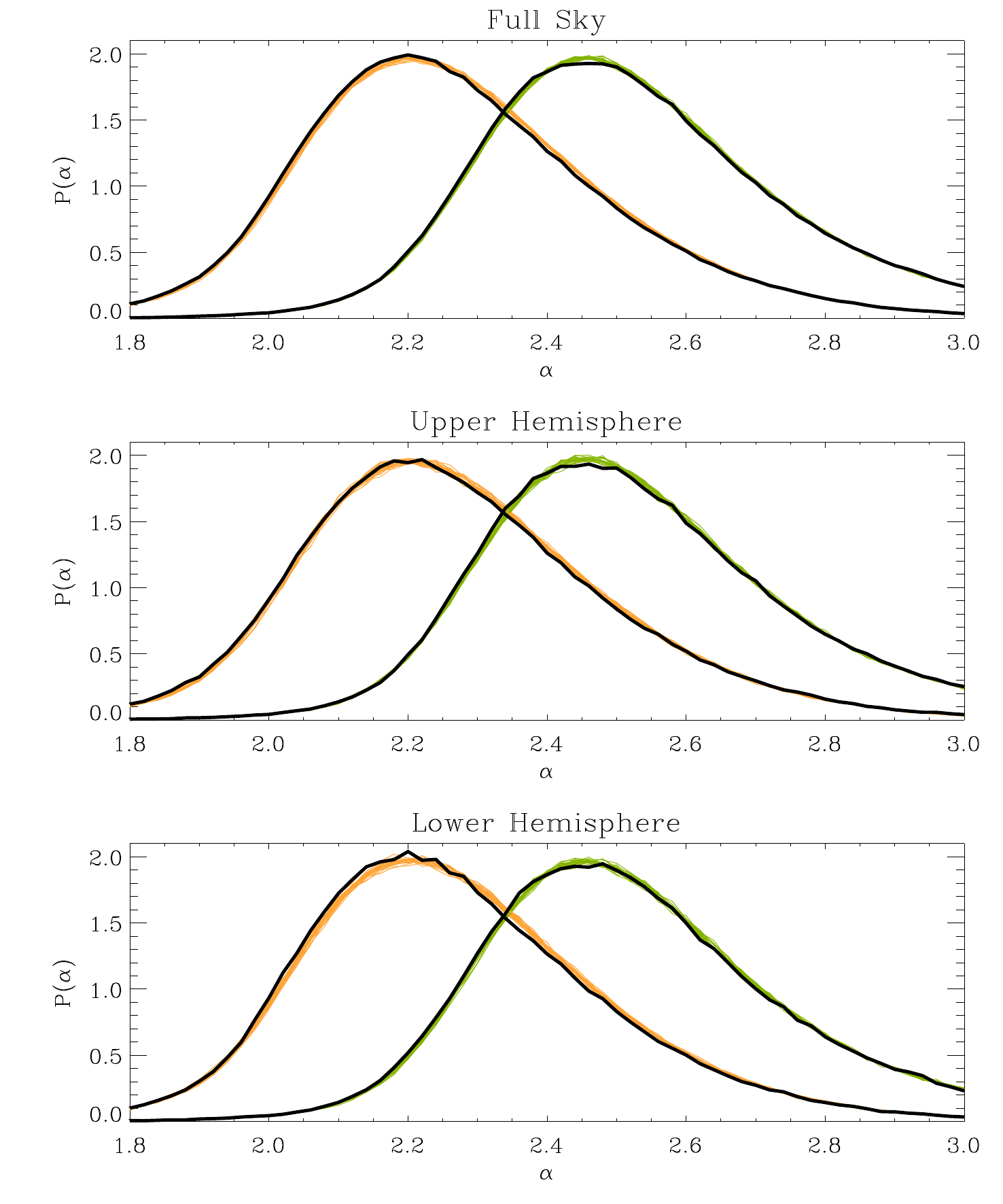}

\caption{Same as figure \ref{fig4} but the second smallest scaling range  $r_{2}$ and the 
 $l-$interval $\Delta l = [120,300]$. } \label{fig5}
\end{figure}

\begin{table}
\begin{tabular}{lcccc}
\hline \hline
$\Delta l$ &  Full Sky  &  Upper & Lower \\ 
 &    &  Hemisphere & Hemisphere \\ \hline
$\langle \alpha(r_{2}) \rangle$: & $(S/\%)$ & $(S/\%)$ & $(S/\%)$ \\ \\
$[2,1024]$       & 7.73 / $>$ 99.8  & 4.53 / $>$99.8 & 1.87 / 96.0 \\ \\
$[2,20]$       & 0.14 / 56.6 & 3.54 / $>$99.8 & 3.44 / $>$99.8 \\
$[20,60]$     & 0.88 / 80.6 & 1.84 / 96.4       & 1.08 / 85.2 \\
$[60,120]$     & 0.26 / 60.4 & 0.32 / 64.8       & 0.64 / 71.6 \\
$[120,300]$   & 6.97 / $>$99.8 & 5.36 / $>$99.8     & 0.92 / 83.0 \\ \hline

\multicolumn{2}{l}{$\sigma_{\alpha(r_{2})}$:} & & \\ \\
$[2,1024]$       & 4.16 / $>$99.8 & 3.77 / $>$99.8       & 0.25 / 61.8 \\ \\
$[2,20]$       & 0.48 / 69.2 & 0.48 / 69.8       & 0.19 / 58.0 \\
$[20,60]$     & 1.70 / 95.2 & 3.18 / $>$99.8       & 1.02 / 84.8 \\
$[60,120]$     & 0.88 / 80.0 & 2.35 / 98.8 & 1.25 / 88.2 \\
$[120,300]$   & 3.54 / $>$99.8 & 1.03 / 83.4       & 3.69 / $>$99.8 \\ \hline

\multicolumn{2}{l}{$\chi^2_{\langle \alpha(r_{2}) \rangle,\sigma_{\alpha(r_{2})}}$:}  & & \\ \\
$[2,1024]$       & 24.55/$>$99.8 & 14.44 / $>$99.8       & 0.94 / 84.4 \\ \\
$[2,20]$       & 0.90 / 85.2 & 7.67 / $>$99.8 & 8.47 / 99.8 \\
$[20,60]$     & 0.82 / 83.4 & 4.03 / 99.2       & 0.31 / 50.4 \\
$[60,120]$     & 0.51 / 61.4 & 3.63 / 98.6 & 1.00 / 85.2 \\
$[120,300]$   & 19.62 / $>$99.8 & 17.17 / $>$99.8       & 4.15 / 99.2 \\  \hline  \hline

\end{tabular}
\caption{Deviations $S$ and empirical probabilities $p$  of the mean, standard deviation
and their $\chi^2$-combination as derived for the scaling indices at the second smallest scale $r_{2}$.
The results of the ILC7 map are shown for the different $l$-bands as well as for the full sky 
and the upper and lower hemispheres. Corresponding to the small scale $r_{2}$ the largest values for $S$ 
are calculated for small scale non-Gaussianities in the $l$-range 
$[120,300]$ and for the scale-independent NGs, where the phases of 
all $l$'s ($\Delta l = [2,1024]$) are included. } \label{table1}
\end{table}

\begin{table}
\begin{tabular}{lcccc}
\hline \hline
$\Delta l$ &  Full Sky  &  Upper & Lower \\ 
 &    &  Hemisphere & Hemisphere \\ \hline
$\langle \alpha(r_{10}) \rangle$: & $(S/\%)$ & $(S/\%)$ & $(S/\%)$ \\ \\
$[2,1024]$       & 3.75 / $>$99.8 & 3.53 / $>$99.8 & 1.72 / 95.4 \\ \\
$[2,20]$       & 0.64 / 74.2 & 3.24 / $>$99.8 & 3.41 / $>$99.8 \\
$[20,60]$     & 0.67/ 74.2 & 1.41 / 91.6       & 2.04 / 98.0 \\
$[60,120]$     & 0.01 / 50.5 & 2.28 / 99.0       & 2.19 / 98.6 \\
$[120,300]$   & 2.45 / 99.4 & 3.58 / $>$99.8       & 1.38 / 92.2 \\ \hline

\multicolumn{2}{l}{$\sigma_{\alpha(r_{10})}$:} & & \\ \\
$[2,1024]$       & 0.66 / 74.4 & 3.60 / $>$99.8 & 2.90 / $>$99.8 \\ \\
$[2,20]$       & 0.84 / 80.0 & 3.09 / $>$99.8       & 1.79 / 96.4 \\
$[20,60]$     & 2.27 / 98.6 & 2.94 / 99.8       & 0.13 / 55.0 \\
$[60,120]$     & 0.77 / 79.0 & 1.63 / 94.6 & 0.47 / 67.6 \\
$[120,300]$   & 0.60 / 73.6 & 1.61 / 95.8       & 0.81 / 79.6 \\ \hline

\multicolumn{2}{l}{$\chi^2_{\langle \alpha(r_{10}) \rangle,\sigma_{\alpha(r_{10})}}$:}  & & \\ \\
$[2,1024]$       & 1.46 / 90.4 & 9.83 / $>$99.8 & 3.15 / 98.0 \\ \\
$[2,20]$       & 0.21 / 54.8 & 7.10 / $>$99.8 & 6.77 / 99.8 \\
$[20,60]$     & 2.74 / 97.2 & 5.27 / 99.6       & 0.29 / 73.6 \\
$[60,120]$     & 0.38 / 50.2 & 2.09 / 94.2 & 0.43 / 75.8 \\
$[120,300]$   & 0.26 / 57.2 & 2.23 / 96.2       & 0.19 / 60.4 \\  \hline  \hline

\end{tabular}
\caption{Same as table \ref{table1} but for the scaling indices at the largest scale $r_{10}$.
The largest values for $S$ are found for large scales non-Gaussianities in the $l$-range 
$[2,20]$.} \label{table2}
\end{table}


\begin{table}
\begin{tabular}{lcccc}
\hline \hline
$\Delta l$ &  Full Sky  &  Upper & Lower \\ 
 &    &  Hemisphere & Hemisphere \\ \hline

$\chi^2_{\langle \alpha \rangle}$: & $(S/\%)$ & $(S/\%)$ & $(S/\%)$ \\ \\
$[2,1024]$       & 5.73 / $>$99.8 & 9.35 / $>$99.8 & 0.33 / 55.2 \\ \\
$[2,20]$       & 0.97 / 95.0 & 4.57 / 99.6 & 4.01 / 99.2 \\
$[20,60]$     & 1.81 / 94.2 & 2.57 / 97.4 & 2.42 / 97.0 \\
$[60,120]$     & 1.41 / 99.0 & 1.53 / 99.6 & 0.91 / 83.8 \\
$[120,300]$   & 3.17 / 92.8 & 10.53 / $>$99.8 & 1.19 / 87.8 \\  \hline

$\chi^2_{\sigma_{\alpha}}$: & & & \\ \\
$[2,1024]$       & 5.50 / $>$99.8 & 11.50 / $>$99.8 & 0.66 / 79.6 \\ \\
$[2,20]$       & 0.32 / 52.8 & 4.03 / 98.6 & 4.04 / 99.6 \\
$[20,60]$     & 2.15 / 95.8 & 4.00 / 99.8 & 2.18 / 96.4 \\
$[60,120]$     & 1.40 /98.2 & 3.26 / 99.4 & 2.01 / 95.6 \\
$[120,300]$   & 3.10 / 99.0 & 8.90 / $>$99.8 & 1.90 / 95.8 \\  \hline

$\chi^2_{\langle \alpha \rangle,\sigma_{\alpha}}$: & & & \\ \\
$[2,1024]$       & 1.89 / 94.2 & 8.38 / $>$99.8 & 3.03 / 98.8 \\ \\
$[2,20]$       & 0.73 / 77.4 & 5.64 / $>$99.8 & 6.01 / 99.8 \\
$[20,60]$     & 1.60 / 92.8 & 3.42 / 99.2 & 1.49 / 91.0 \\
$[60,120]$     & 0.26 / 52.4 & 2.15 / 96.6 & 0.53 / 75.6 \\
$[120,300]$   & 1.68 / 92.8 & 5.34 / 99.8 & 0.22 / 63.2 \\  \hline \hline

\end{tabular}
\caption{Same as table  \ref{table1} but for the scale-independent $\chi^2$-statistics. Also for this 
statistics the largest values for $S$ are found for the largest $ \Delta l = [2,20]$ 
and smallest scales $ \Delta l = [120,300]$ and for the scale-independent NGs.   } \label{table3}
\end{table}

\begin{figure*}
\centering
\includegraphics[width=8.25cm, keepaspectratio=true]{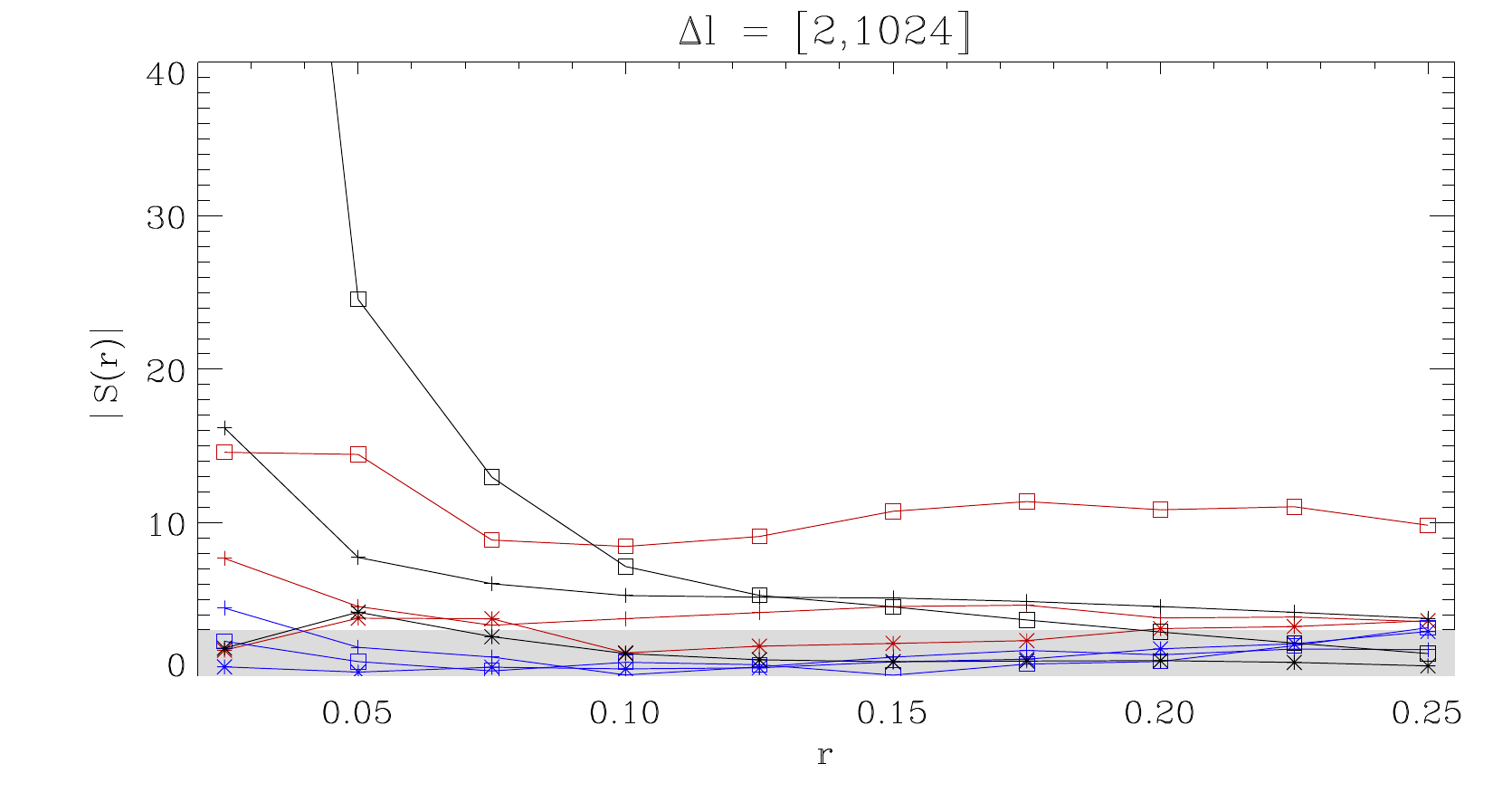}
\includegraphics[width=8.25cm, keepaspectratio=true]{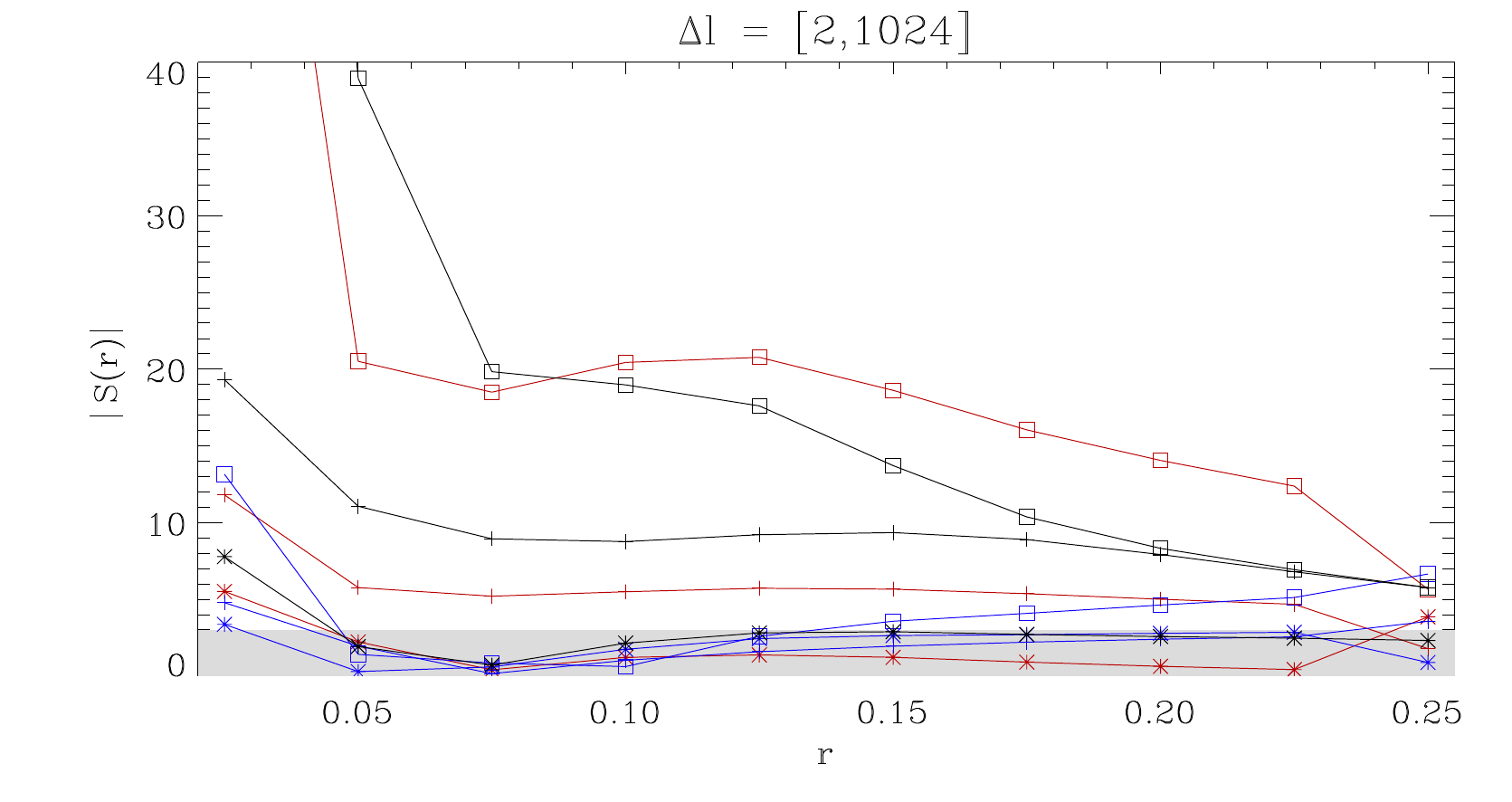}
\includegraphics[width=8.25cm, keepaspectratio=true]{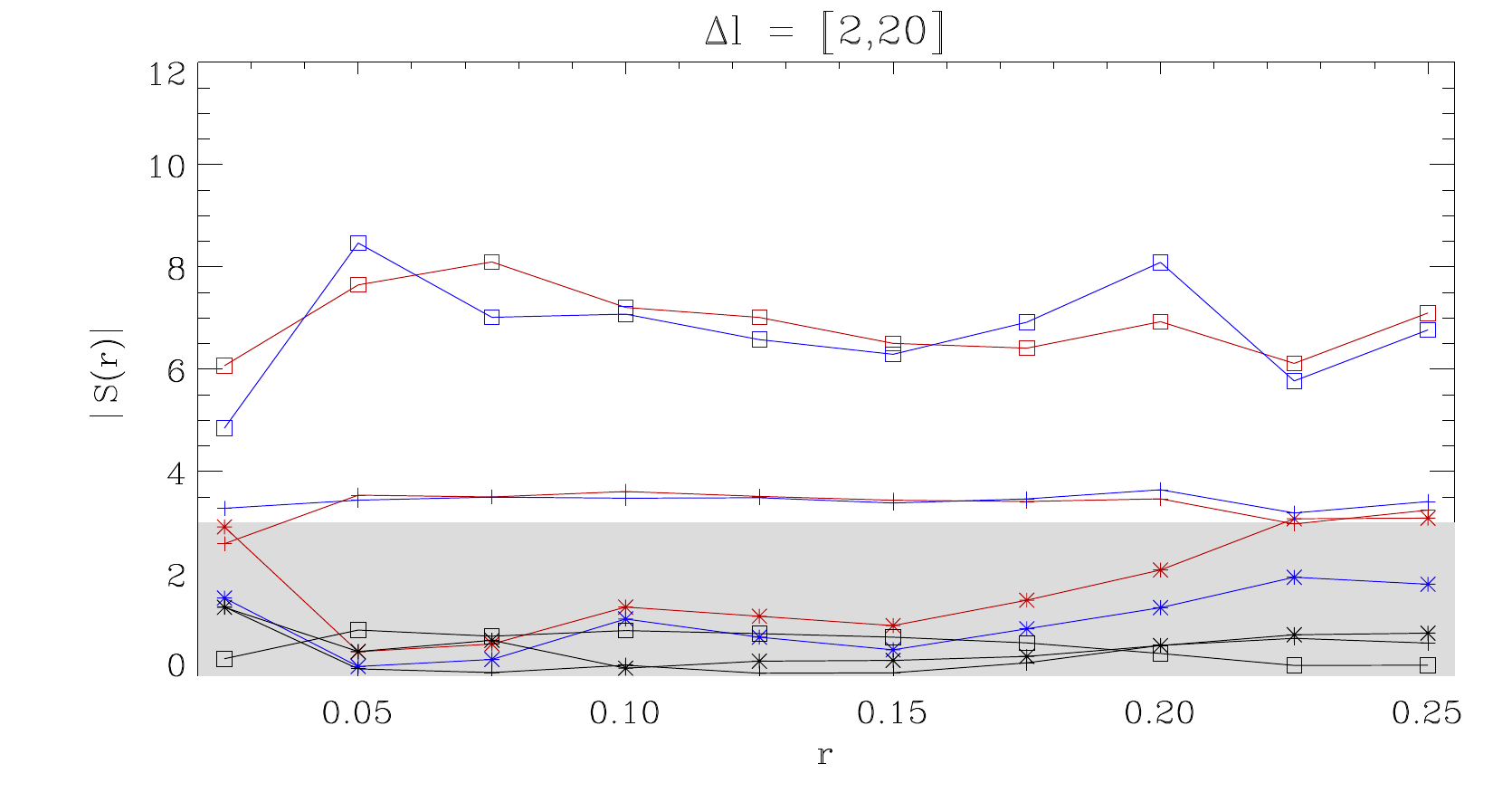}
\includegraphics[width=8.25cm, keepaspectratio=true]{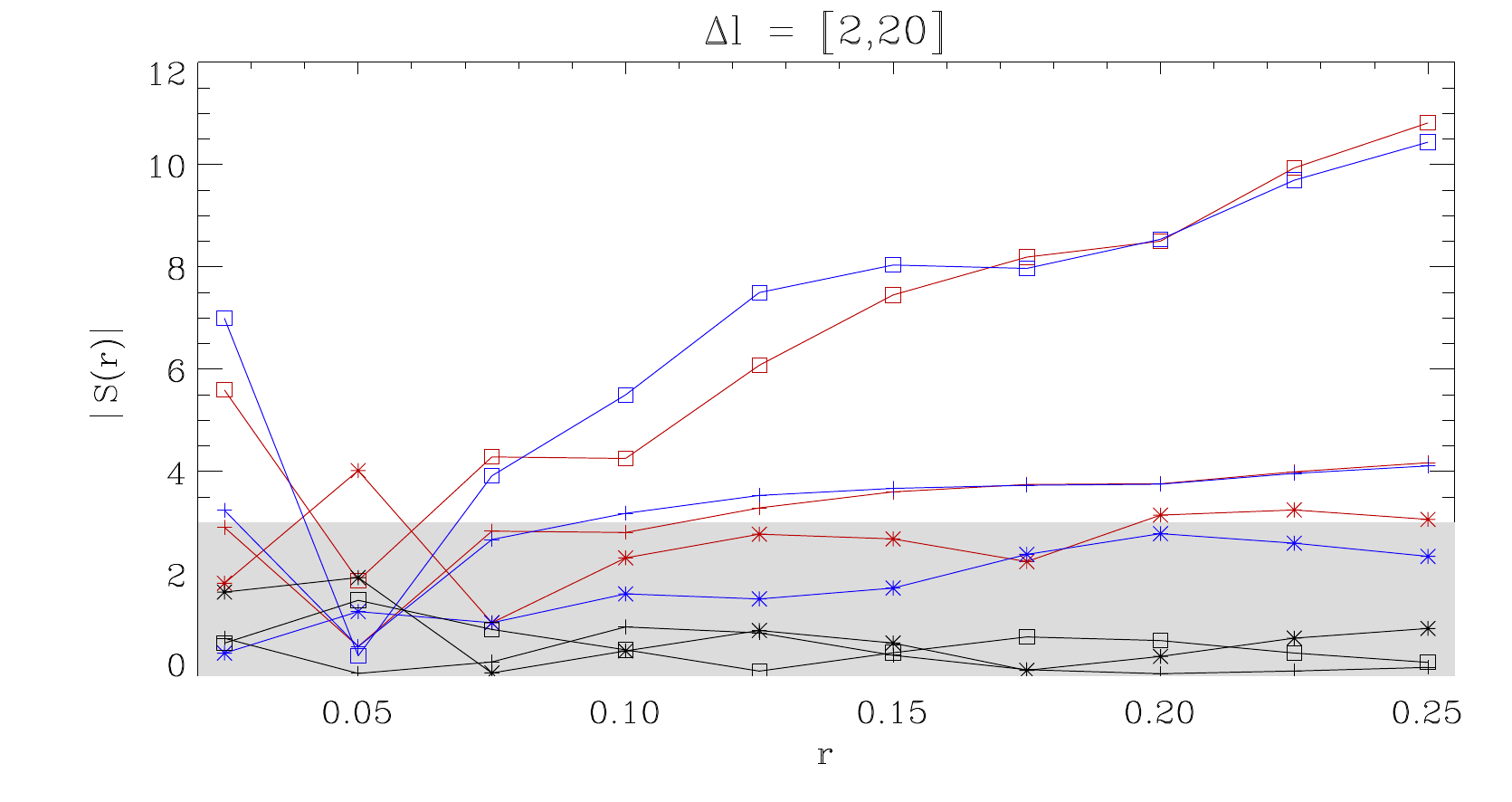}

\caption{Deviations $|S(r)|$ for the ILC7 (left) and NILC5 (right) map as a function of 
                the scale parameter $r$ for the full sky (black) and 
                the upper (red) and lower (blue) hemisphere. The plus signs denote 
                the results for the mean $\langle \alpha (r_{k}) \rangle$, the star-signs for the standard deviation  $\sigma_{\alpha (r_{k})} $
                and the boxes for the $\chi^2$-combination of $\langle \alpha (r_{k}) \rangle$ and $\sigma_{\alpha (r_{k})} $. The 
                shaded region indicates the $3\sigma$ significance interval. } \label{fig6}
\end{figure*}

To test for NGs and asymmetries in the ILC7 map and the NILC5 map, 
we compare the different surrogate maps in the following way:\\ 
For each scale we calculate the mean $\langle \alpha(r_k) \rangle$ and standard 
deviation $\sigma_{\alpha (r_k)}$ of the map of scaling indices 
$\alpha(\theta, \phi; r_k) $ of the full sky and a set of 768 rotated hemispheres. 
The northern pole of the different hemispheres is located at every pixel centre of the 
full sky with $N_{side} = 8$ in the HEALpix\footnote{http://healpix.jpl.nasa.gov/} \citep{gorski05a} pixelisation scheme.
The differences of the two classes of surrogates are quantified by 
the $\sigma$-normalized deviation $S$
\begin{equation}
S(Y) = \frac{Y_{{\rm surro1}}- \langle Y_{{\rm surro2}} \rangle}{ \sigma_{Y_{{\rm surro2}}}}
\end{equation}
with, $Y= \langle \alpha(r_k) \rangle, \sigma_{\alpha(r_k)}, \chi^2$. 
Every hemisphere of the set of $768$ hemispheres delivers one deviation value $S$,
which is then plotted on a sky map at that pixel position where the z-axis of the rotated hemisphere pierces the sky.  
Fig. \ref{fig3} shows the deviations $S$ for the mean value $S(\langle \alpha(r_k) \rangle), k=2, 6, 10$ for the ILC7 map 
as derived from the comparison of the different classes of surrogates for the scale-independent surrogate test and for the four selected $l$-ranges. 
The following striking features become immediately obvious:\\
First, various deviations representing features of non-Gaussianity and asymmetries can be found in the $S$-maps for the ILC7 map. 
These features can nearly exactly be reproduced when the  NILC5 map is taken as input map (results not shown). \\
Second, we find for the scale-independent surrogate test (leftmost column in figs. \ref{fig3}) 
large isotropic deviations for the scaling indices calculated for the smallest scale shown in the figure. 
The negative values for $S$ indicate that
the mean of the scaling indices for the first order surrogate is smaller than for the second order surrogate maps.
This systematic trend can be interpreted such that there's more structure detected in the first order surrogate 
than in the second order surrogate maps. Obviously, the random shuffle of all phases has destroyed a significant amount of 
structural information at small scales in the maps.\\  
Third, for the scale-dependent analysis we obtain for the largest scales ($\Delta l = [2,20] $) highly significant signatures for 
non-Gaussianities and ecliptic hemispherical asymmetries at the largest $r-$values (second column in figs. \ref{fig3}). 
These results are perfectly consistent  with those obtained for the  WMAP 5 yr ILC map and the foreground 
removed maps generated by \cite{tegmark03a} on the basis of the WMAP 3 year data (see \cite{raeth09a}). 
The only difference between this study and our previous one is that
we now obtain higher absolute values for $S$ ranging now from $-4.00 < S < 3.72$ for the ILC7 map
and  $-4.36 < S < 4.50$ for  the NILC5 map as compared to $-3.87 < S < 3.51$ for the WMAP 5 yr ILC map.
Thus, the cleaner the map becomes due to better signal-to-noise ratio and/or improved map making techniques the
higher the significances of the detected anomalies, which suggests that the signal is of intrinsic CMB origin.\\
Fourth,  we also find for the smallest considered scales ($\Delta l = [150,300] $)  
large isotropic deviations for the scaling indices calculated for a small scaling range $r$ very similar to those 
observed for the scale-independent test.\\
Fifth, we do not observe very significant anomalies for the two other bands 
($\Delta l = [20,60] $ and $\Delta l = [60,120] $) being considered in this study. 
Thus, the results obtained for the scale independent surrogate test can clearly be interpreted
as a superposition of the signals identified in the two $l$-bands covering the largest ($\Delta l = [2,20] $) and 
smallest $\Delta l = [120,300] $) scales.
Let us investigate the observed anomalies in more details. We begin with a closer look at the most significant 
deviations.  
Fig. \ref{fig4} shows the probability densities derived for the full sky and for (rotated) hemispheres
for the scaling indices at the largest scaling range $r_{10}$ for the 
first and second order surrogates for the $l$-interval $\Delta l = [2,20] $.
We recognize the systematic shift of the whole density distribution 
towards higher values for the upper hemisphere and to lower values for the lower hemisphere.
As these two effects cancel each other for the full sky, we do no longer see significant differences
in the probability densities in this case.
Since the densities as a whole are shifted, the significant differences between first and second order surrogates 
found for the moments cannot be attributed to some salient localizable features leading to an excess 
(e.g. second peak) at very low or high values in otherwise very similar $P(\alpha)$-densities. Rather, the shift to 
higher (lower) values for the upper (lower) hemisphere must be interpreted as a global trend indicating that
the first  order surrogate map has less (more) structure than the respective set of second order surrogates.
The seemingly counterintuitive result for the upper hemisphere is on the other hand consistent with a linear 
hemispherical structure analysis by means of a power spectrum analysis, where also a lack of power in the 
northern hemisphere and thus a pronounced hemispherical asymmetry was detected \citep{hansen04a, hansen09a}.
However, it has to be emphazised that the effects contained in the power spectrum are -- by construction -- exactly
preserved in both classes of surrogates, so that the scaling indices measure effects that can solely be induced by 
HOCs thus being of a new, namely non-Gaussian, nature. 
Interestingly though, the linear and nonlinear 
hemispherical asymmetries seem to be correlated with each other.\\
Fig.  \ref{fig5}  is very similar to fig.  \ref{fig4} and shows the probability densities for the scaling indices 
calculated for the second smallest scaling range $r_{2}$ for the 
first and second order surrogates for the $l$-interval $\Delta l = [120,300] $.
The systematic shift towards smaller values for the first order surrogate for both hemispheres 
and thus for the full sky is visible. It is interesting to note that all densities derived from the ILC7 and NILC5 map
differ significantly from each other. These differences can be attributed to e.g. the smoothing of the ILC7 map. 
However, the systematic differences between first and second order surrogates induced by the phase manipulations
prevailed in all cases -- irrespective of the input map.\\
The results for the deviations $|S(r)|$ for the full sky and rotated upper and lower hemisphere 
are shown for all considered $l$-ranges and all scales $r$ in figs.  \ref{fig6} .
The corresponding values for $r_{2}$ and $r_{10}$ are listed in the tables   \ref{table1}  and  \ref{table2}. 
In table \ref{table3}  we further summarize the results for the scale-independent $\chi^{2}$-measures 
$\chi^2_{\langle \alpha \rangle}$, $\chi^2_{\sigma_{\alpha}}$ and $\chi^2_{\langle \alpha \rangle,\sigma_{\alpha}}$.

\begin{figure*}
\centering

\includegraphics[width=3.4cm, keepaspectratio=true]{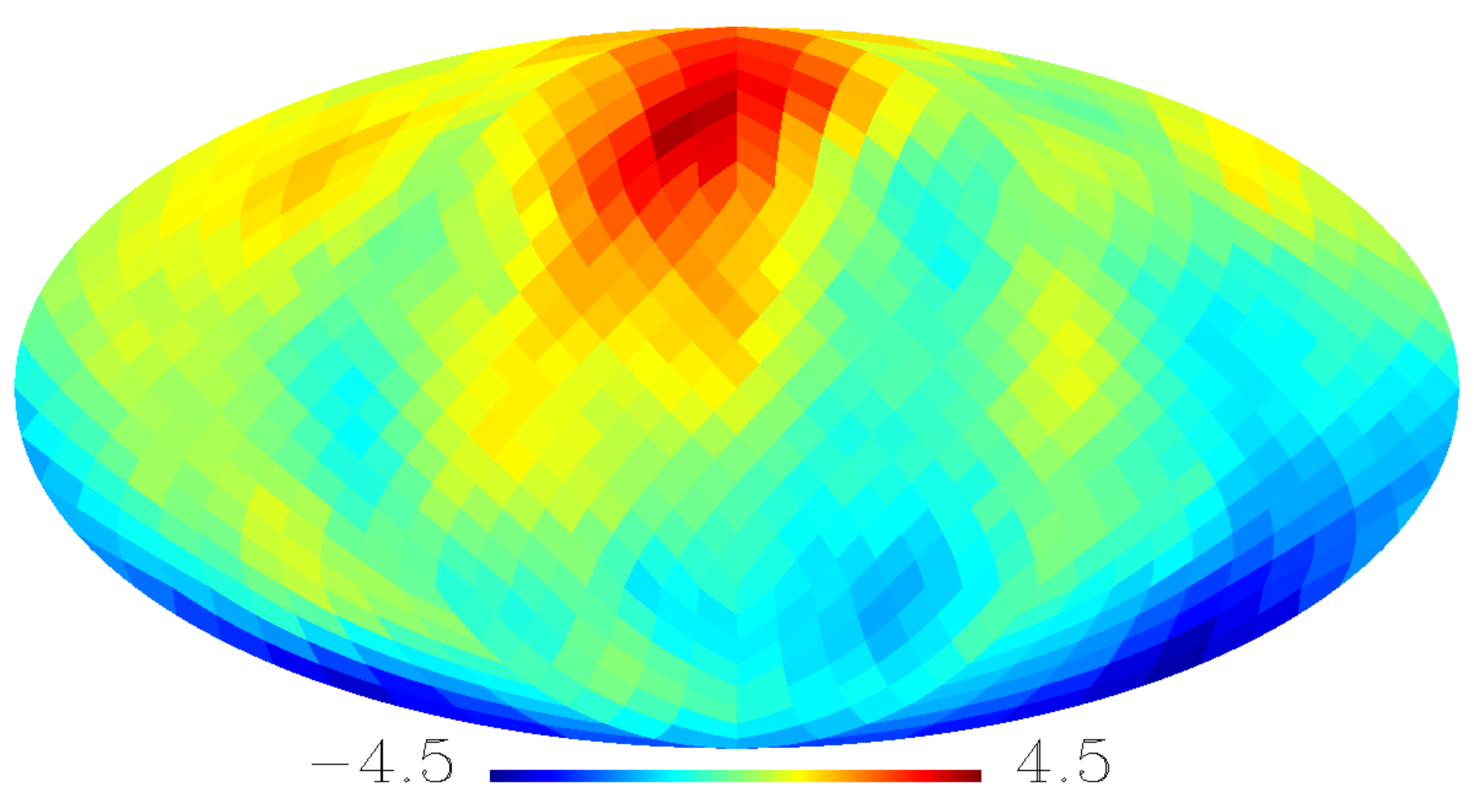}
\includegraphics[width=3.4cm, keepaspectratio=true]{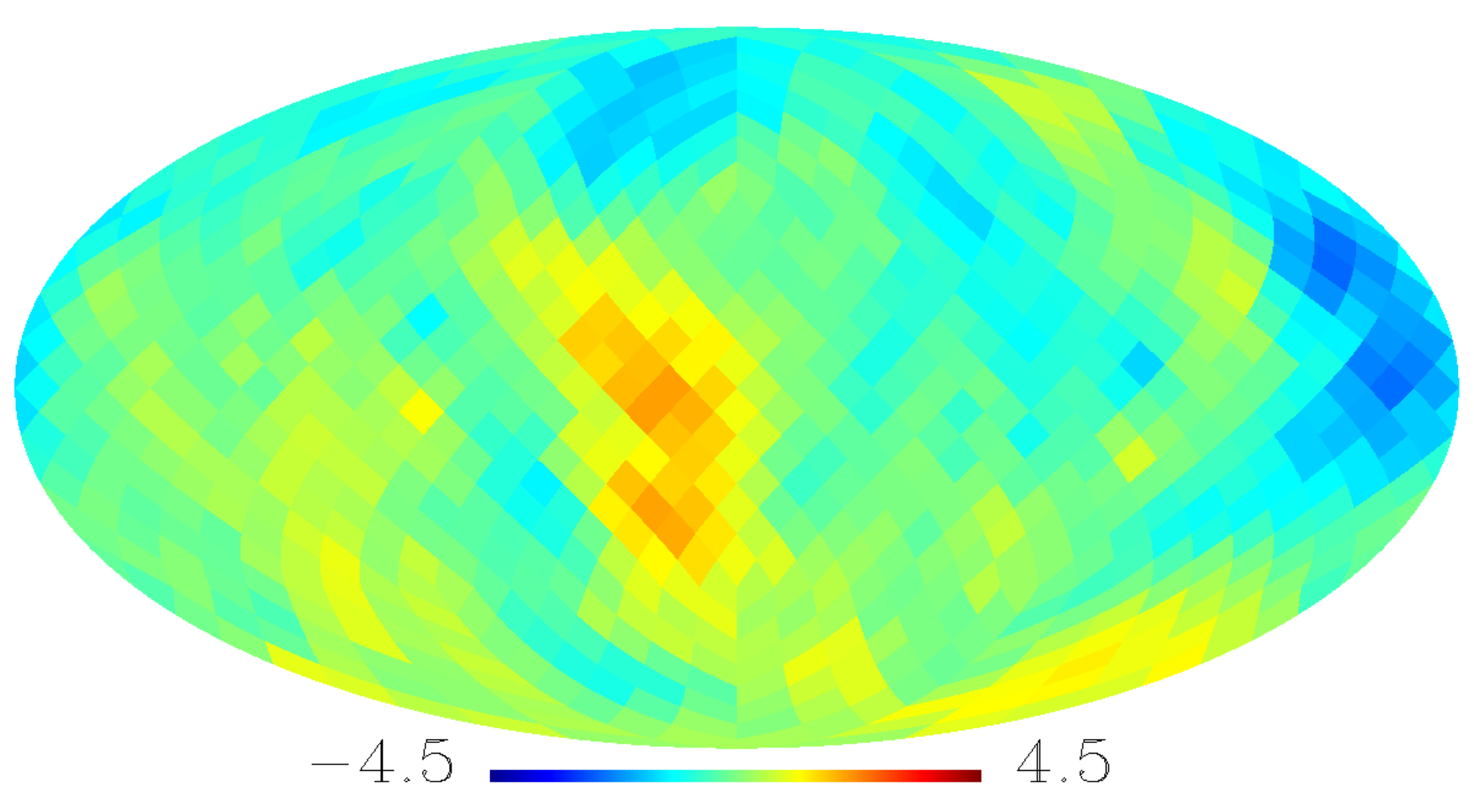} \hspace{0.25cm}
\includegraphics[width=3.4cm, keepaspectratio=true]{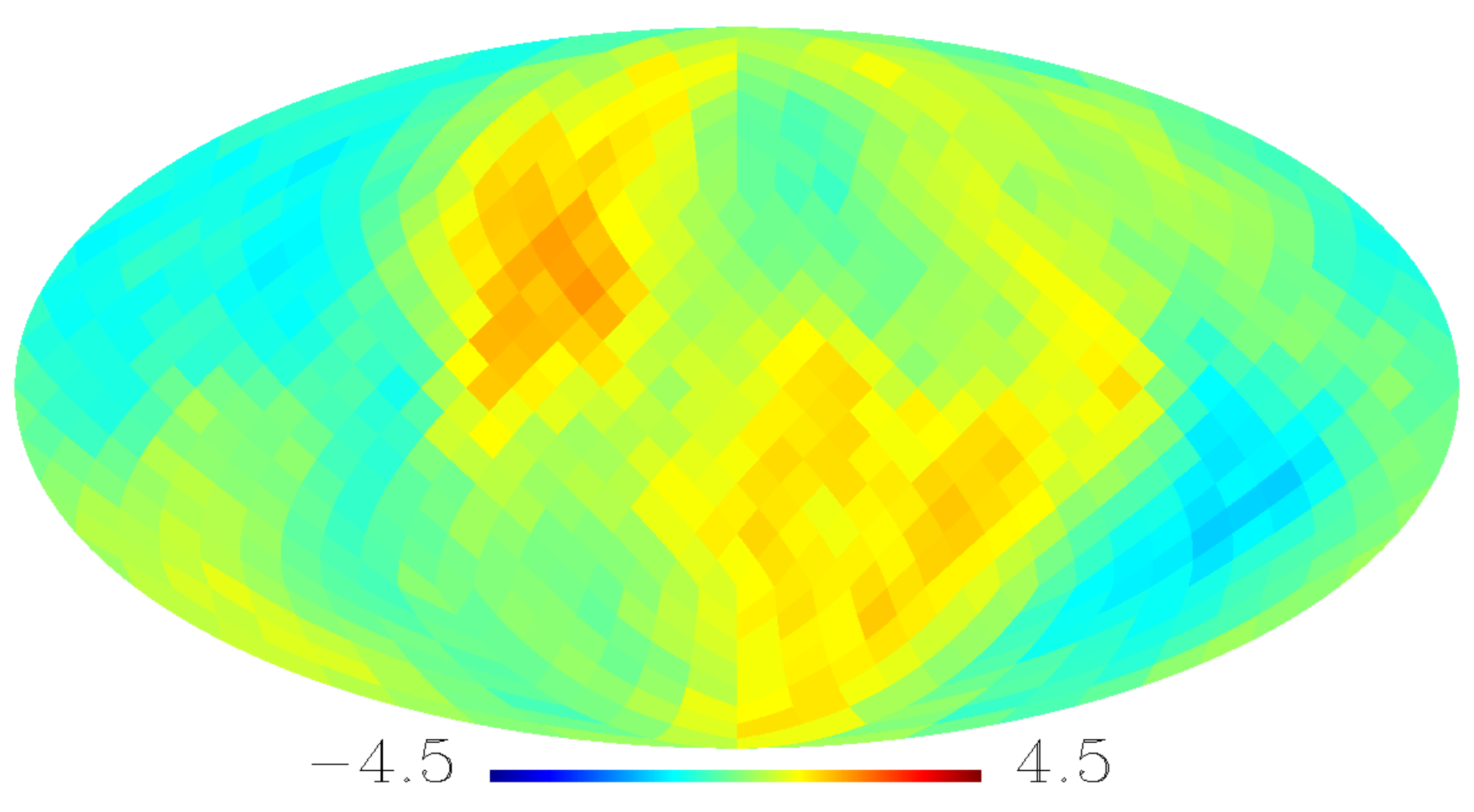}
\includegraphics[width=3.4cm, keepaspectratio=true]{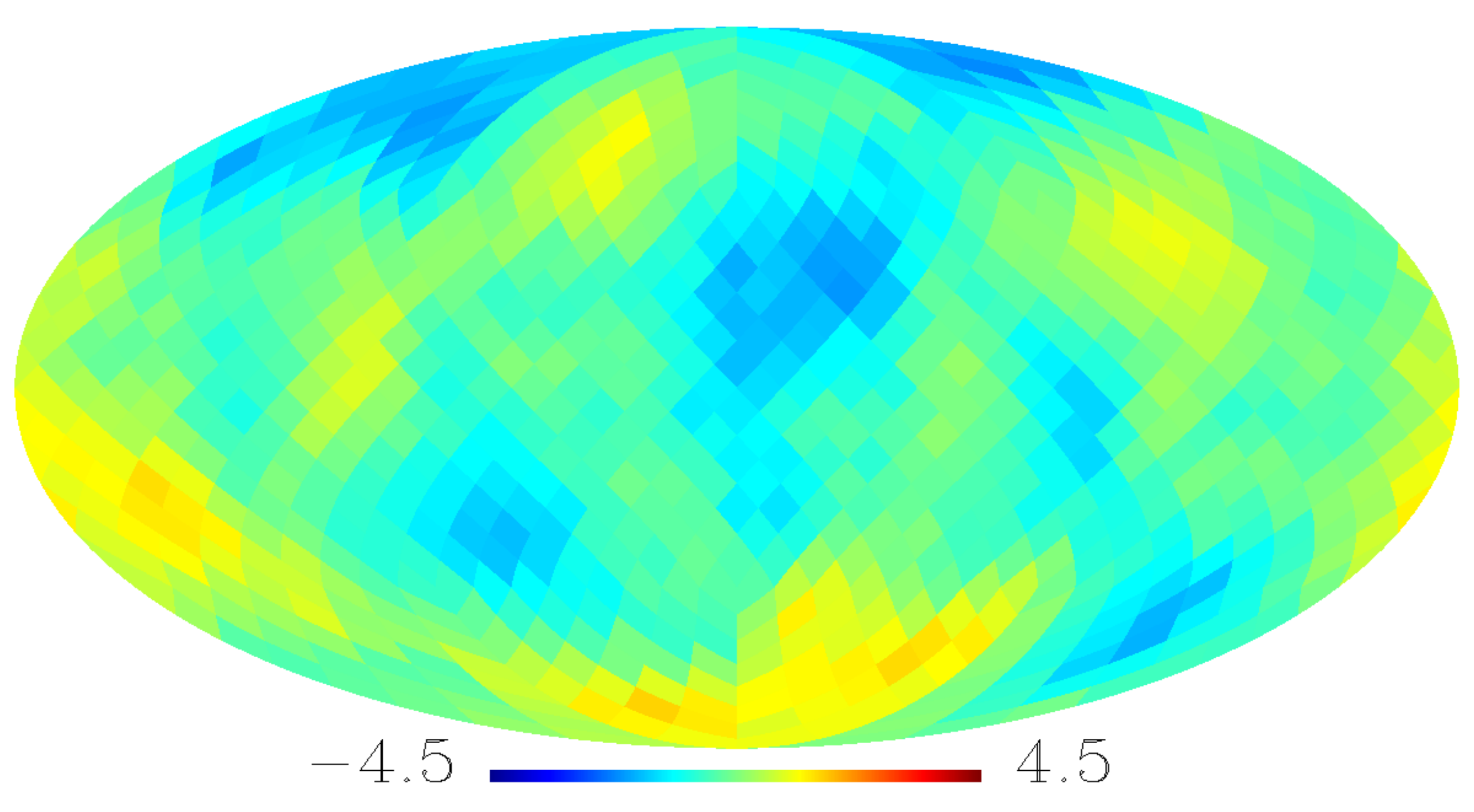}
\includegraphics[width=3.4cm, keepaspectratio=true]{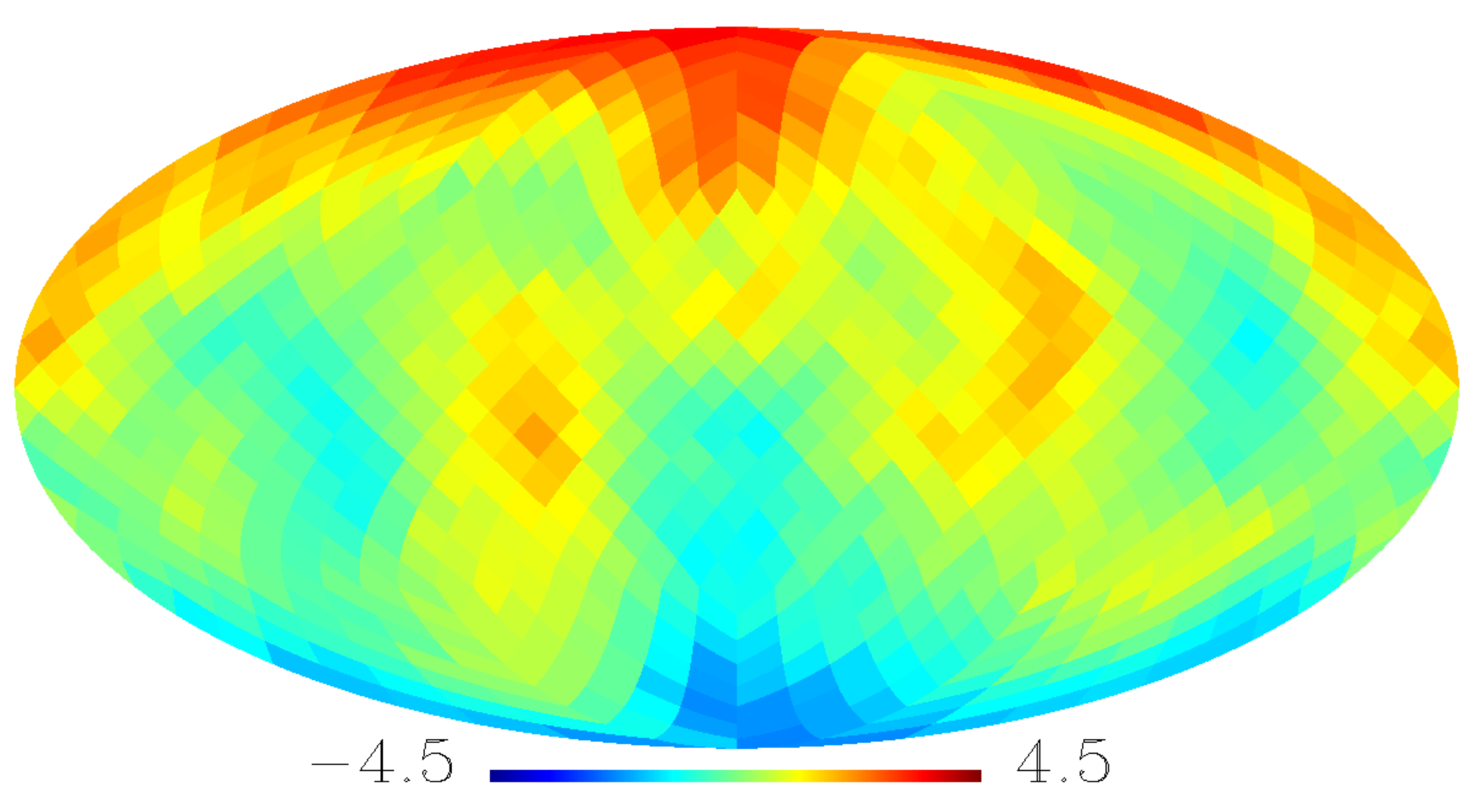}

\includegraphics[width=3.4cm, keepaspectratio=true]{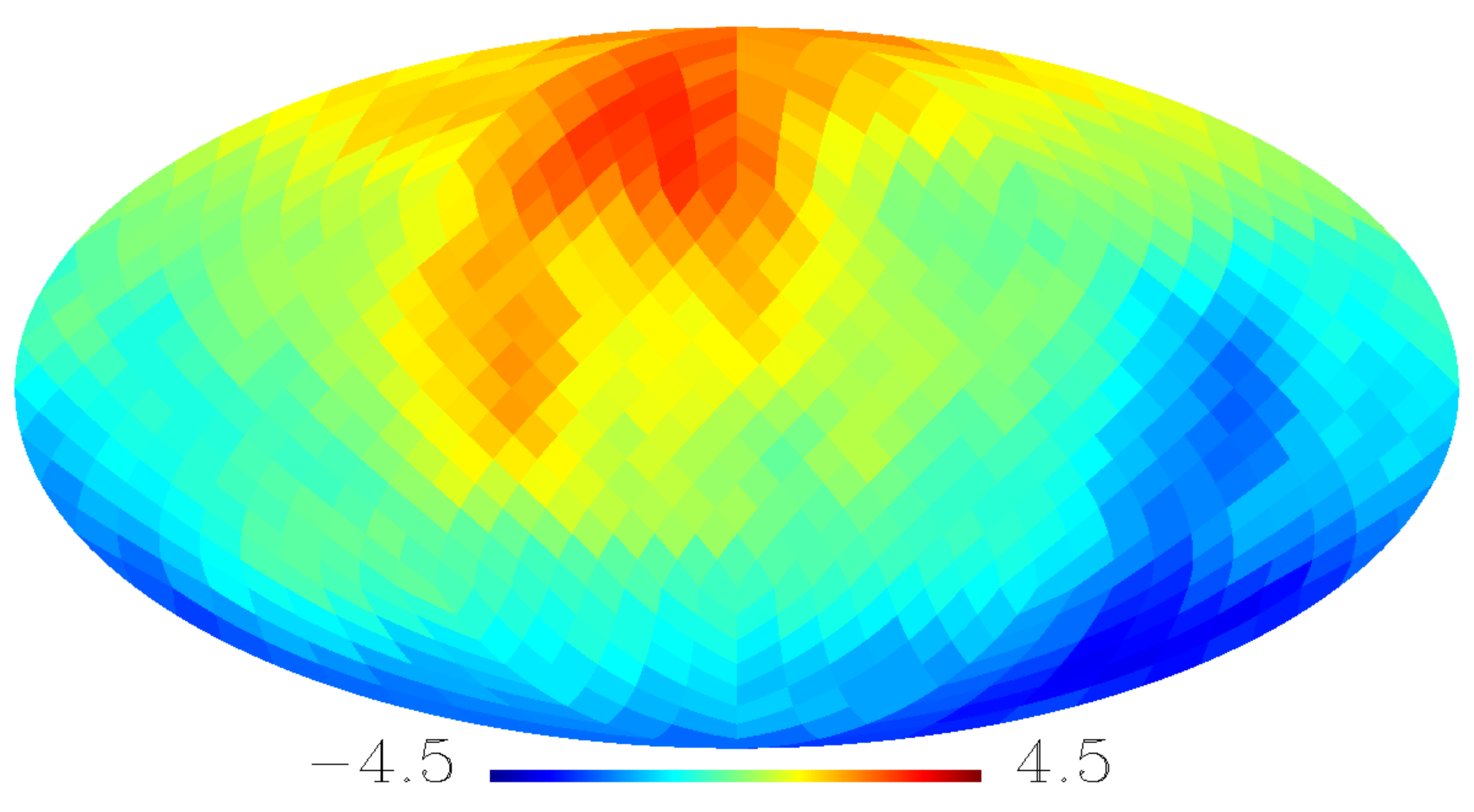}
\includegraphics[width=3.4cm, keepaspectratio=true]{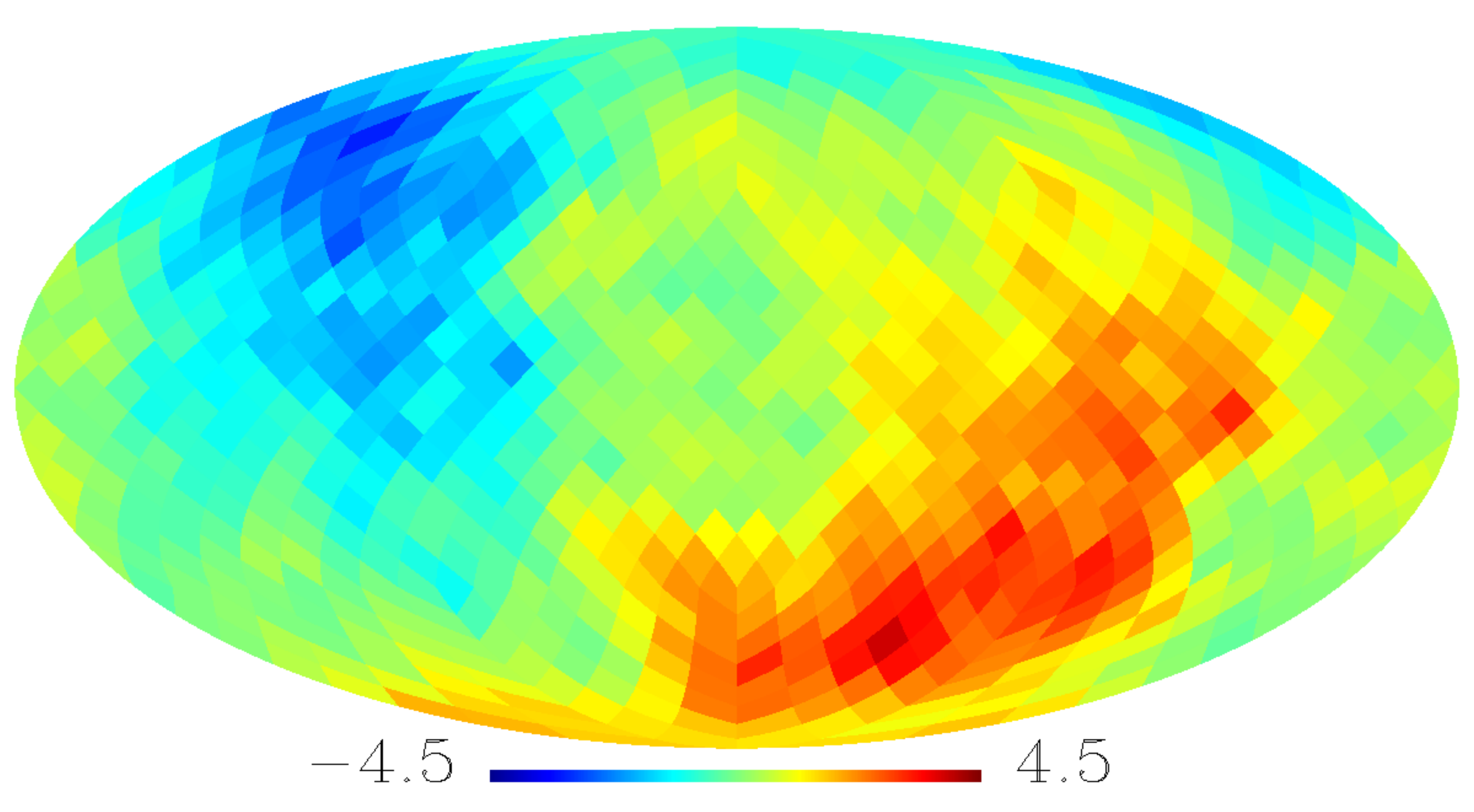} \hspace{0.25cm}
\includegraphics[width=3.4cm, keepaspectratio=true]{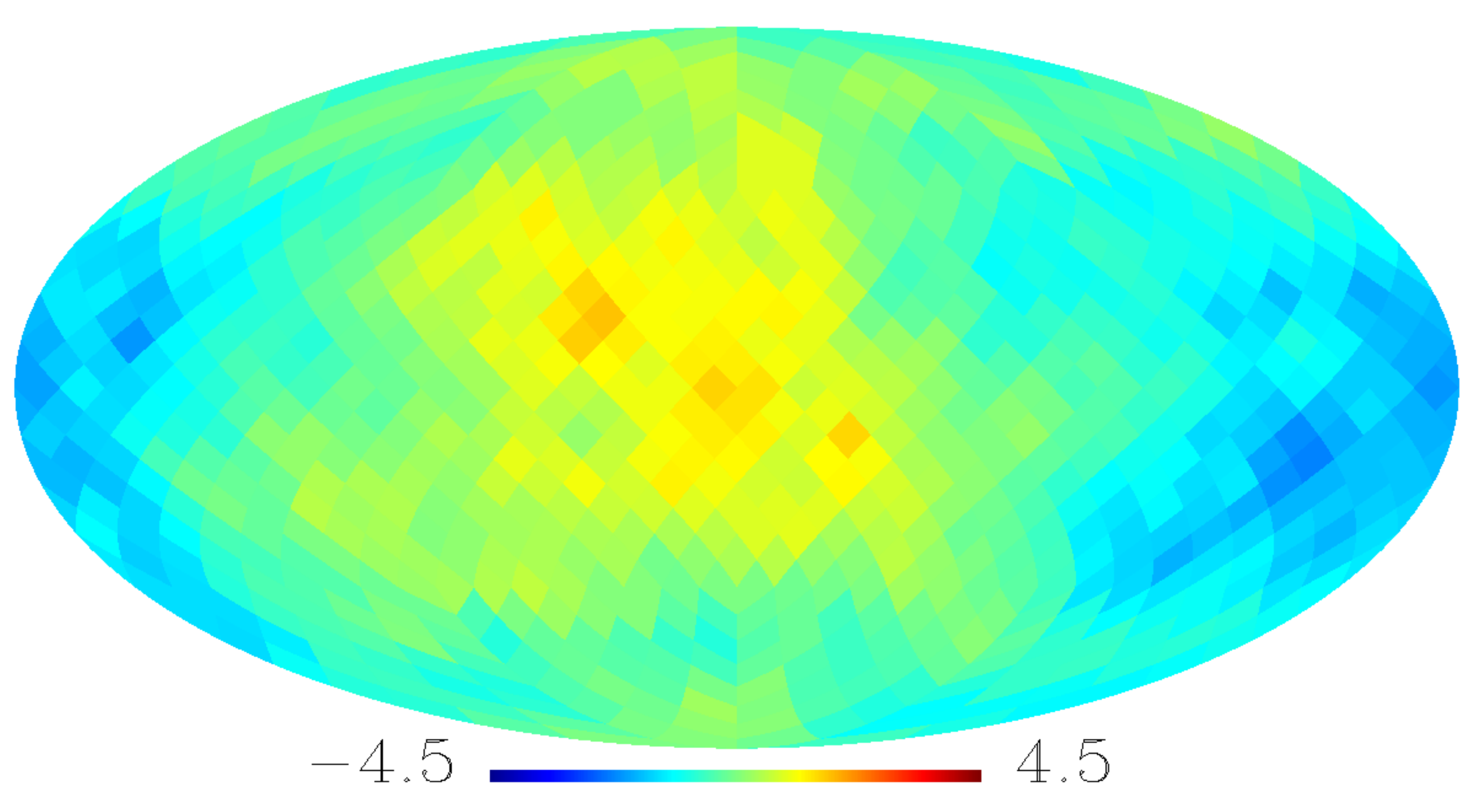}
\includegraphics[width=3.4cm, keepaspectratio=true]{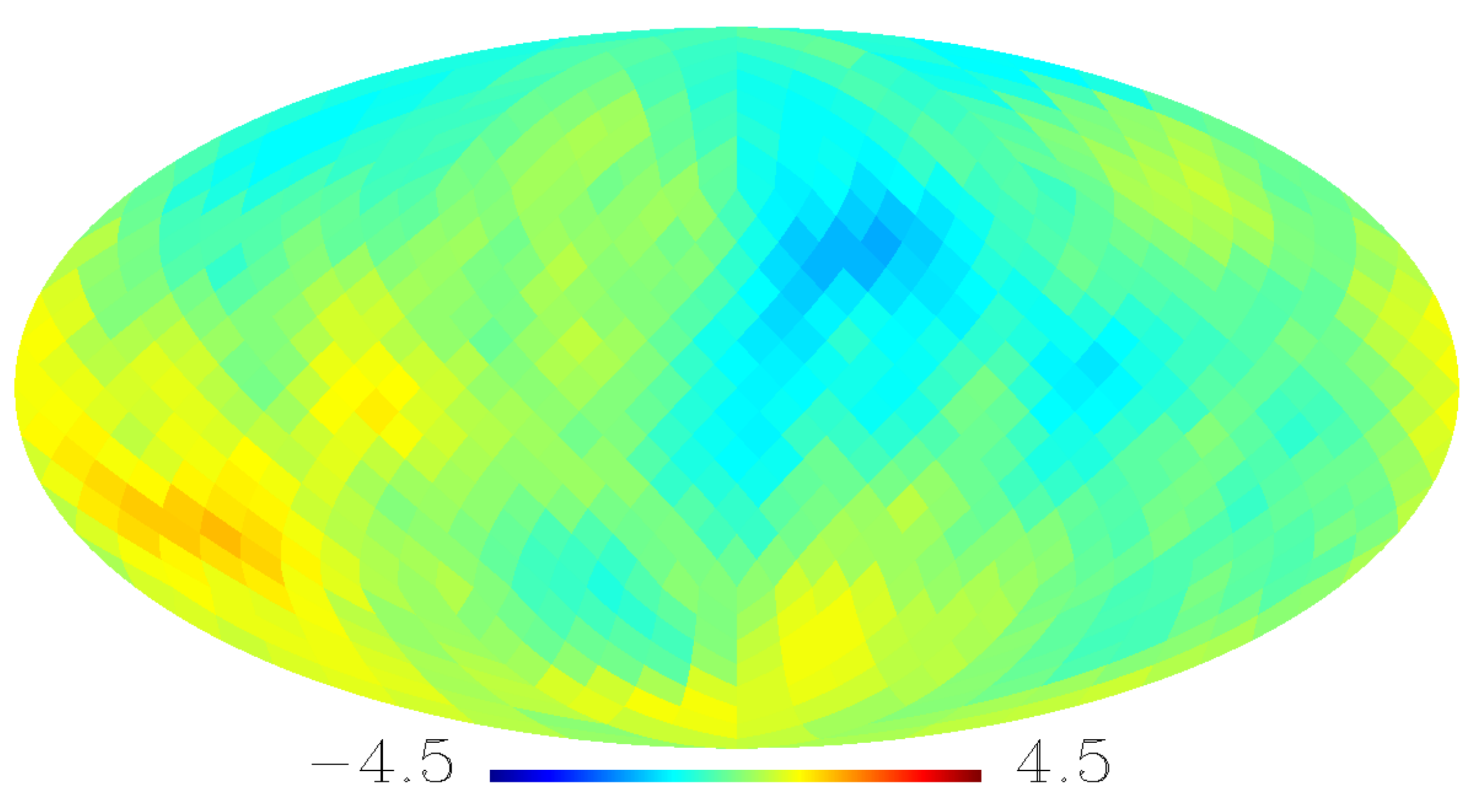}
\includegraphics[width=3.4cm, keepaspectratio=true]{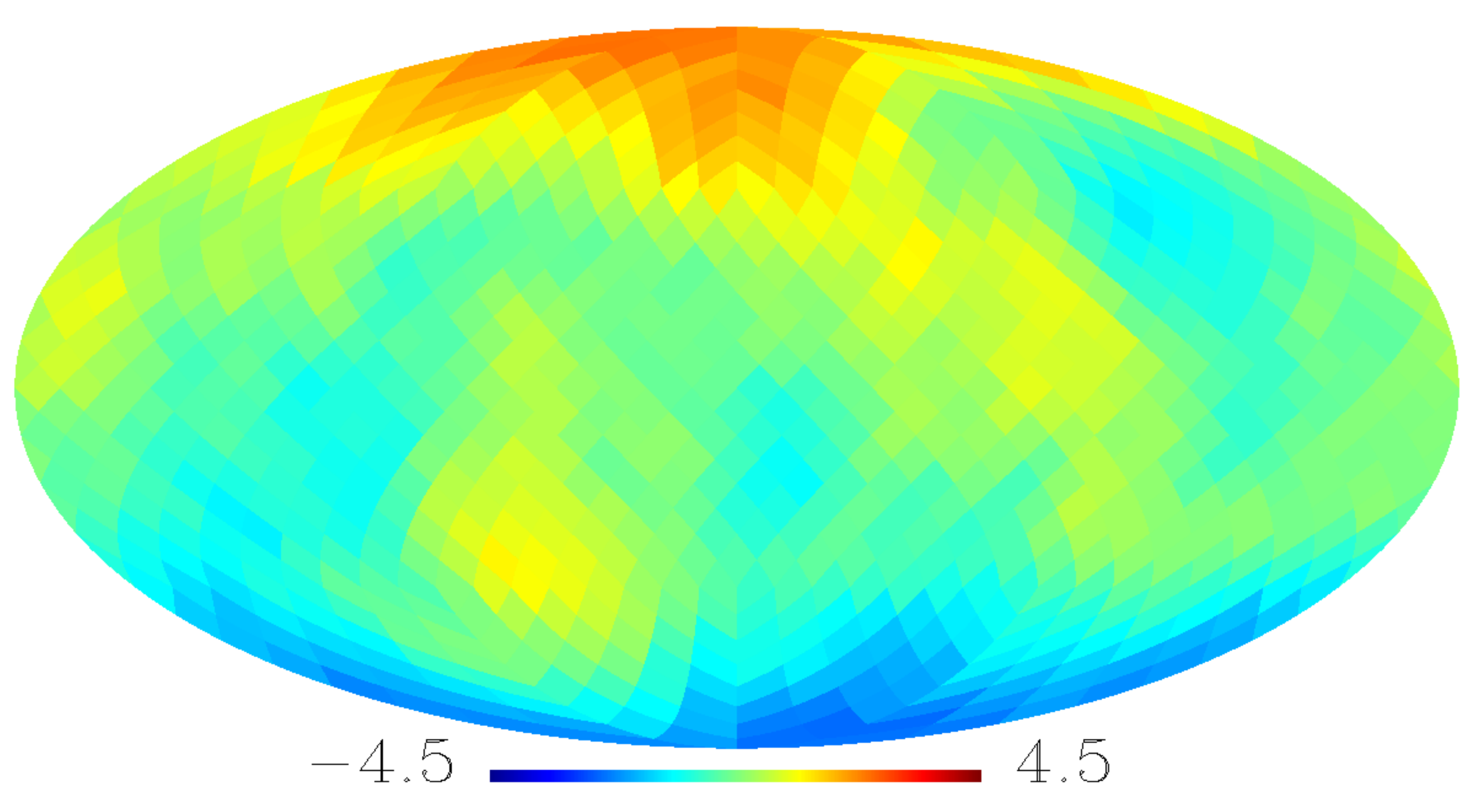}

\includegraphics[width=3.4cm, keepaspectratio=true]{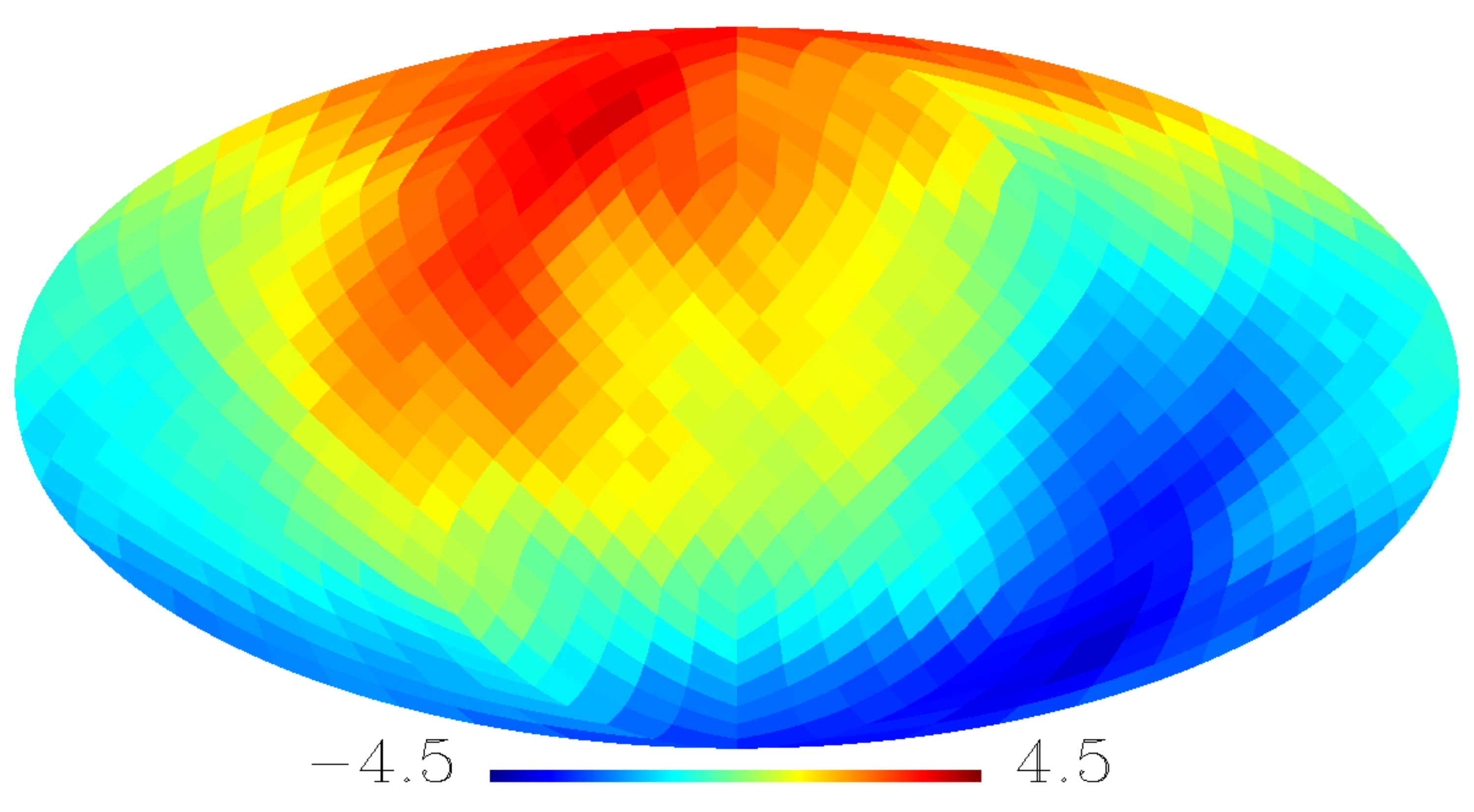}
\includegraphics[width=3.4cm, keepaspectratio=true]{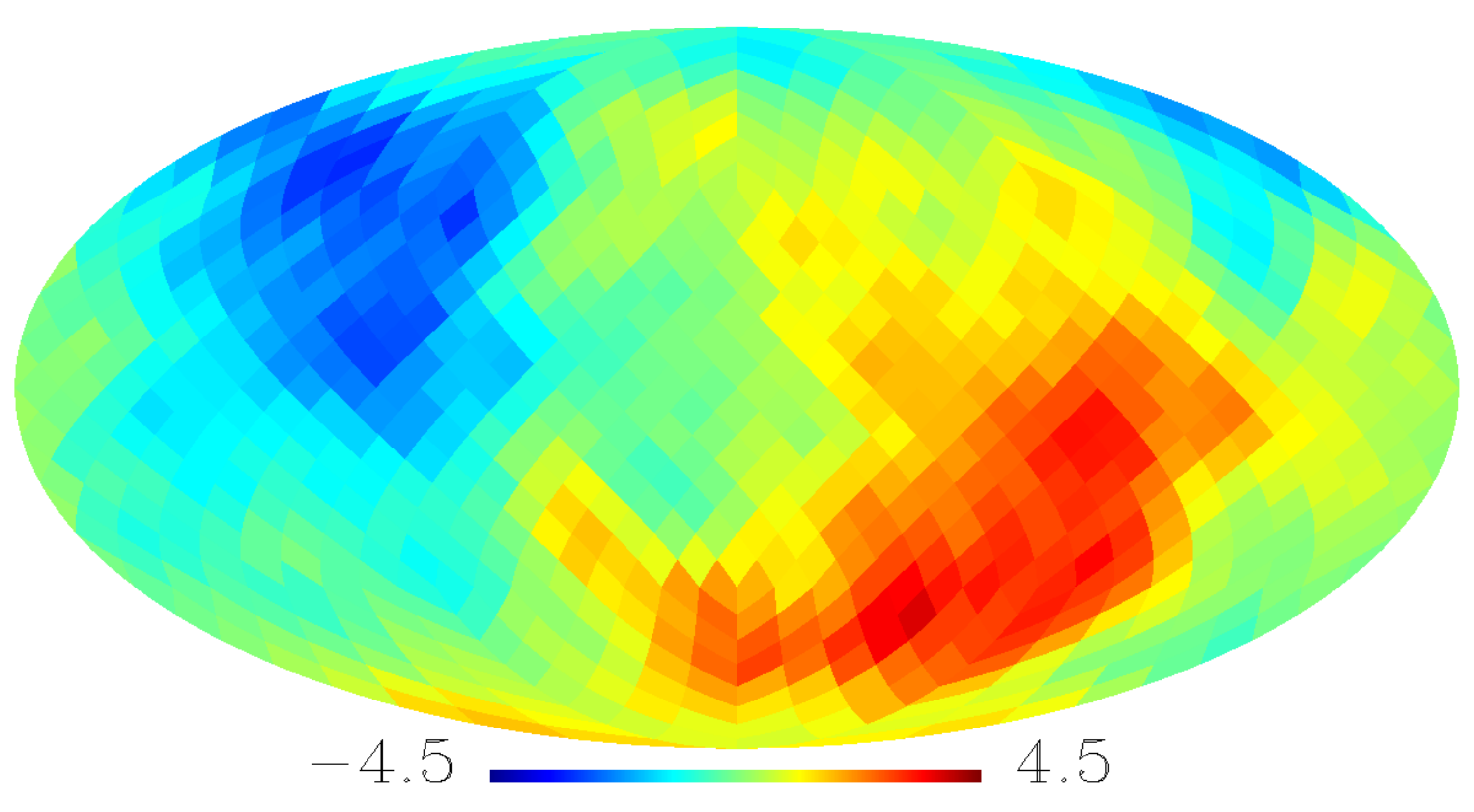} \hspace{0.25cm}
\includegraphics[width=3.4cm, keepaspectratio=true]{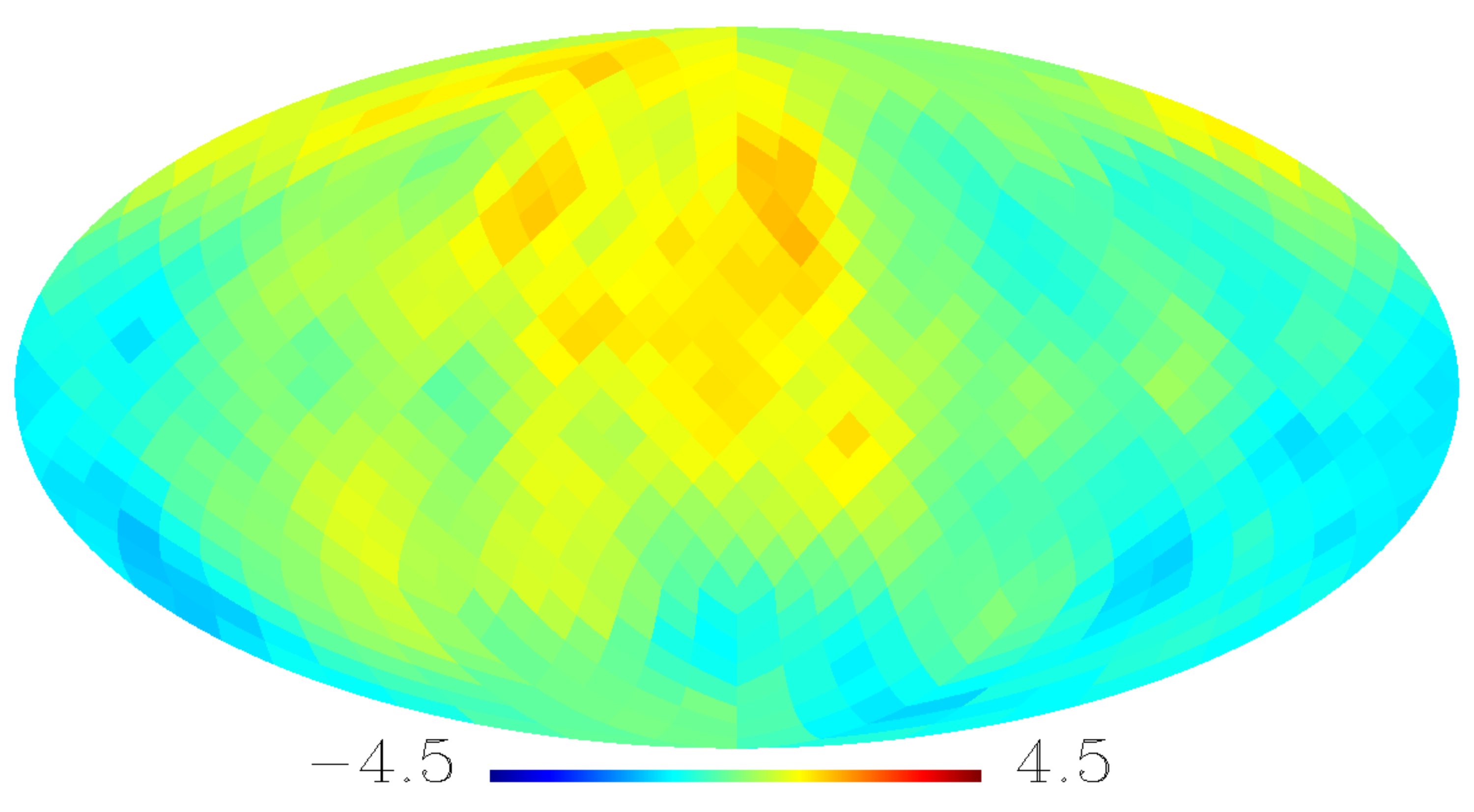}
\includegraphics[width=3.4cm, keepaspectratio=true]{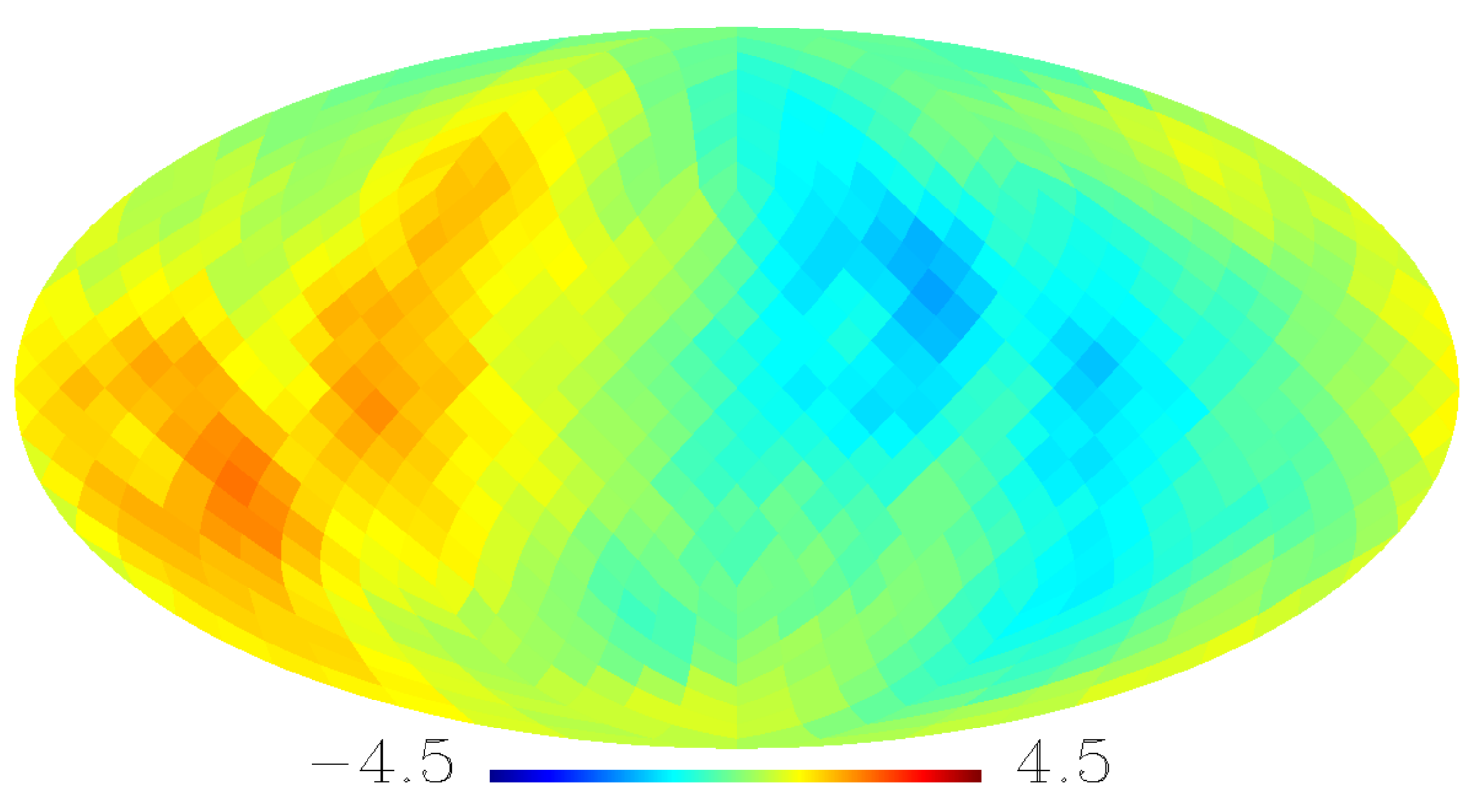}
\includegraphics[width=3.4cm, keepaspectratio=true]{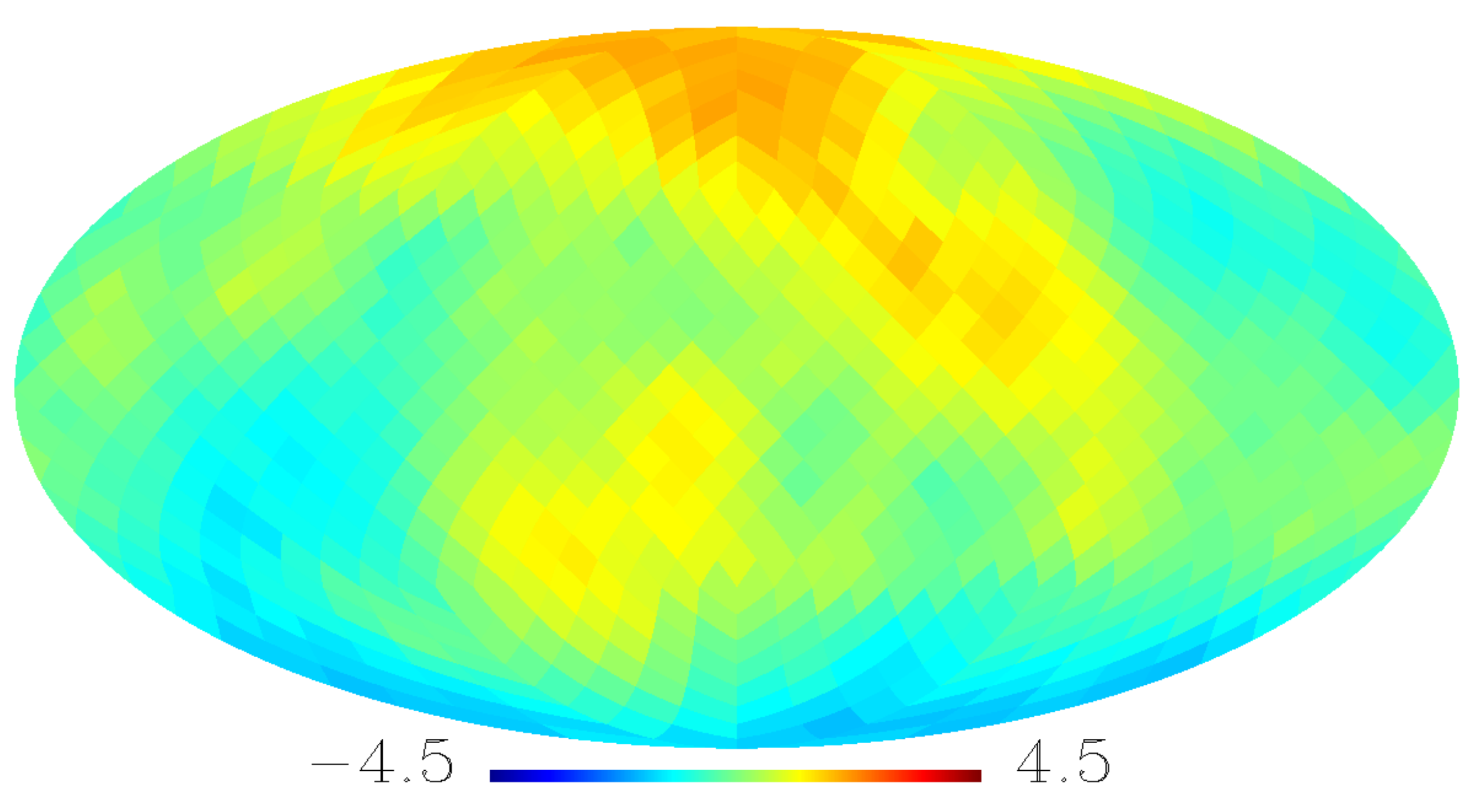}

\caption{Deviations $S(\langle \alpha (r_{k}) \rangle)$ for the three scales $r_{k}, k=2,6,10$ 
(from top to bottom) for $\Delta l = [2,20]$. The results are shown for (from left to right) the UILC7 map, 
the difference map  7yr ILC - 6yr  ILC map, the asymmetric beam map,
the coadded V and W-band from a standard simulation and the simulated ILC-like  map 
(for more detailed information about the different maps see text).
} \label{fig8}
\end{figure*}

\begin{figure*}
\centering

\includegraphics[width=3.4cm, keepaspectratio=true]{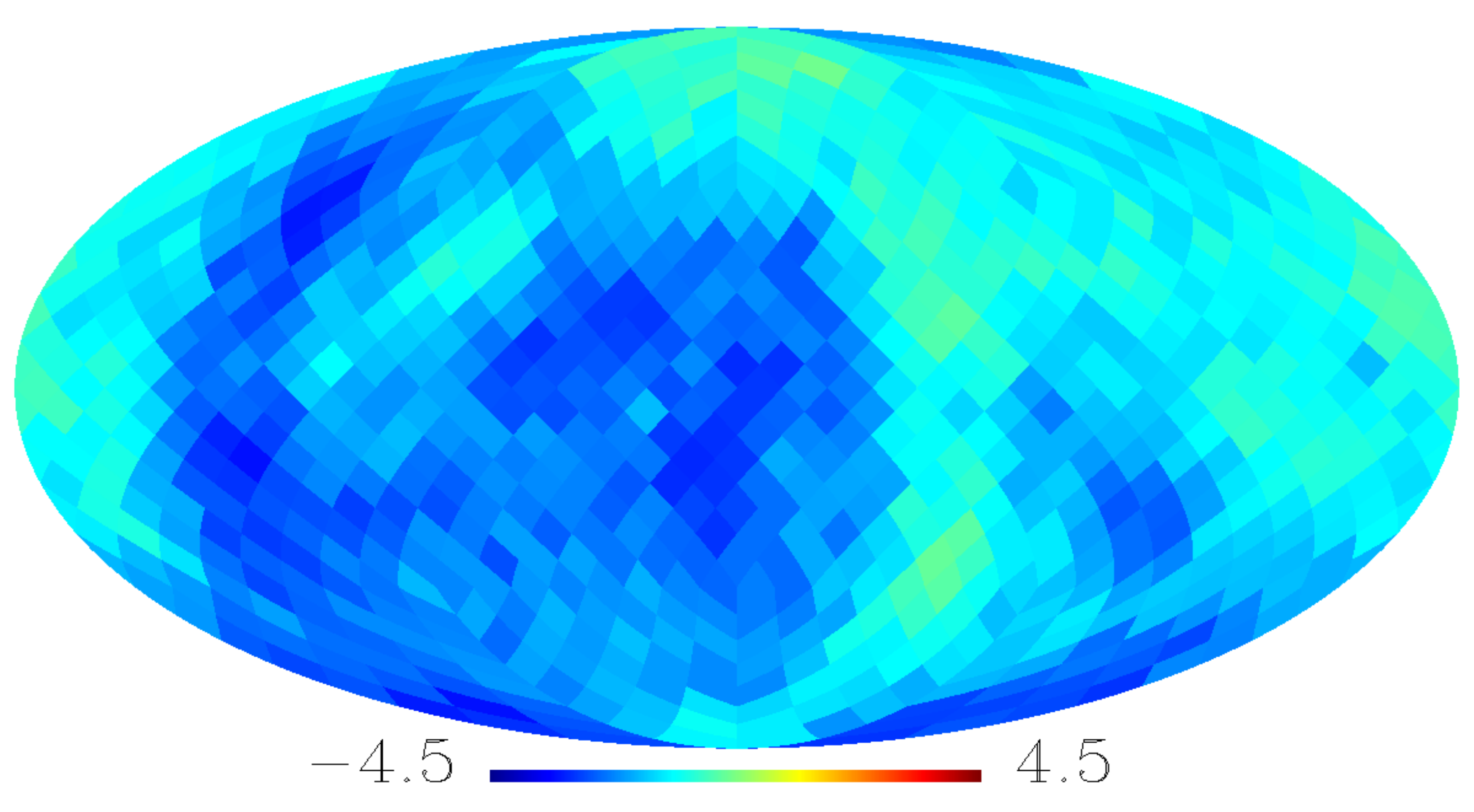}
\includegraphics[width=3.4cm, keepaspectratio=true]{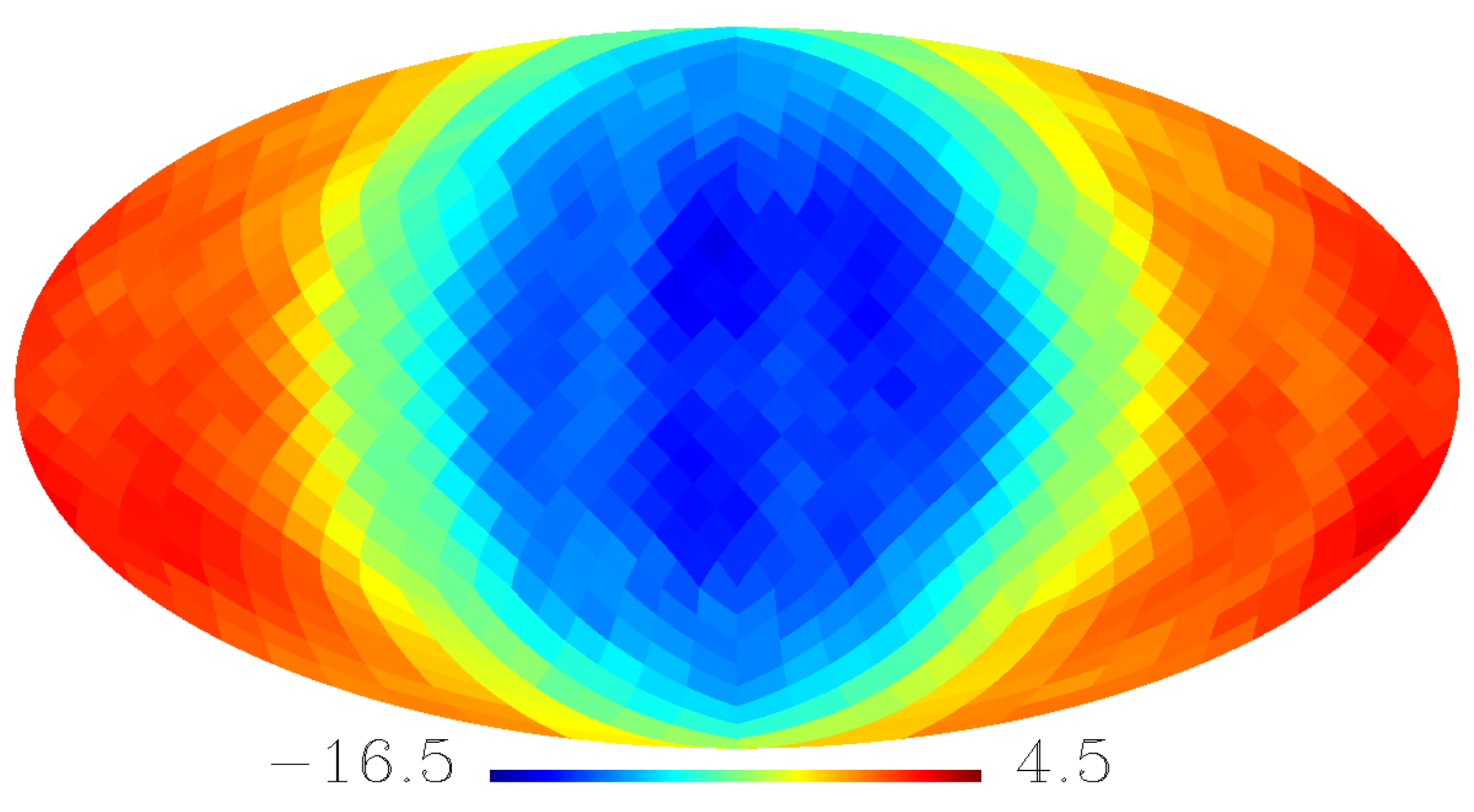} \hspace{0.25cm}
\includegraphics[width=3.4cm, keepaspectratio=true]{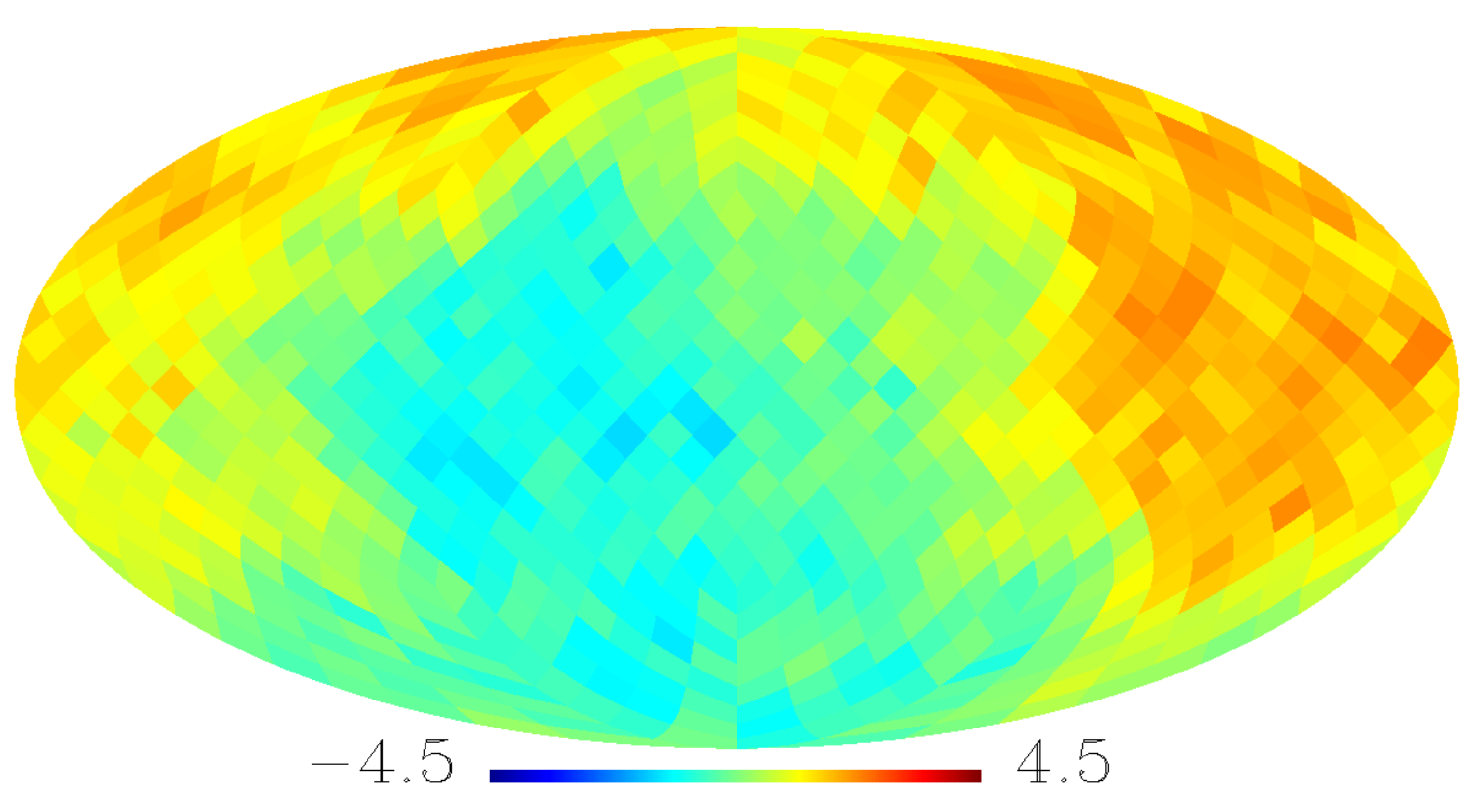}
\includegraphics[width=3.4cm, keepaspectratio=true]{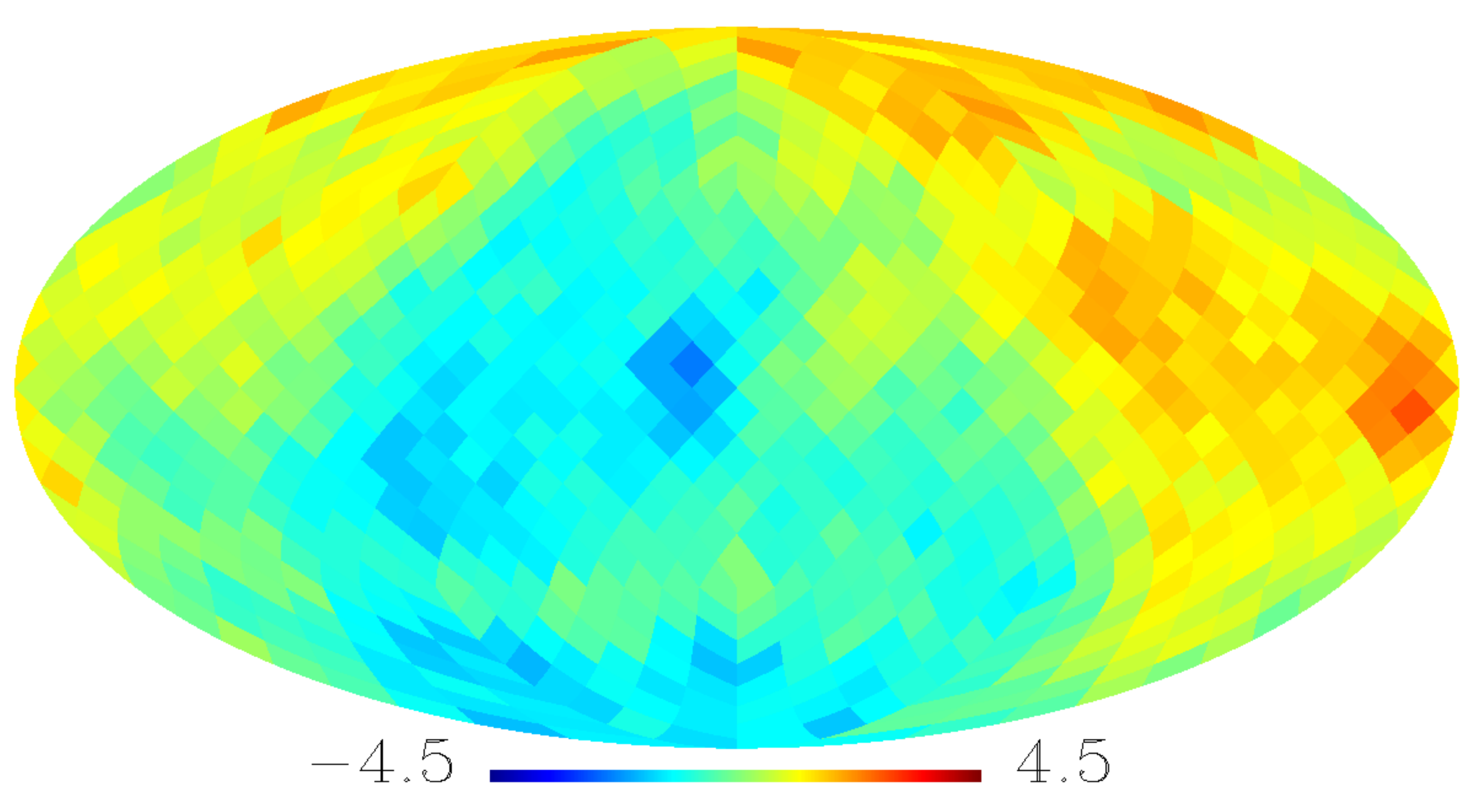}
\includegraphics[width=3.4cm, keepaspectratio=true]{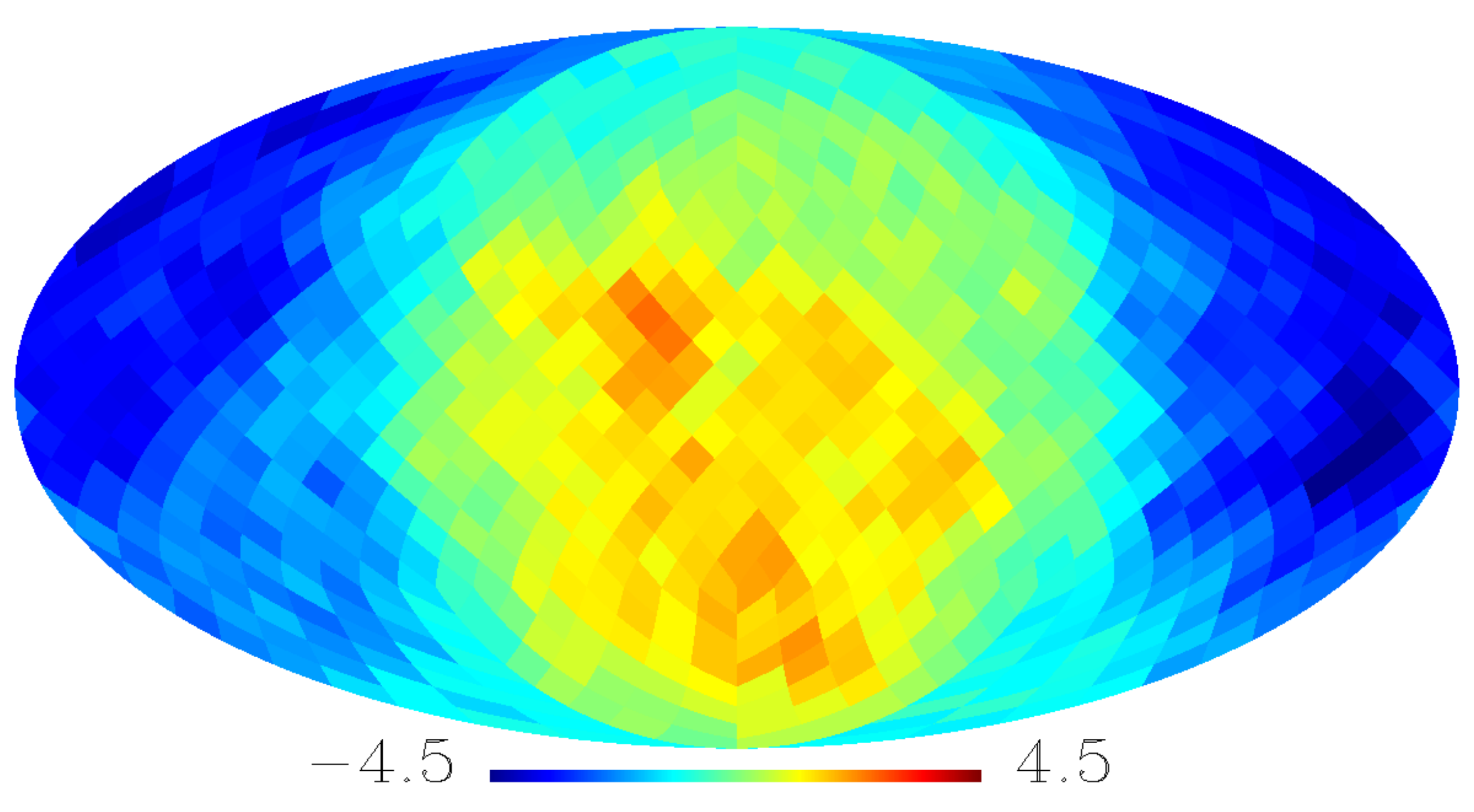}

\includegraphics[width=3.4cm, keepaspectratio=true]{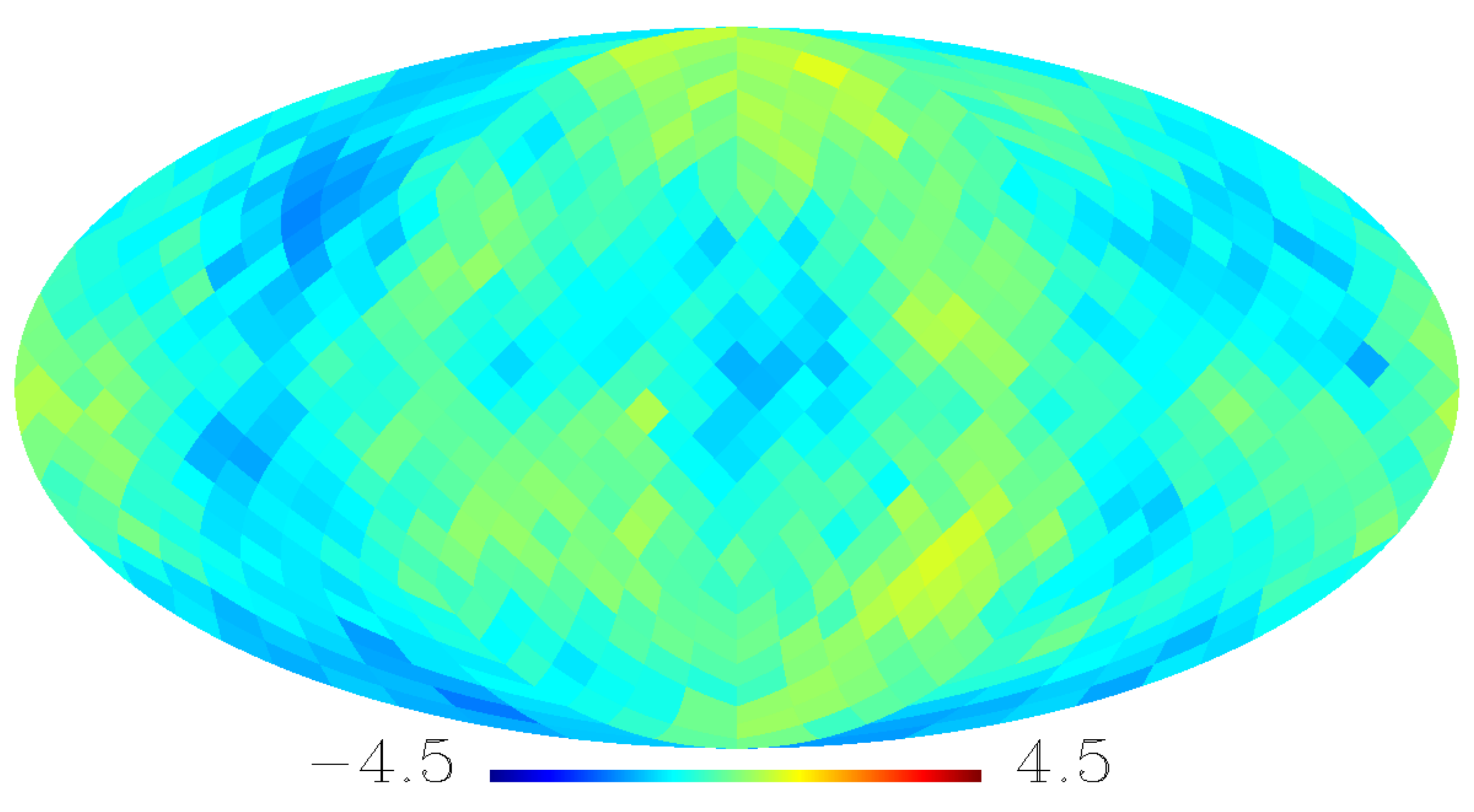}
\includegraphics[width=3.4cm, keepaspectratio=true]{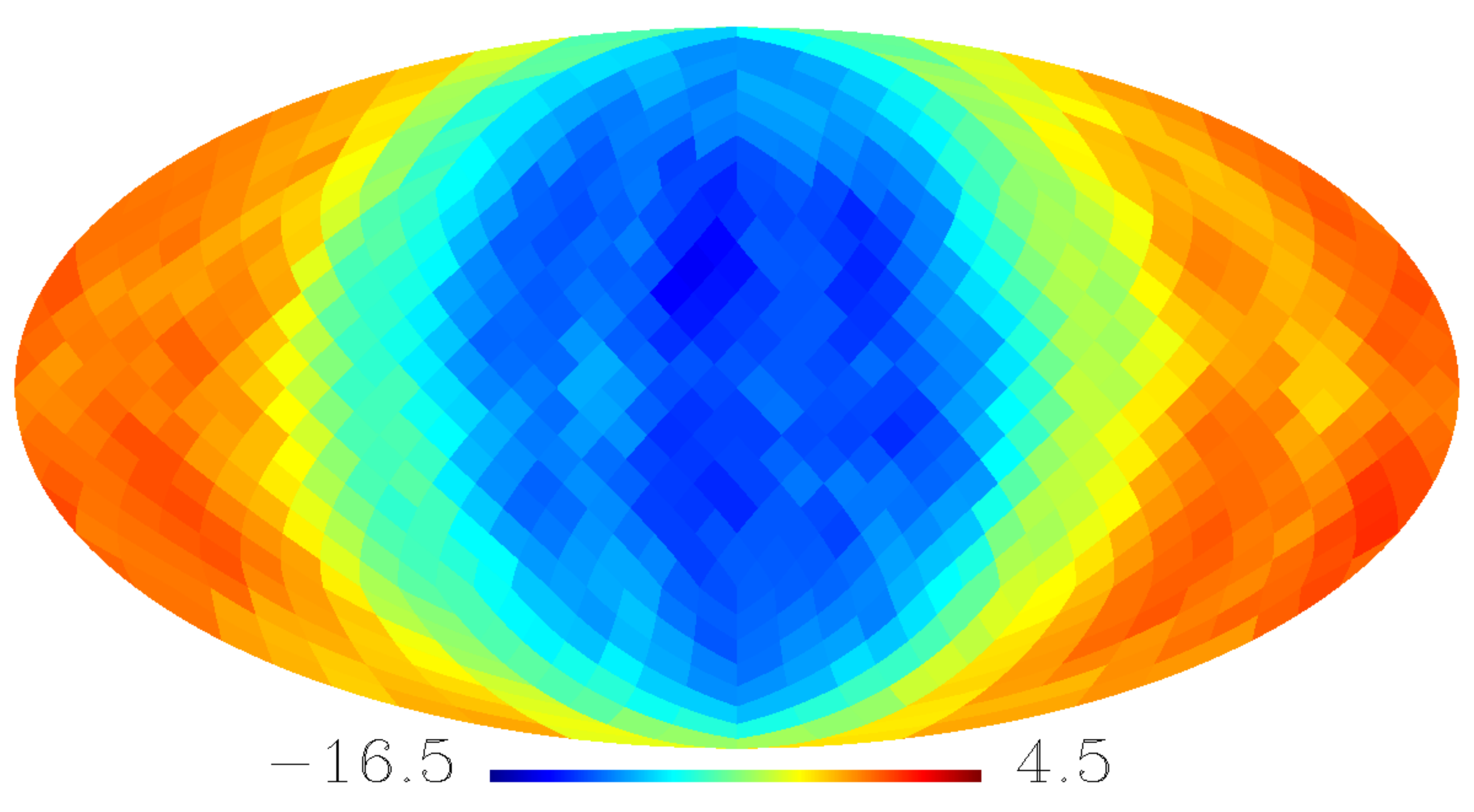} \hspace{0.25cm}
\includegraphics[width=3.4cm, keepaspectratio=true]{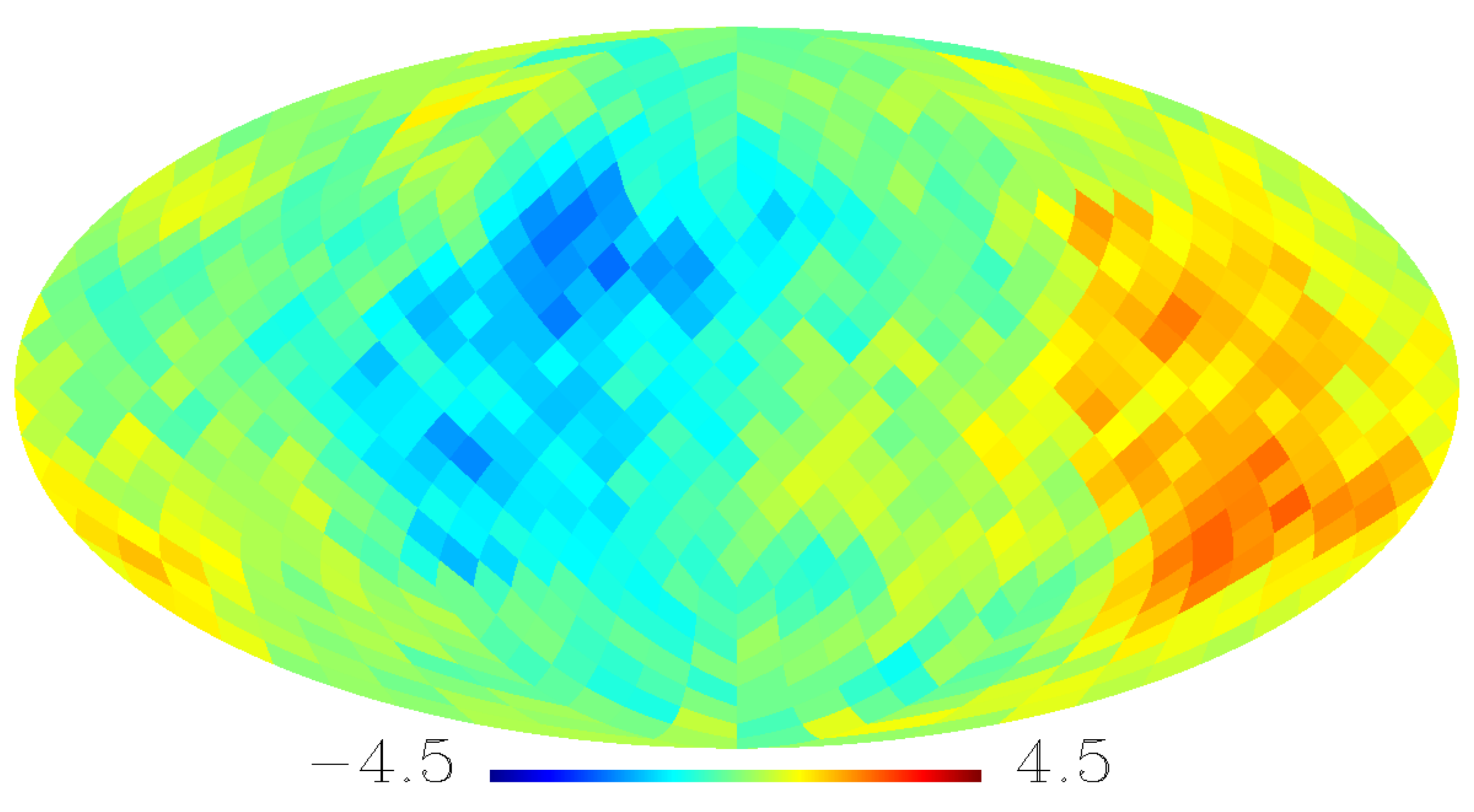}
\includegraphics[width=3.4cm, keepaspectratio=true]{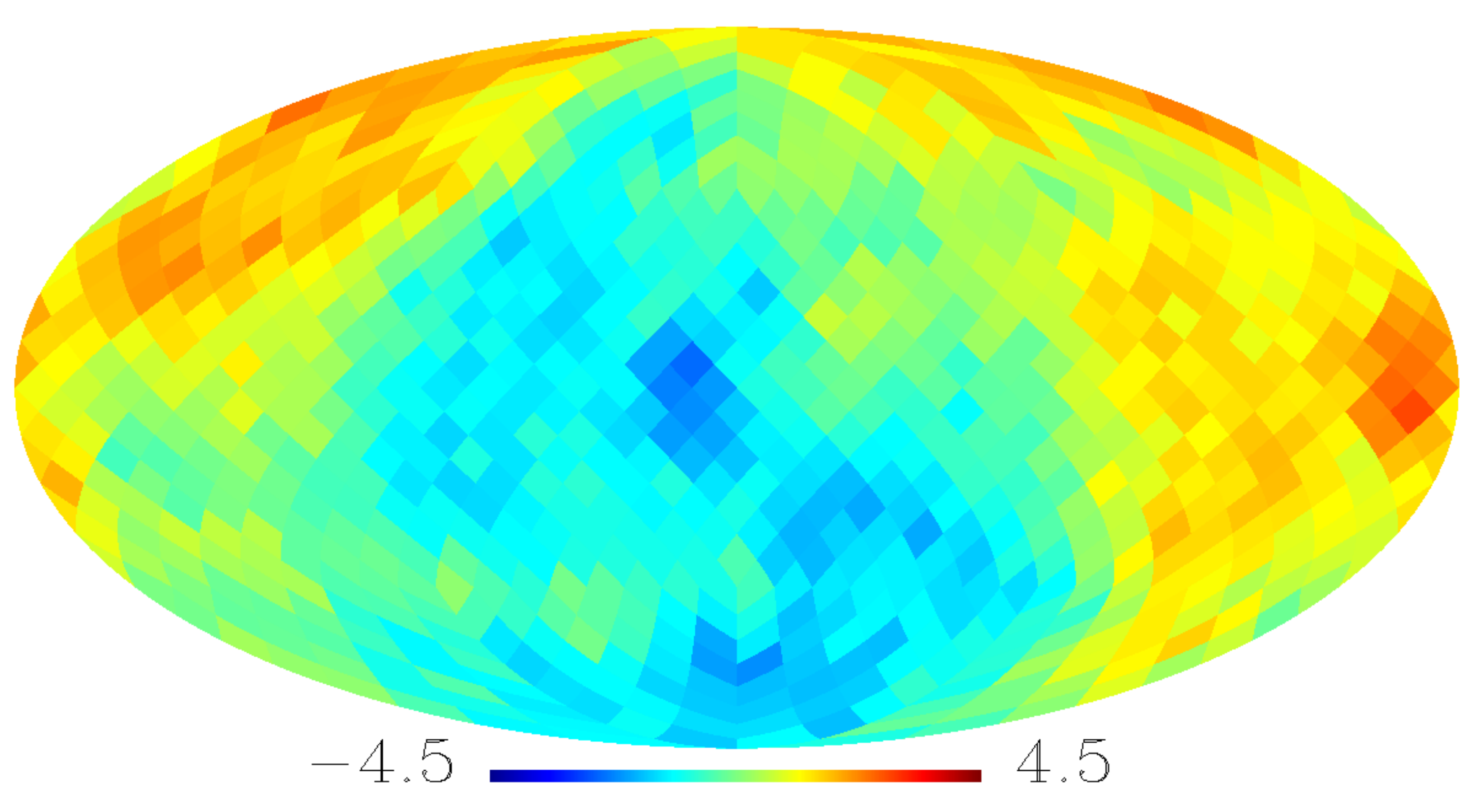}
\includegraphics[width=3.4cm, keepaspectratio=true]{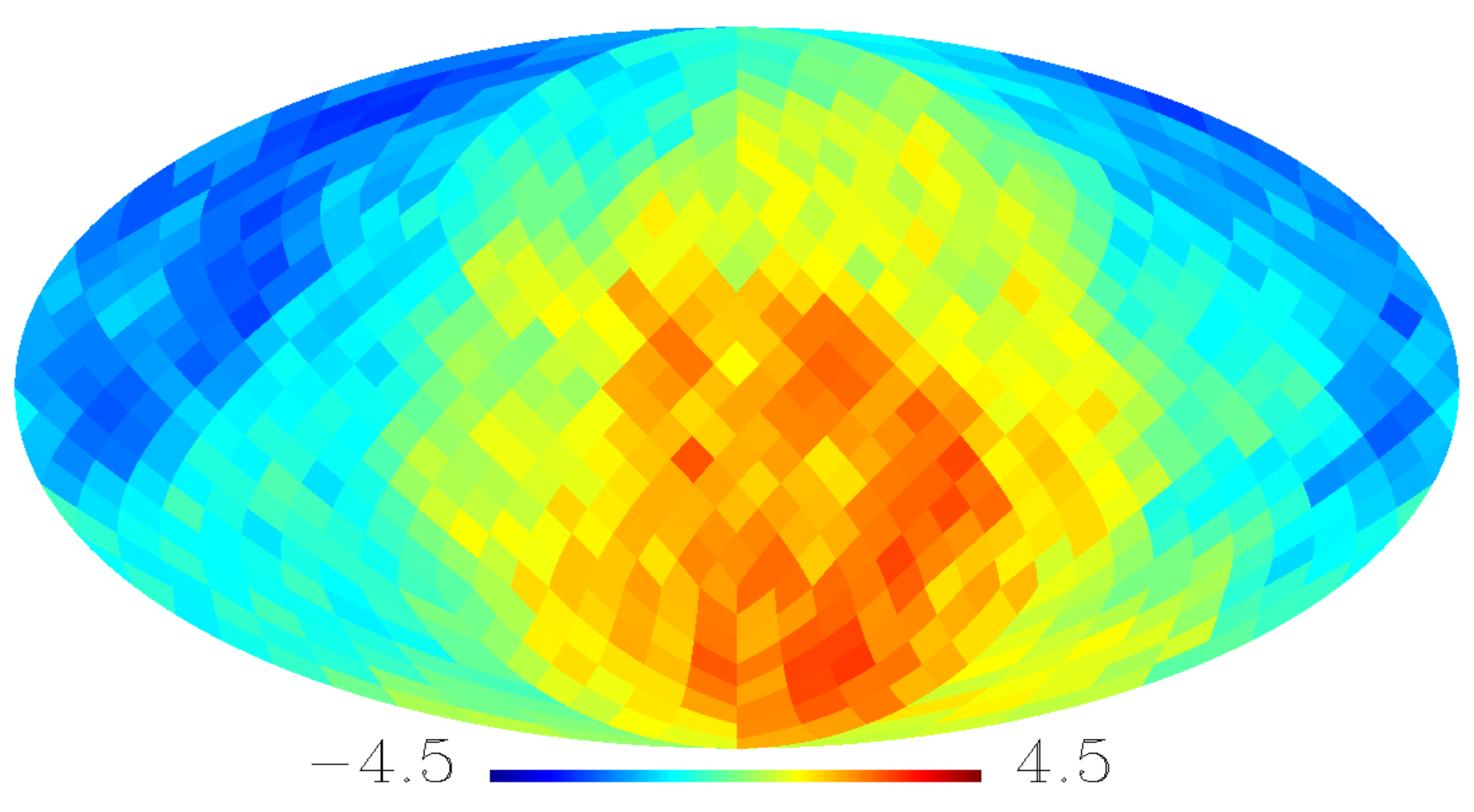}

\includegraphics[width=3.4cm, keepaspectratio=true]{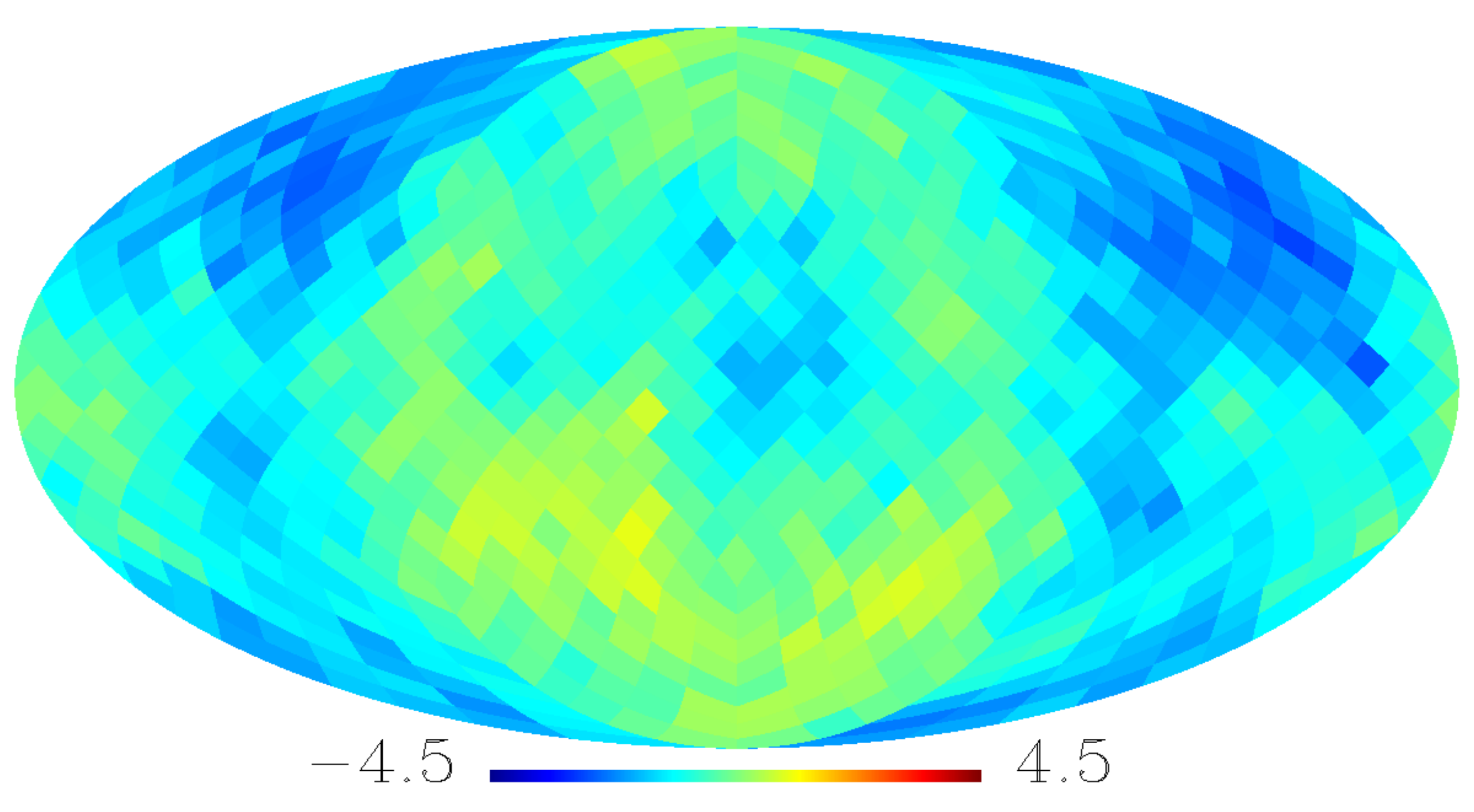}
\includegraphics[width=3.4cm, keepaspectratio=true]{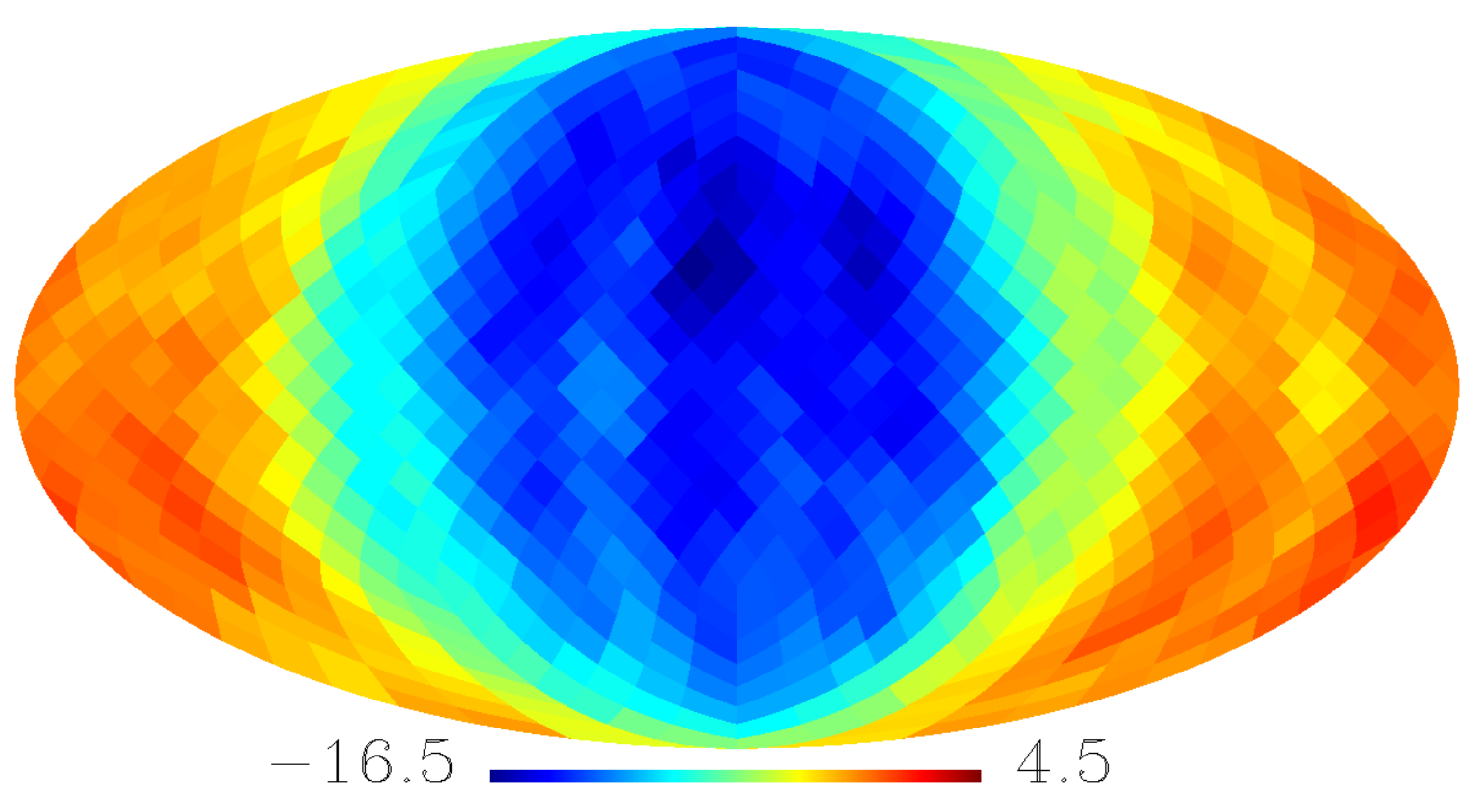} \hspace{0.25cm}
\includegraphics[width=3.4cm, keepaspectratio=true]{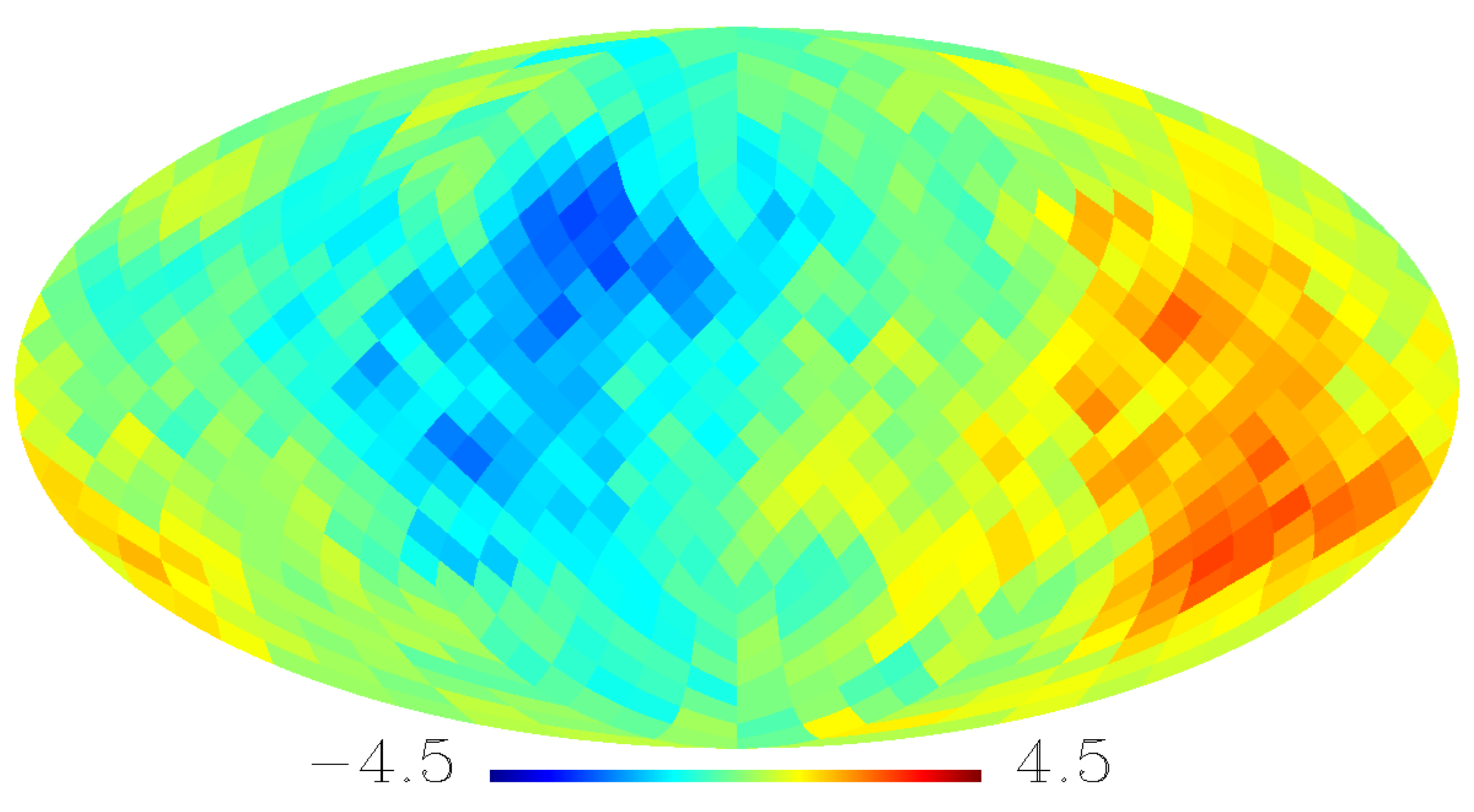}
\includegraphics[width=3.4cm, keepaspectratio=true]{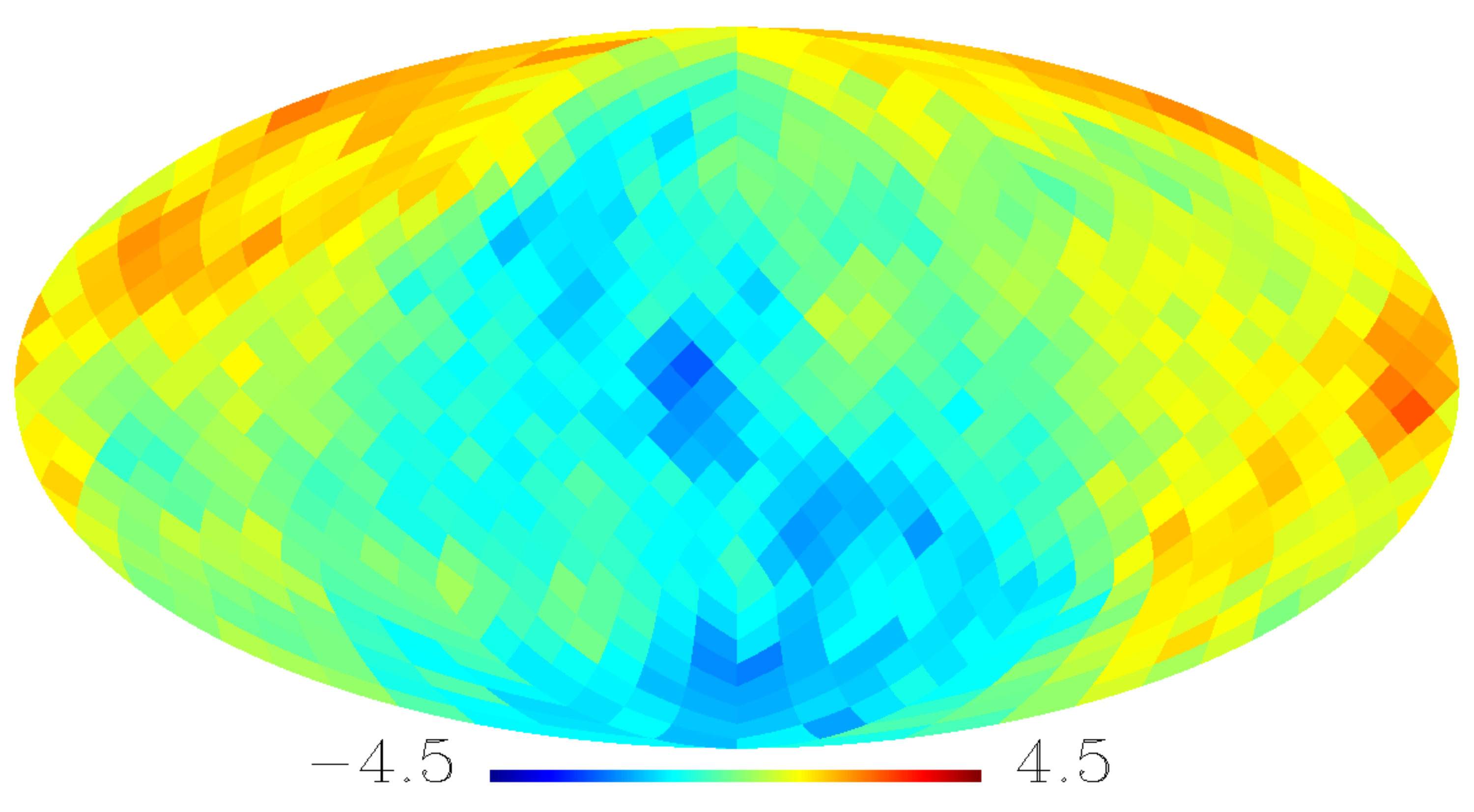}
\includegraphics[width=3.4cm, keepaspectratio=true]{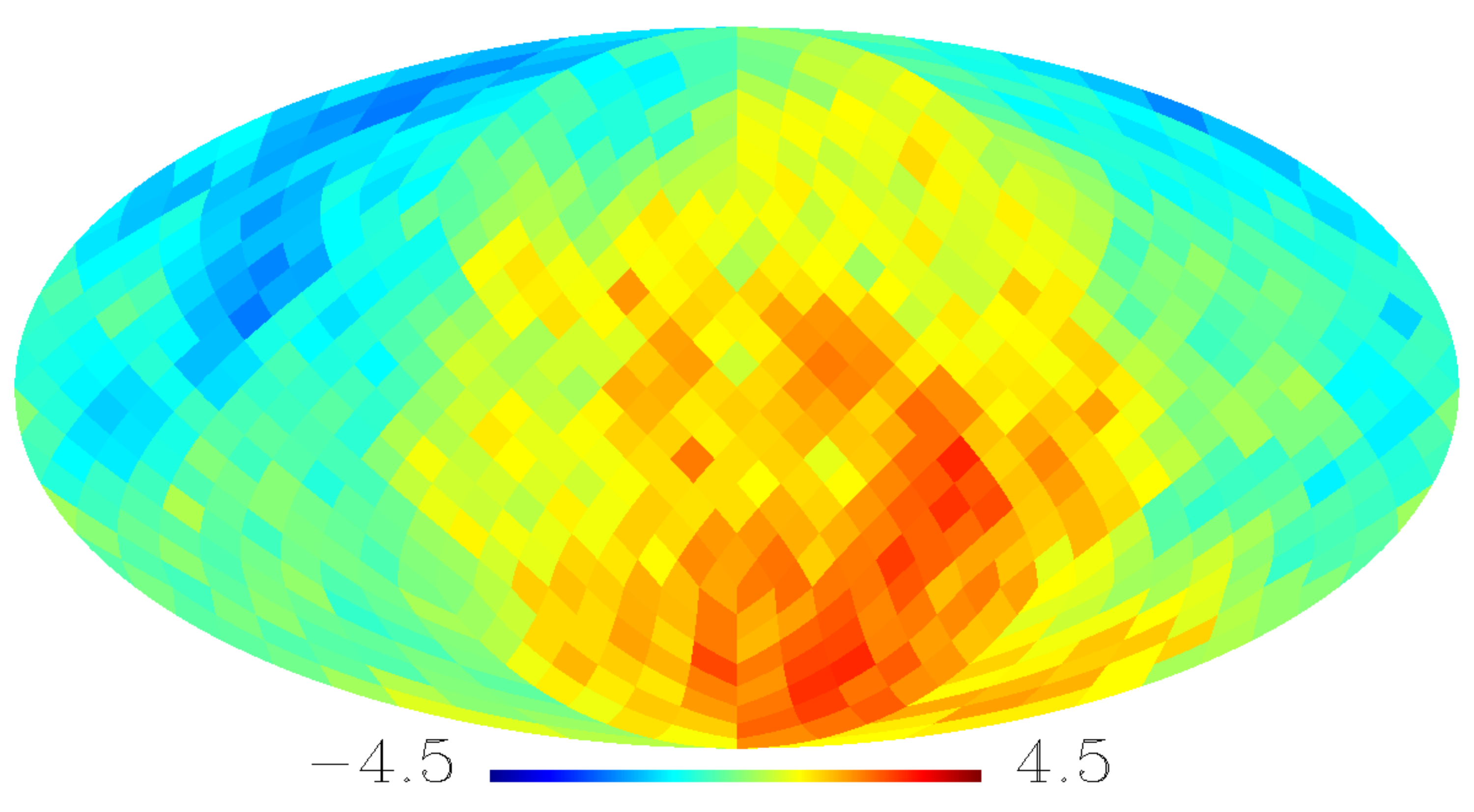}

\caption{Same as Fig. \ref{fig8} but for $\Delta l = [120,300]$. Note that the scale for the color coding has significantly changed for the
difference map (second column).} \label{fig9}
\end{figure*}

The main results which were already briefly discussed on the basis of figs. 1 become much more apparent 
when interpreting fig. \ref{fig6} and tables   \ref{table1}  to  \ref{table3}. 
We find stable $3.7 - 12 \sigma$ deviations for all $r$-values for $S(\langle \alpha (r_{k}) \rangle)$ 
and the scale-independent surrogate test when considering 
the full sky. This yields $S$-values of $S(\langle \alpha (r_{2}) \rangle)=7.73$ (ILC7 map) and  
$S(\langle \alpha (r_{2}) \rangle)=11.06$ (NILC5 map) for the scaling indices calculated for the small value $r_2$ and 
$S(\langle \alpha (r_{10}) \rangle)=3.75$ (ILC7 map) and  
$S(\langle \alpha (r_{10}) \rangle)=5.77$ (NILC5 map) for the largest radius $r_{10}$.
This stable $r$-independent effect leads to very high values of  the deviations $S$ for the 
scale-independent $\chi^2$-statistics $S(\chi^2_{\langle \alpha  \rangle})$, where we
find $S(\chi^2_{\langle \alpha  \rangle})=5.73$ (ILC7 map) and  
$S(\chi^2_{\langle \alpha  \rangle})=27.93$ (NILC5 map). It is interesting to compare these results
with those obtained for the diagonal $\chi^2$-statistics. In this case we find 
$S(\chi^2_{\langle \alpha  \rangle})=57.32$ (ILC7 map) and  
$S(\chi^2_{\langle \alpha  \rangle})=119.16$ (NILC5 map), which is up to an order of magnitude larger
than the values for the full $\chi^2$-statistics.
These results are very remarkable, since they represent -- to the best of our knowledge -- by far 
the most significant detection of non-Gaussianities in the WMAP data to date.
Note that we used here only the mean value of the distribution of scaling indices, 
which is a robust statistics not being sensitive to contributions of some spurious outliers. Further, the scale-independent 
statistics $\chi^2_{\langle \alpha  \rangle}$ calculated for the full sky represents a rather
unbiased statistical approach.\\
The hemisperical asymmetry for NGs on large 
scale ($\Delta l = [2,20]$) finds its reflection 
in the results of $S(r)$. While we calculate significant and stable deviations $S$ for the
upper and lower hemispheres separately (red and blue lines) in fig. \ref{fig6}, the results 
for the full sky (black lines) are not significant, because the deviations detected in the 
two hemispheres are complementary and thus cancel each other. Therefore, we obtain 
only for the hemispheres high values for $S$ ranging from $S=3.24$ up to $S = 7.10$ 
($S=4.11$ up to $S = 10.82$) for the ILC7 (NILC5) map when considering the statistics 
derived from the scaling indices for the largest scales $r_{10}$ and $S=4.01$ up to $S = 9.76$
for the scale-independent $\chi^2$-statistics.\\ 
For the smallest scales considered so far ($\Delta l=[120,300]$) we also find significant 
deviations from non-Gaussianity being much more isotropic and naturally more 
pronounced at smaller scaling ranges $r < 0.15$. Thus,  we obtain $S = 6.97$ (ILC7 map) and $S=5.30$ (NILC5 map)
for $S(\langle \alpha (r_{2}) \rangle)$ considering the full sky. For the scale-independent $\chi^2$-statistics the most 
significant signatures of NGs are detected for the respective upper hemispheres  ranging from $S=5.16$ to $S=10.53$.
To test whether all these signatures are of intrinsic cosmic origin or more likely due to foregrounds or 
systematics induced by e.g. asymmetric beams 
or map making, we performed the same surrogate and scaling indices analysis for the five additional maps 
described in Section 2. Figs. \ref{fig8} and   \ref{fig9} show the significance maps for the 
two $l$-ranges $\Delta l = [2,20]$ and $\Delta l = [120,300]$, for which we found 
the most pronounced signatures in the ILC7 and NILC5 map.
For the large scale NGs we find essentially the same results for the UILC7 map. 
The difference map, shows some signs of NGs and asymmetries, especially for large $r$-values.
A closer look reveals, however, that both the numerator and denominator in the equation for $S$
are an order of magnitude smaller than the values obtained for the ILC7 (NILC5) maps. Thus the signal 
of the difference map can be considered to be subdominant.
And even if it were not subdominant, the signal coming from the residuals would rather diminish the 
signal in the ILC map than increase its significance, because the foreground signal is spatially 
anticorrelated with the CMB-signal.
Both the asymmetric beam map 
and the simulated coadded VW-map do not show any significant signature for NGs and asymmetries. 
Finally, the simulated  ILC map does show some signs of (galactic) north-south asymmetries which become smaller and
therefore insignificant for increasing $r$, where we find the largest signal in the CMB maps.\\
For the small scale NGs  ($\Delta l = [120,300]$) we also find that the UILC7-map 
yields similar results as the ILC7 and NILC5 
map with smaller significance. Once again the asymmetric beam map 
and the simulated coadded VW-map do not show significant signature for NGs and  asymmetries. 
This is not the case for the simulated  ILC map. Here, we find highly significant signatures
for NGs and asymmetries, which show some similarities with significance patterns observed 
in the ILC7 (NILC5) map. 
Even much more striking features are detected in the difference map, where we find 
deviations as high as $|S| \approx 15$ forming a very peculiar pattern in the significance maps
for all $r$. One of us (G.R.) named this pattern 'Eye of Sauron', which we think is a nice and adequate association.
It is worth noticing that we found the same pattern when analyzing other difference maps, e.g. year 7 - year 1 or year 2 - year 1.\\â
To better understand, where these features may come from we had a closer look at the zeroth, first and second order
surrogate maps. 
It became immediately obvious that for the difference maps the fluctuations are systematically smaller
in the regions in the galactic plane used for the ILC-map making than in the rest of the sky. 
This effect persists in the first order surrogate map and is only destroyed in the second order surrogates.
This more (less) structure in first order surrogate map leads to lower (higher) values for the scaling indices, which can 
qualitatively explain the observed patterns in the significance maps.\\
A much more detailed study of these high $l$ effects and their possible origins is part of our current work but is beyond the 
scope of this paper. The results for the difference map shown here point, however, already towards a very
interesting application of the surrogate technique.  It may become a versatile tool to define criteria of  the 
cleanness of maps in the sense of e.g. absence of artificially induced (scale-dependent) NGs in the map of the residual signal.
Such a criterion may then in turn be implemented in the map making procedure so that ILC-like maps
are not only minimizing the overall quadratic error in the map, but also e.g. the amount of unphysical 
NGs of the foregrounds. 

\section{Conclusions}
 
To the best of our knowledge this work represents the first 
comprehensive study of scale-dependent 
non-Gaussianities in full sky CMB data as measured
with the WMAP satellite. By applying the method of surrogate maps, which 
explicitly relies on the scale-dependent shuffling of Fourier phases while 
preserving all other properties of the map, we
find highly significant signatures of non-Gaussianities for very large scales
and for the $l$-interval covering the first peak in the power spectrum.
In fact, our analyses yield by far the most significant evidence 
of non-Gaussianities in the CMB data to date.
Thus, it is no longer the question whether there are
phase correlations in the WMAP data. It is rather to be figured out what the 
origin of these scale-dependent non-Gaussian signatures is. The checks on 
systematics we performed so far revealed that no clear candidate can be found 
to explain the low-$l$ signal, which we take to be cosmological at high significance. 
These findings would strongly disagree
with predictions of isotropic cosmologies with single field slow roll inflation.\\
The picture is not that clear for the 
signatures found at smaller scales, i.e. at higher $l$'s. In this case we found that NGs can also easily be 
induced by the ILC map making procedure so that it is difficult to disentangle 
possible intrinsic anomalies from effects induced by the preprocessing of 
the data. More tests are required to further pin down the origin of the detected
high $l$ anomalies and to probably uncover yet unknown systematics being responsible 
for the low $l$ anomalies.  Another way of ruling out effects of unknown systematics is to perform 
an independent observation preferably via a different instrument as we are now able to do with the 
Planck satellite.\\  
In any case our study has shown that the method of surrogates in conjunction with sensitive 
higher order statistics offers the potential
to become an important tool not only for the detection of scale-dependent non-Gaussianity
but also for the assessment of possibly induced artefacts leading to NGs in the residual map 
which in turn may have important consequences for the map making procedures.

\section*{Acknowledgments}

Many of the results in this paper have been derived using the HEALPix \citep{gorski05a} software 
and analysis package. We acknowledge use of
the Legacy Archive for Microwave Background Data Analysis (LAMBDA). 
Support for LAMBDA is provided by the NASA Office of Space Science.


\label{lastpage}

\end{document}